\documentclass[letterpaper,onecolumn,12pt]{article}
\usepackage{amsmath}
\usepackage{amssymb}
\usepackage{amsfonts}
\usepackage[bookmarks=true, bookmarksnumbered=true,allbordercolors={1 1 1}]{hyperref}
\usepackage[open=true, numbered=true]{bookmark}
\usepackage{placeins}
\usepackage[cal=boondoxo]{mathalfa}
\usepackage{tikz}
\usetikzlibrary{fit,positioning}
\usepackage{lscape}
\usepackage{longtable}
\usepackage{epstopdf}
\usepackage{float}
\usepackage{setspace}
\usepackage{euscript}
\usepackage{textcomp}
\usepackage{algorithm}
\usepackage[end]{algpseudocode}
\algnewcommand\algorithmicto{\textbf{to }}

\usepackage[nameinlink]{cleveref}
\Crefname{figure}{Fig.}{Figures}

\usepackage{etoolbox}
\patchcmd{\thebibliography}{\section*{\refname}}{}{}{}

\newtheorem{theorem}{Theorem}[section]

\newtheorem{lemma}[theorem]{Lemma}
\newcommand*{\QEDA}{\hfill\ensuremath{\blacksquare}}

\begin{document}

\title{\LARGE{High-dimensional macroeconomic forecasting using message passing algorithms}\thanks{This paper has significantly improved based on suggestions from two anonymous referees, an Associate Editor, and the Editor (Todd Clark), whom I gratefully acknowledge. I would like to thank Fabio Canova, Eric Ghysels, George Kapetanios, Gary Koop, Massimiliano Marcellino, Geert Mesters, Davide Pettenuzzo, Giorgio Primiceri, Giuseppe Ragusa, Giovanni Ricco, Lucrezia Reichlin, Frank Schorfheide, Herman van Dijk, Hal Varian and Mike West for useful comments and/or stimulating discussions. I would also like to thank Joshua Chan, and Maria Kalli/Jim Griffin for sharing MATLAB codes that replicate time-varying parameter specifications suggested by these authors. Additionally, I would like to acknowledge helpful comments and questions from participants at the 2017 Deutsche Bundesbank Forecasting Workshop; the 2017 Norges Bank conference on ``Big data, Machine Learning and the Macroeconomy''; the 10th ECB Workshop on Forecasting Techniques: ``Economic Forecasting with Large Datasets''; the BayesComp / ISBA 2018 conference; the 3rd annual Now-Casting.com Workskhop; the 2018 Budapest School for Central Bank Studies; the 2019 Econometric Institute workshop ``Machine Learning Meets Econometrics''; and the 2019 Joint Research Centre European Commission workshop ``Big Data and Economic Forecasting''. Finally, I would like to acknowledge useful feedback from seminar participants at various Universities and central banks.}}
\author{Dimitris Korobilis\thanks{%
Adam Smith Business School, Glasgow, G12 8QQ, United Kingdom.
\href{mailto: dikorobilis@googlemail.com}{%
\nolinkurl{dikorobilis@googlemail.com}}} \\
\textit{University of Glasgow}}

\date{ }

\maketitle

\begin{abstract}
\noindent This paper proposes two distinct contributions to econometric analysis of large information sets and structural instabilities. First, it treats a regression model with time-varying coefficients, stochastic volatility and exogenous predictors, as an equivalent high-dimensional static regression problem with thousands of covariates. Inference in this specification proceeds using Bayesian hierarchical priors that shrink the high-dimensional vector of coefficients either towards zero or time-invariance. Second, it introduces the frameworks of factor graphs and message passing as a means of designing efficient Bayesian estimation algorithms. In particular, a Generalized Approximate Message Passing (GAMP) algorithm is derived that has low algorithmic complexity and is trivially parallelizable. The result is a comprehensive methodology that can be used to estimate time-varying parameter regressions with arbitrarily large number of exogenous predictors. In a forecasting exercise for U.S. price inflation this methodology is shown to work very well. 
\end{abstract}

\medskip

\emph{Keywords:} high-dimensional inference; factor graph; Belief Propagation; Bayesian shrinkage; time-varying parameter model

\medskip

\emph{JEL Classification:}\ C11, C22, C52, C55, C61

\newpage
\doublespacing
\section{Introduction}
As a response to the increasing linkages between the macroeconomy and the financial sector, as well as the expanding interconnectedness of the global economy, empirical macroeconomic models have increased both in complexity and size. For that reason, estimation of modern models that inform macroeconomic decisions -- such as linear and nonlinear versions of dynamic stochastic general equilibrium (DSGE) and vector autoregressive (VAR) models -- many times relies on Bayesian inference via powerful Markov chain Monte Carlo (MCMC) methods.\footnote{See Herbst and Schorfheide (2015) and Koop and Korobilis (2010) for detailed discussion of Bayesian computation in DSGE and VAR models, respectively.} However, existing posterior simulation algorithms cannot scale up to very high-dimensions due to the computational inefficiency and the larger numerical error associated with repeated sampling via Monte Carlo; see Angelino et al. (2016) for a thorough review of such computational issues from a machine learning and high-dimensional data perspective. In that respect, while Bayesian inference is a natural probabilistic framework for learning about parameters by utilizing all information in the data likelihood and prior, computational restrictions might make it less suitable for supporting real-time decision-making in very high dimensions.

This paper introduces to the econometric literature the framework of factor graphs (Kschischang et al., 2001) for the purpose of designing computationally efficient, and easy to maintain, Bayesian estimation algorithms. The focus is not only on ``faster'' posterior inference broadly interpreted, but on designing algorithms that have such low complexity that are future-proof and can be used in high-dimensional econometric problems with possibly thousands or millions of coefficients. While a graph, in general, is a structure that allows the representation of objects that are related in some sense\footnote{The most popular use of graphs in economics is to represent networks of agents, banks, social networks etc; see Jackson (2008).}, a factor graph representation of a high-dimensional vector of model parameters, in particular, depicts how each of its scalar elements is connected with each other based on the functional form of their joint posterior distribution. As a result, the factor graph representation provides a visual tool for the decomposition of a high-dimensional joint posterior distribution into smaller, tractable parts. By doing so, factor graphs can be used to design parallel versions of MCMC algorithms, as well as efficient iterative algorithms called \textit{message passing algorithms} -- the latter being the concept of interest in this paper.\footnote{Message passing algorithms are dynamic programming methods designed for efficiently performing large computations by distributing calculations among a number of simpler processors. Readers working with High-Performance Clusters (HPC) might be familiar with the related concept of message passing interface (MPI) which is a standardized means for exchanging data/commands between multiple processors in a computer cluster.}

Having the factor graph as the starting point, interest lies in an estimation strategy called the sum-product algorithm which is not well known in mainstream statistics, despite the fact that it is computationally powerful (Wand, 2017, p. 137-138). The sum-product algorithm is a general rule in factor graphs that allows to iteratively approximate marginal (posterior) distributions. When applied to a parametric problem with arbitrary likelihood and prior functions, the so-called Generalized Approximate Message Passing (GAMP) algorithm introduces further Gaussian and quadratic approximations to the possibly complicated expressions derived by the sum-product iterative algorithm. Proposed by Rangan (2011), GAMP is an extension of the popular Approximate Message Passing (AMP) algorithm of Donoho et al. (2009). The GAMP algorithm has desirable properties, namely, high-dimensional scalability, parallelizability, and effortless maintenance. Therefore, the first task of this paper is to analyze the concept of message passing algorithms in general; simplify the jargon stemming from signal processing, computing science, and similar literatures that have introduced such algorithms; and show how GAMP, in particular, can lead to efficient posterior inference in very high-dimensions.

At the same time, a second important task is to provide compelling evidence that the proposed algorithm is relevant for modeling macroeconomic variables. For that reason, I utilize a regression model setting with time-varying coefficients, stochastic volatility, and exogenous predictors. Regression models featuring time-varying parameters (TVPs) have been popular in economics at least since the seminal work of Cooley and Prescott (1976). More recently, there has been a systematic effort to introduce efficient MCMC algorithms for flexible estimation and shrinkage in Bayesian TVP models; see Belmonte et al. (2014), Chan et al. (2012), Giordani and Kohn (2008), Groen et al. (2013), Kalli and Griffin (2014), Koop and Potter (2007), Kowal et al. (2018), Nakajima and West (2013), Ro\v{c}kov\'{a} and McAlinn (2018) and Stock and Watson (2007) among others. These are examples of carefully designed MCMC algorithms that result in flexible joint modeling of structural instabilities and parameter shrinkage, but that may not be scalable to very high dimensions due to their reliance on repeated sampling via Monte Carlo. 

As a consequence, a novel empirical contribution introduced in this paper is to estimate a time-varying parameter regression model by using an observationally equivalent high-dimensional static regression form, and to address computational concerns by using message passing inference. With $T$ observations and $p$ predictors, the TVP model can be written as a static regression with the same $T$ observations but $(T+1)p$ covariates -- where the product $(T+1)p$ can easily be in the order of tens of thousands in standard macroeconomic applications. This static representation of the time-varying parameter model is anything but new, however, its estimation in the past has been exclusively tackled by specifying an additional hierarchical random walk (or some times stationary autoregressive) model for all time-varying parameters. This hierarchical form allows for inference using state-space methods and at the same time it can be interpreted as an informative shrinkage prior that makes estimation of this high-dimensional problem feasible. Instead I propose to completely drop this ``random-walk prior'' and the resulting state-space representation, and estimate the time-varying parameter model as a high-dimensional static regression with the assistance of a flexible Bayesian hierarchical shrinkage prior inspired by Tipping (2001). That way, by casting the TVP regression model into equivalent static form, standard shrinkage principles can be used in order to determine by how much coefficients evolve over time, or whether their value is zero and they are completely irrelevant. Most importantly, the use of the low-complexity GAMP algorithm ensures that the static form of the TVP regression with $(T+1)p$ covariates can be estimated quickly. The benefits of this algorithm and modeling strategy are illustrated using a forecasting exercise for monthly U.S. inflation that extends Stock and Watson (1999) to the TVP setting. The static form of the TVP regression estimated with GAMP is contrasted with powerful but slow MCMC algorithms for TVP models, such as Chan et al. (2012) and Kalli and Griffin (2014). The proposed approach, by incorporating a larger number of predictors and by shrinking coefficients flexibly, does perform significantly better compared to competitors in out-of-sample forecasting.

In the next section I introduce the general framework of factor graphs on random variables (parameters) and with the help of a toy example I show how this framework allows for efficient calculation of marginal distributions. Next, in Section 3 I introduce the TVP regression setting, rewrite the likelihood in static regression form and specify a shrinkage ``sparse Bayesian learning'' (SBL) prior. Under the given functional forms for the likelihood and prior, I proceed to derive a GAMP algorithm for this particular problem. In Section 4 the benefits of the proposed high-dimensional modeling approach are evaluated in a forecasting exercise for U.S. price inflation. Section 5 concludes the paper.

\section{Factor graphs and the sum-product algorithm}
\begin{figure}[H]
\centering
{\footnotesize
\begin{tikzpicture}
\tikzstyle{main}=[circle, minimum size = 8mm, thick, draw =black!100, node distance = 8mm]
\tikzstyle{second}=[rectangle, minimum size = 2mm, thick, draw =black!100, node distance = 6mm]
\tikzstyle{connect}=[-, thick]
\tikzstyle{box}=[rectangle, draw=black!100]
\matrix[row sep=4mm, column sep=8mm] {     
    & 
    &  \node[second, fill = black!100] (alpha2) [label=above:$f_{b}\left( x_{1} \text{,} x_{2} \right)$] { };
    &
    & \node[second, fill = black!100] (alpha3) [label=above:$f_{c}\left( x_{2} \text{,} x_{3} \right)$] { };
    &\\
    & \node[main] (beta1) [label=center:$x_{1}$] { }; 
    & 
    & \node[main] (beta2) [label=center:$x_{2}$] { };
    &  
    & \node[main] (beta3) [label=center:$x_{3}$] { };     \\ 
    & \node[second, fill = black!100] (alpha1) [label=below:$f_{a}\left( x_{1} \right)$] { };
    &
    &
    &
    & \node[second, fill = black!100] (alpha4) [label=below:$f_{d}\left( x_{3} \right)$] { };   \\   
          };
  \path (alpha1) edge [connect] (beta1)
        (beta1) edge [connect] (alpha2)
        (alpha2) edge [connect] (beta2)
        (beta2) edge [connect] (alpha3)
        (alpha3) edge [connect] (beta3)
        (beta3) edge [connect] (alpha4);                
\end{tikzpicture}
}
\caption{\emph{Simple factor graph representation of the decomposition of joint function $p \left(x_{1},x_{2},x_{3} \right)$.}}  \label{fig:example_factor_graph}
\end{figure}
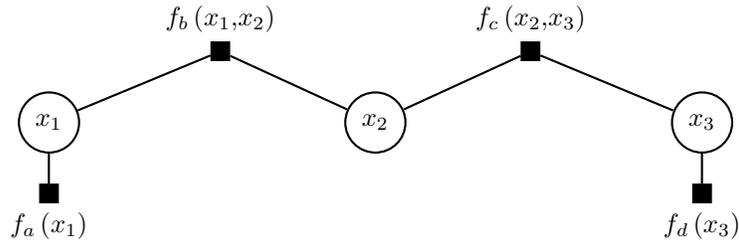

A factor graph represents the way a global function of several variables can be decomposed into a product of simpler functions (``factors''). Consider a generic example with \emph{discrete} random variables $x=\left(x_{1}, x_{2}, x_{3} \right)$ and a joint mass function $p$ that we can decompose, say, as
\begin{equation}
p\left(x_{1}, x_{2}, x_{3} \right) = f_{a} \left(x_{1}\right) f_{b} \left(x_{1},x_{2}\right) f_{c} \left(x_{2},x_{3} \right) f_{d} \left(x_{3} \right), \label{factor_dec}
\end{equation}
where $f_{a},f_{b},f_{c},f_{d}$ are the factors that have known functional forms.\footnote{In the next section, the discrete random variables $x$ are replaced by continuous model parameters, and the factors/functions are conditional or marginal probability distributions over these parameters.} This simple example can be depicted using the factor graph of \autoref{fig:example_factor_graph}, where circles denote the place of random variables in the graph and filled boxes denote the factors/functions.\footnote{In graph theory, symbols like the boxes and the circles in this example are called nodes or vertices. Nodes that depend to each other are connected with a solid line, and each connected pair of nodes is called an ``edge''. }

Consider now calculation of the marginal distribution of $x_{i}$. This is a computationally demanding task due to the fact that it involves integration (summation, in the discrete variable case) over all variables other than $x_{i}$
\begin{equation}
p\left(x_{i}\right)  = \sum_{x \setminus x_{i}} p \left(x_{1},x_{2},x_{3} \right), \label{summation}
\end{equation}
where $x \setminus x_{i}$ denotes the set $x$ with the element $x_{i}$ removed. As an example, if the variables in $x$ have two states (e.g. they are binary variables), then the above sum would only require $2^3$ operations. However, for high number of states and/or variables computational requirements proliferate substantially. Nevertheless, if $p(x_{1},x_{2},x_{3})$ is replaced with the expression in \eqref{factor_dec} it can be seen that not each variable is coupled to every other one, and this feature can be exploited in order to simplify the summation. For example, in the case of variable $x_{1}$, \autoref{fig:example_factor_graph} depicts that it is directly connected to $x_{2}$ and the factors $f_{a}(x_{1})$ and $f_{d}(x_{1},x_{2})$, but it is only indirectly connected to $x_{3}$ and the remaining factors. Put differently, we can simplify \eqref{summation} via identity \eqref{factor_dec} as follows
\begin{eqnarray}
p\left(x_{1}\right) & = & \sum_{x_2} \sum_{x_3} f_{a} \left(x_{1}\right) f_{b} \left(x_{1},x_{2}\right) f_{c} \left(x_{2},x_{3} \right) f_{d} \left(x_{3} \right) \label{simpl1} \\
& = & f_{a} \left(x_{1}\right) \sum_{x_2}  f_{b} \left(x_{1},x_{2}\right) \sum_{x_3} f_{c} \left(x_{2},x_{3} \right) f_{d} \left(x_{3} \right). \label{simpl2}
\end{eqnarray}
The second line of the equation above implies less algorithmic operations compared to the expression in the first line.

It should be clear at this point that the role of the factor graph representation is to allow to pin down the full path of influence that each variable $x_{i}$ exerts on other variables. As a consequence, by having this path of influence, only the required factors $f_{j}$ can be used when calculating marginal distributions, which increases computational efficiency. This is where the concept of \emph{message passing} formalizes such an efficient procedure for computing marginals. Each variable node passes messages to the next variable, where these messages are real-valued functions showing the influence that this variable exerts on all other variables. In the remainder of this Section message passing inference is introduced and the sum-product algorithm is derived, such that simplifications similar to the ones in equations \eqref{simpl1}-\eqref{simpl2} are formalized mathematically. Subsequently, in Section 3 the results of this toy example with three discrete random variables (parameters) can be generalized to a high-dimensional regression setting with possibly millions of parameters. More detailed introductions to these concepts can be found in popular machine learning textbooks, such as Barber (2012) and Bishop (2006). A recent introduction of message passing inference in factor graphs from a statistician's perspective is provided in Wand (2017).

Denote with $\mu_{x_{i} \rightarrow f_{j}}$ the message sent from variable $x_{i}$ to function $f_{j}$, and with $\mu_{f_{j} \rightarrow x_{i}}$ the message sent from factor node $f_{j}$ to variable node $x_{i}$, where $i=1,2,3$ and $j=a,b,c,d$ in our simple example with three variables and four factors. The message sent from variable $x_{i}$ to factor node $f_{j}$ is equal to the product of all messages arriving to node $x_{i}$ except from the message coming from the target node $f_{j}$:
\begin{equation}
\mu_{x_{i} \rightarrow f_{j}} = \prod_{k \in N(x_{i}), k \neq j} \mu_{f_{k} \rightarrow x_{i}}, \label{sum_product1}
\end{equation}
where $N(x_{i})$ is the set of neighboring (factor) nodes to $x_{i}$. Similarly, the message sent from factor node $f_{j}$ to variable node $x_{i}$ is given by the sum over the product of the factor function $f_{j}$ itself and all the incoming messages, except the messages from the target variable node $x_{i}$:
\begin{equation}
\mu_{f_{j} \rightarrow x_{i}} = \sum_{x \setminus x_{i}} f_{j}\left(x\right) \prod_{l \in N(x_{i}), l \neq i} \mu_{x_{l} \rightarrow f_{j}}. \label{sum_product2}
\end{equation}
Due to the form of the equation above, algorithms that are designed to iterate between \eqref{sum_product1} and \eqref{sum_product2} are called \emph{sum-product} algorithms; see also respective equations for the regression model in the next section.

In the special case where $x_{i}$ is an external node (as is the case with $x_{1}$ and $x_{3}$ in this example) it holds that $\mu_{x_{i} \rightarrow f_{j}}=1$. Similarly, if $f_{j}$ is an external factor node (see $f_{a}(x_{1})$ and $f_{d}(x_{3})$ in \autoref{fig:example_factor_graph}) it holds that $\mu_{f_{j} \rightarrow x_{i}}=f_{j}\left(x_{i}\right)$. Equations \eqref{sum_product1}-\eqref{sum_product2} define the iterations of the so-called sum-product algorithm (also called Belief Propagation; see Pearl, 1982), that allows calculation of marginal distributions (also called ``beliefs'' in computing science and the Bayesian networks literature). Upon convergence, it can be shown\footnote{It is beyond the scope of this paper to derive and prove the algorithm, and the reader is referred to the excellent machine learning books of Barber (2012) and Bishop (2006).} that
\begin{equation}
p\left(x_{i}\right) \propto \prod_{m \in N(x_{i})}  \mu_{f_{m} \rightarrow x_{i}}, \label{belief}
\end{equation}
that is, the marginal distribution of variable $x_{i}$ is simply the product of all messages received only from factor nodes that are connected to $x_{i}$.

Consider for example calculation of $p\left( x_{2} \right)$. Starting from the left of the graph, the messages emitted to node $x_{2}$ are:
\begin{eqnarray}
\mu_{f_{a} \rightarrow x_{1}} & = & f_{a} \left(x_{1} \right), \label{external_node_eg} \\
\mu_{x_{1} \rightarrow f_{b}} & = & \mu_{f_{a} \rightarrow x_{1}} =  f_{a} \left(x_{1} \right), \\
\mu_{f_{b} \rightarrow x_{2}} & = & \sum_{x_{1}} f_{b}\left(x_{1}x_{2}\right) \mu_{x_{1} \rightarrow f_{b}},
\end{eqnarray}
where the first identity holds because $f_{a} \left(x_{1}\right)$ is an external factor node, the second identity is a result of equation  \eqref{sum_product1}, and the third identity is a result of \eqref{sum_product2}. Similarly, the messages that arrive to $x_{2}$ stating from the right of the graph are
\begin{eqnarray}
\mu_{f_{d},x_{3}} & = & f_{d} \left(x_{3} \right), \\
\mu_{x_{3} \rightarrow f_{c}} & = & \mu_{f_{c} \rightarrow x_{3}} =  f_{d} \left(x_{3} \right), \\
\mu_{f_{c}\rightarrow x_{2}} & = & \sum_{x_{3}} f_{c}\left(x_{2},x_{3}\right) \mu_{x_{3}\rightarrow f_{c}},
\end{eqnarray}
where again the first identity results from the fact that $f_{d}\left( x_{3},x_{4} \right)$ is an external factor node, the second results from equation \eqref{sum_product1} and the third from equation \eqref{sum_product2}. Therefore, the marginal distribution of $x_{2}$ is now
\begin{equation}
p\left( x_{2} \right) \propto \mu_{f_{b} \rightarrow x_{2}} \times \mu_{f_{c} \rightarrow x_{2}}. \label{final_prod}
\end{equation}
Using similar arguments we can derive $p\left(x_{1} \right)$ and $p\left( x_{3} \right)$.

In this particular example, the formula derived in \eqref{final_prod} might seem redundant as for a wide class of distributions $p\left( \bullet \right)$, one can simply calculate the marginal distribution of $x_{2}$ using numerical integration. However, in high dimensions with many random variables, the sum-product rule can provide us with scalable and parallel posterior inference algorithms that can be several times faster compared to conventional algorithms that iterate sequentially (e.g. Gibbs sampler). It can be shown that the sum-product (Belief Propagation) algorithm is a special case of the more general expectation propagation algorithms that have been very popular in Bayesian machine learning; see Vehtari et al. (2018). Finally, note at this point that there is no mention about how to approximate the summations in \eqref{sum_product2}, which will not necessarily be tractable. Given the sum-product formula, there are several algorithms that would allow for the approximation of the required messages which are functions of the factors $f_{j}$. For example, Wand (2017) develops message passing inference inspired by the variational Bayes method. In the next section I adopt a recently developed algorithm (Generalized Approximate Message Passing) that performs Normal approximations to the functions implied by the sum-product iterations.

\section{Econometric Methodology}
\subsection{Time-varying parameter regression}
The starting point is the following time-varying parameter (TVP) regression with stochastic volatility of the form
\begin{equation}
y_{t}  = x_{t} \beta_{t} + \varepsilon_{t}, \label{tvp_reg1}
\end{equation}
subject to an initial condition for $\beta_{t}$ at $t=0$ (denoted as $\beta_{0}$), where $y_{t}$ is the $t^{th}$ observation on the variable of interest, $t=1,...,T$, $x_{t}$ is a $1 \times p$ vector of predictors (possibly including lags of $y_{t}$), $\beta_{t}$ is a $p \times 1$ vector of coefficients, and $\varepsilon_{t} \sim N \left( 0 , \sigma_{t}^2 \right)$ with $\sigma_{t}^2$ the time-varying variance parameter. It is desirable to estimate the initial condition in this model, rather than assume it is knonw. For that reason, following Fr\"{u}hwirth-Schnatter and Wagner (2010), this model can be written using an equivalent non-centered parametrization that allows to split the parameter $\beta_t$ into a part that is constant (which is equivalent to its initial condition $\beta_{0}$), and an ``add-on'' time-varying part with initial condition fixed to zero. The equivalent specification is
\begin{equation}
y_{t}  = x_{t} \widetilde{\beta} + x_{t} \widetilde{\beta}_{t} + \varepsilon_{t}, \label{tvp_reg1_modified}
\end{equation}
where now $\widetilde{\beta}_{t}$ has initial condition zero and it holds that $\beta_{t} = \widetilde{\beta} + \widetilde{\beta}_{t}$. As shown in Belmonte et al. (2014) this parametrization allows to use shrinkage priors to determine whether a variable has constant coefficient (by only shrinking the time-varying part), or it is completely irrelevant for modeling $y$ (by shrinking both the constant and time-varying parts to zero). More details of this approach are provided in the Online Appendix, Section D.1.

The TVP regression can be written in the following equivalent static regression form 
\begin{equation}
y =  \EuScript{X} \beta + \varepsilon, \label{reg1}
\end{equation}
where $y = \left[y_{1},...,y_{T} \right]^{\prime}$ and $\varepsilon = \left[\varepsilon_{1},...,\varepsilon_{T} \right]^{\prime}$ are column vectors stacking the observations $y_{t}$ and $\varepsilon_{t}$ respectively, $\beta = \left[\widetilde{\beta}^{\prime},\widetilde{\beta}_{1}^{\prime},...,\widetilde{\beta}_{T}^{\prime} \right]^{\prime}$ is a $(T+1)p \times 1$ vector, and 
\begin{equation}
 \EuScript{X} = \left[
\begin{array}{cccccc}
x_{1} & x_{1} & 0_{1\times p} &... &0_{1\times p} & 0_{1\times p}\\
x_{2} & 0_{1\times p} & x_{2}&... & 0_{1\times p}& 0_{1\times p}\\
\vdots & \vdots & \ddots & \ddots& \ddots& \vdots\\
x_{T-1} & 0_{1\times p} & 0_{1\times p} &... & x_{T-1}&0_{1\times p} \\
x_{T} & 0_{1\times p} & 0_{1\times p}&... & 0_{1\times p} &x_{T} \\
\end{array}
\right],
\end{equation}
is a $T \times (T+1)p$ matrix. It is evident that the first $p$ columns of $\EuScript{X}$ specify a constant parameter regression and its remaining columns add ``time-dummies'' to that regression. The Gram matrix $\left( \EuScript{X}^{\prime} \EuScript{X} \right)$ is of rank $T$ and the $q = (T+1)p$, in total, regression coefficients in \eqref{reg1} cannot be estimated with OLS. For that reason, following a long-standing tradition in engineering, economists tend to assume that $\beta_{t}$ (similarly for $\widetilde{\beta}_{t}$ in the non-centered parametrization) typically follows a random walk of the form $\beta_{t} = \beta_{t-1} + \eta_{t}$, where $\eta_{t} \sim N\left(0,Q\right)$ for some $p \times p$ symmetric, positive-definite covariance matrix $Q$. This random walk regression for $\beta_{t}$ allows to write the full time-varying parameter regression model in familiar state-space form, and also provides the additional information needed to estimate $\beta_{t}$ using data $y$ and $\EuScript{X}$. By doing so, estimation typically relies on Markov chain Monte Carlo methods by means of a simulation smoother; see Primiceri (2005) for a representative example. From a Bayesian point of view this additional information can be viewed as a conditional hierarchical prior of the form $p\left(\beta_{t} \vert \beta_{t-1} \right) \sim N\left(\beta_{t-1}, Q\right)$ that provides appropriate level of shrinkage. Put differently, equation \eqref{reg1} alone can be seen as an ill-posed problem where OLS does not have a unique solution and regularization is imperative for estimation.

In this paper I adopt this shrinkage view of the time-varying parameter regression model and propose an alternative inference strategy. That is, inference is done without reference to the useful but rather informative and subjective conditional hierarchical prior for $\beta_{t}$ given $\beta_{t-1}$ outlined above. Instead, the time-varying parameters are recovered by estimating directly equation \eqref{reg1} using data-based hierarchical shrinkage priors. In particular, I follow Tipping (2001) and define the following independent hierarchical prior for each element $\beta_{i}$ of the vector $\beta$, $i=1,2,...,(T+1)p$,
\begin{eqnarray}
p \left( \beta_i | \alpha_i  \right) & = & N \left( 0, \alpha_{i}^{-1} \right),  \label{SBL1} \\
p \left(\alpha_i \right) & = & Gamma \left(\underline{a},\underline{b} \right).  \label{SBL2}
\end{eqnarray}
This conditionally Normal prior for $\beta_{i}$ and Gamma prior for the precision parameter $\alpha_i$ is a scale mixture of Normal representation of a Student-t prior. Tipping (2001) calls this heavy-tailed prior a sparse Bayesian learning (SBL) prior, and I adopt this name henceforth; see also Korobilis (2013) for a detailed explanation why such hierarchical priors have good shrinkage properties. I follow Tipping (2001) and present all empirical results using the uniform hyperpriors (over a logarithmic scale) $\underline{a}=\underline{b}=1\times 10^{-10}$.

Two additional comments are in order regarding this time-varying parameter regression. First, the number of columns of $ \EuScript{X}$ is $q = (T+1)p$, therefore, the number of coefficients grows rapidly. For example, with 700 monthly observations and only 100 predictors, we end up with 70,100 regression coefficients. As a consequence, it is imperative to choose a fast estimation algorithm that approximates the parameter posterior, and this is where the scalability of message passing algorithms comes into play. Second, there is no mention yet of inference on $\sigma_{t}^{2}$, as this issue is covered later in this section after the GAMP inference algorithm is outlined. In a nutshell, estimation of stochastic volatility $\sigma_{t}^{2}$ also follows the same shrinkage principles defined for $\beta_{t}$. That is, it is shown that we can write estimation of $\sigma_{t}^{2}$ as a high-dimensional regression problem, without having to assume any kind of first-order Markov dependence to $\sigma_{t-1}^{2}$.

\subsection{A factor graph representation of Bayesian regression}
At this point we have all the necessary ingredients in order to cast the static form of the time-varying parameter regression in equation \eqref{reg1} into a factor graph form.\footnote{For the sake of brevity, notation for prior, posterior and likelihood distributions is generic, that is, there is no reference to their exact functional forms. Exact details and parametric formulas can be found in the Online Appendix, Section B.} Consider first an independent (but not necessarily i.i.d) prior for $\beta$, denoted $p\left( \beta \right)  = \prod _{i=1}^{q} p\left( \beta_{i} \right)$, and the resulting posterior from Bayes Theorem
\begin{eqnarray}
p \left( \beta \vert y \right) & \propto & p \left( y \vert \beta \right) p \left( \beta \right) \label{Bayes1} \\
& = & \prod_{t=1}^{T} p \left( y_{t} \vert \beta \right) \prod_{i=1}^{q} p\left( \beta_{i} \right).  \label{Bayes2}
\end{eqnarray}
The exact marginal posterior of $\beta_i$, $i=1,...,q$ is of the form
\begin{eqnarray}
p\left( \beta_i | y \right) & = & \int p \left( \beta \vert y \right) d\beta_{j \neq i}, \\
& \propto & \int p \left( y \vert \beta \right) p \left( \beta \right) d\beta_{j \neq i}, \\
& = & p \left( \beta_i \right)  \int p \left( y \vert \beta \right) \prod_{j=1,j \neq i} ^{q} p\left( \beta_j \right) d \beta_{j \neq i}, \label{ex_marg_post}
\end{eqnarray}
where $d\beta_{j \neq i}$ denotes integration over the whole set of $q-1$ parameters $\beta_j$ for $j \neq i$. Therefore, the formula above requires integration over a $(q-1)$-dimensional integral, a numerical problem that can become computationally infeasible for a high-dimensional vector $\beta$.

\begin{figure}
\centering
{\footnotesize
\begin{tikzpicture}
\tikzstyle{main}=[circle, minimum size = 8mm, thick, draw =black!100, node distance = 8mm]
\tikzstyle{second}=[rectangle, minimum size = 2mm, thick, draw =black!100, node distance = 6mm]
\tikzstyle{third}=[ minimum size = 2mm, thick, draw =white!100, node distance = 24.5mm]
\tikzstyle{fourth}=[rectangle,minimum size = 2mm, thick, draw =black!100, node distance = 22mm]
\tikzstyle{fifth}=[circle, minimum size = 6mm, thick, draw = black!20, node distance = 6mm]
\tikzstyle{connect}=[-, thick]
\tikzstyle{connect2}=[dashed, thick]
\tikzstyle{box}=[rectangle, draw=black!100]
\matrix[row sep=0.5mm,column sep=8mm] {
    \node[second, fill = black!100] (alpha1) [label=above:$p\left( \beta_1 \right)$] { }; & \node[main] (beta1) [label=center:$\beta_1$] { }; &     &     \\
    &     &     \node[fourth, fill = black!100] (gamma1) [right = of beta1,label=above:$p\left(y_1|\beta \right)$] { }; &  \node[fifth, fill = black!20] (delta1) [label=center:$y_1$] { }; \\
    \node[second, fill = black!100] (alpha2) [label=above:$p\left( \beta_2 \right)$] { }; & \node[main] (beta2) [label=center:$\beta_2$] { }; &     &     \\
    &     &     \node[fourth, fill = black!100] (gamma2) [right = of beta2,label=above:$p\left(y_2|\beta \right)$] { }; &  \node[fifth, fill = black!20] (delta2) [label=center:$y_2$] { }; \\
    \node[second, fill = black!100] (alpha3) [label=above:$p\left( \beta_3 \right)$] { }; & \node[main] (beta3) [label=center:$\beta_3$] { }; &     &     \\
    & & & \\ & & & \\ & & & \\ & & & \\ & & & \\ & & & \\ & & & \\ & & & \\ & & & \\   & & & \\ & & & \\ & & & \\
\node[third] (g1) [label=center:$...$] { }; & \node[third] (g2) [label=center:$...$] { }; & \node[third] (g3) [right=of g2,label=center:$...$] { }; & \node[third] (g4) [label=center:$...$] { }; \\
    & & & \\ & & & \\ & & & \\ & & & \\ & & & \\ & & & \\ & & & \\ & & & \\ & & & \\
    &     &     \node[fourth, fill = black!100] (gamma3) [right = of beta2,label=below:$p\left(y_T|\beta \right)$] { }; &  \node[fifth, fill = black!20] (delta3) [label=center:$y_T$] { }; \\
    \node[second, fill = black!100] (alpha4) [label=above:$p\left( \beta_q \right)$] { }; & \node[main] (beta4) [label=center:$\beta_q$] { }; &     &     \\
    };

  \path (alpha1) edge [connect] (beta1)
        (alpha2) edge [connect] (beta2)
        (alpha3) edge [connect] (beta3)
        (alpha4) edge [connect] (beta4)
        (beta1) edge [connect] (gamma1)
        (beta1) edge [connect] (gamma2)
        (beta1) edge [connect] (gamma3)
        (beta2) edge [connect] (gamma1)
        (beta2) edge [connect] (gamma2)
        (beta2) edge [connect] (gamma3)
        (beta3) edge [connect] (gamma1)
        (beta3) edge [connect] (gamma2)
        (beta3) edge [connect] (gamma3)
        (beta4) edge [connect] (gamma1)
        (beta4) edge [connect] (gamma2)
        (beta4) edge [connect] (gamma3)
        (gamma1) edge [connect2] (delta1)
        (gamma2) edge [connect2] (delta2)
        (gamma3) edge [connect2] (delta3);
\end{tikzpicture}
}
\caption{\emph{Factor graph representation for the high-dimensional regression model.}}  \label{fig:factor_graph}
\end{figure}
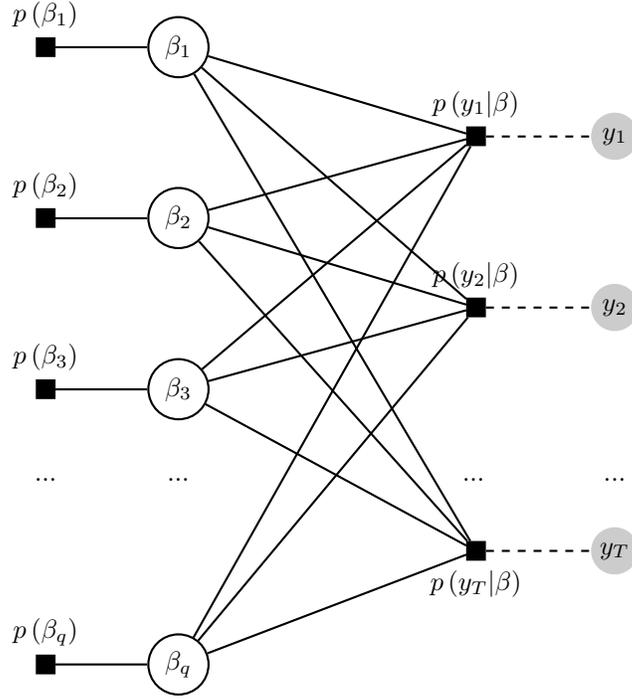

We can now call the framework of factor graphs in order to factorize efficiently the marginal posteriors of $\beta$. The factor graph representation of the regression model is depicted in Figure \ref{fig:factor_graph}. Based on this figure, the marginal posterior of $\beta_i$, presented in equation \eqref{ex_marg_post}, can be defined as the product of incoming messages at node $\beta_i$ in the graph
\begin{equation}
p\left( \beta_i | y \right) = \mu_{ p(\beta_i) \rightarrow \beta_i } \prod_{t=1}^{T} \mu_{ p(y_{t} | \beta) \rightarrow \beta_i }. \label{marginal}
\end{equation}
Similar to equation \eqref{external_node_eg} in the example of Section 2, the message $\mu_{ p(\beta_i) \rightarrow \beta_i }$ is an external factor node and for that reason it is equal to the prior $p \left( \beta_i \right)$. Generalizing the example sum-product rule derived in equations \eqref{sum_product1} - \eqref{sum_product2} of the previous section, we can write the messages from $p(y_{t} | \beta)$ $\forall t$ to $\beta_{i}$ using the following expression
\begin{equation} \label{message1}
\mu_{ p(y_{t} | \beta) \rightarrow \beta_i } = \int p \left(y_t | \beta \right) \prod_{j=1,j \neq i} ^{p} \mu_{ \beta_j \rightarrow p(y_{t} | \beta) } d\beta_{j \neq i}.
\end{equation}
In the decomposition above, the message from node $\beta_j$ to function (factor) $p \left( y_t | \beta \right)$ is the product of all incoming messages to node $\beta_i$, excluding the message coming from $p \left( y_t | \beta \right)$ itself
\begin{equation} \label{message2}
\mu_{ \beta_j \rightarrow p(y_{t} | \beta) } = p \left( \beta_j \right) \prod_{s=1, s \neq t}^{T}  \mu_{ p(y_{s} | \beta) \rightarrow \beta_j }.
\end{equation}

We can see in equations \eqref{message1}-\eqref{message2} that in order to obtain the message $\mu_{ p(y_{t} | \beta) \rightarrow \beta_i }$ we need $\mu_{ \beta_j \rightarrow p(y_{t} | \beta) }$ and vice-versa. Therefore, one can simply update both equations iteratively using the following iterative sum-product scheme
\begin{eqnarray} \label{eqn:BP1}
\mu_{ p(y_{t} | \beta) \rightarrow \beta_i }^{(r+1)} & = & \int p \left(y_t | \beta \right) \prod_{j=1,j \neq i} ^{q} \mu_{ \beta_j \rightarrow p(y_{t} | \beta) }^{(r)} d\beta_{j \neq i},  \\ 
\label{eqn:BP2}
\mu_{ \beta_j \rightarrow p(y_{t} | \beta) }^{(r+1)} & = & p \left( \beta_j \right) \prod_{\mathcal{s}=1, \mathcal{s} \neq t}^{T}  \mu_{ p(y_{\mathcal{s}} | \beta) \rightarrow \beta_j }^{(r)}, 
\end{eqnarray}
where the superscript $(r)$ denotes the $r^{th}$ iteration of the algorithm. In graphs with a tree structure, one iteration of the algorithm above will always recover the exact marginal posteriors for the parameters $\beta_i$. In a factor graph with loops there are no guarantees that the sum-product rule will converge to a good fixed point. However, the sum-product rule can still achieve a good approximation and this is the reason why it is used extensively in applications of coding theory, machine vision, and compressive sensing that have a loopy graph representation (Mooij and Kappen, 2007). Translating these facts into familiar jargon for the static regression in equation \eqref{reg1}, algorithmic convergence is achieved if the correlation of right-hand side predictors is not excessively high. If this is not the case, the joint posterior of the coefficients $\beta$ might also be highly correlated, which would make inference solely based on the marginal posteriors $p(\beta_{i})$ less accurate. In our benchmark time-varying parameter regression in \eqref{reg1}, correlation is by default not excessively high due to the fact that the Gram matrix $ \EuScript{X}^{\prime}  \EuScript{X}$ has a certain block-diagonal structure that allows for a general sparse correlation pattern -- even if within a given block correlation may be high. In the empirical application, predictor variables are mainly principal components or lags thereof, such that correlation within each block is also low. Finally, note that the specific time-decomposition of the likelihood function does not accommodate autoregressive and general time-series models, where the likelihood at time $t$ may be written conditional on past observations. In the empirical application it is found that, despite this approximation, autoregressive coefficients are recovered accurately.\footnote{A simulation exercise in the Online Appendix, Section C.3, generating artificial data from an AR(4) model, also verifies that the proposed GAMP algorithm performs well even if the likelihood function is not i.i.d. Another assumption that affects performance of GAMP is that $\EuScript{X}$ is mean-zero Gaussian; see the discussion in Al-Shoukairi et al. (2018) and references therein. In a time series context this means that GAMP will have better convergence when right hand-side predictors are strictly stationary, although the use of weakly stationary predictors is not excluded.}

\subsection{Generalized Approximate Message Passing}
While the core of any message passing algorithm is fully described by the sum-product iterations, deriving the exact functional form of the messages in equations \eqref{eqn:BP1} and \eqref{eqn:BP2} under the regression likelihood and the Student-t hierarchical prior implies that cumbersome integrations might be necessary. The GAMP algorithm introduces certain Gaussian approximations to the sum-product iterations. Unlike Laplace approximations, that is, Gaussian approximations to parameter posteriors that many times can be poor, the GAMP approximation is fully based on asymptotic results that make it more reliable as the number of predictors grows large. First, when $q \rightarrow \infty$ a central limit theorem (CLT) postulates that the messages $\prod_{j=1,j \neq i} ^{q} \mu_{ \beta_j \rightarrow p(y_{t} | \beta) }$ can be approximated by a Gaussian distribution with respect to the uniform norm.\footnote{This is a result of the Berry-Esseen central limit theorem which states that a sum of random variables converge to a Gaussian density; see a proof of this theorem in Donoho et al. (2011). Given that the sum-product equations involve products of random variables, rather than sums, derivations of GAMP  based on this central limit theorem typically proceed by taking logarithms of equations \eqref{marginal}-\eqref{message2}. The marginal posterior $p \left( \beta_i | y \right)$ is then recovered by performing an exponential transformation of the log messages, and by normalizing so that the posterior integrates to one; see the Online Appendix, Section A, for details.} This result means that messages in \eqref{message1} can be represented to be proportional to a Gaussian distribution. A second approximation involves taking the Taylor-series expansion of terms in the messages, so that the first two moments (mean and variance) of $p \left( \beta_i | y \right)$ can be obtained analytically up to the omission of $O\left(1/q \right)$ terms. Exact derivation of these approximations involves many tedious steps and transformations, and the reader is referred to the Online Appendix for more details. What is important to stress at this point is that both the CLT and Taylor-series approximations vanish as $q \rightarrow \infty$ with $q/T \rightarrow \delta$ for some constant $\delta$; see Rangan (2011) and Rangan et al. (2016) for more details. This is an example of the ``blessing of Big Data'' -- rather than the ``curse of dimensionality'' embedded in many traditional estimation algorithms -- as the GAMP algorithm fully facilitates the large $q$ asymptotics.

Deriving the GAMP algorithm involves several steps and lengthy proofs which are left for the Online Appendix. The final product of all the approximations to the two sum-product update equations \eqref{eqn:BP1} - \eqref{eqn:BP2}, is a simple iterative algorithm that provides an approximation to the mean and variance of $p \left( \beta_{i} \vert y \right)$. The algorithm iterates through computationally trivial scalar multiplications and additions that result in worst case algorithmic complexity of $\mathcal{O}(Tq)$. That is, estimation of the marginal parameter posterior distribution does not involve costly operations such as high-dimensional integration or inversion of large matrices. This feature implies that the algorithm can handle regressions with an excessively large number of predictors with the same ease it can handle smaller regression models. Convergence is achieved when the difference between estimates of the posterior mean of $\beta$ between two consecutive iterations is below a pre-specified tolerance level. Other parameters can be updated by combining the GAMP algorithm with EM updates.\footnote{See Al-Shoukairi et al. (2018) and Zou et al. (2016) for examples of how to derive EM updates for prior hyperparameters.} This feature is explained in the Online Appendix, where it is shown how to update the hyperparameter $\alpha_{i}$ introduced in the hierarchical prior of equation \eqref{SBL2}.

A sketch of the algorithm is provided in Algorithm \autoref{algorithm:GAMP}. This is a simplified version that focuses on estimation of $\beta$ by assuming that the regression variance and prior hyperparameters are all known and fixed. Following the analysis in Section 2, the algorithm can be split into two steps: i) evaluating all messages that leave each variable node $\beta_{j}$ (output), and ii) evaluating all messages that arrive at each variable node $\beta_{j}$ (input). The final product is estimates of the posterior mean and variance of $\beta_{j}$ which are denoted as $\widehat{\beta}_{i}$ and $\widehat{\tau}_{i}^{\beta}$, respectively. At the core of the calculation of the posterior mean and variance are the scalar functions $g_{in}$ and $g_{out}$. Derivation of the exact form of these two functions depends on the form of the prior distribution and the likelihood. Online Appendix, Section B, provides a detailed algorithm in the case of the regression likelihood in equation \eqref{reg1} and the prior in \eqref{SBL1}-\eqref{SBL2}. In any case, Rangan (2011) shows that regardless of the form of the nonlinear scalar functions $g_{in}$ and $g_{out}$, the worst-case complexity of the GAMP algorithm is not affected and is always $\mathcal{O}(Tq)$.

\begin{algorithm}[H]
\caption{\textit{Generalized Approximate Message Passsing (GAMP) with known variance and prior hyperparameters}}\label{algorithm:GAMP}
\begin{algorithmic}[1]
\State Initialize $\widehat{\beta}_{i}^{(0)} = 0$ and $\widehat{\tau}_{i}^{\beta,(0)}= 100$ $\forall i=1,...,q$, and set $\widehat{s}_{t}^{(0)}=0$ $\forall
t=1,...,T$.
    \State $r=1$
    \While{$ \Vert \widehat{\beta}^{\left(r\right)} - \widehat{\beta}^{\left(r-1\right)} \Vert \rightarrow 0$}
    \State 1) \textsc{\underline{Output Messages Step:}} 
    \For{$t = 1$ \algorithmicto $T$}
         \State  \hspace{1cm} $\widehat{c}_{t}^{(r)}  = \sum_{i=1}^{q} \EuScript{X}_{t,i} \widehat{\beta}_{i}^{(r-1)} - \widehat{s}_{t}^{(r-1)}\widehat{\tau}_{t}^{c,(r)}$
         \State  \hspace{1cm} $\widehat{\tau}_{t}^{c,(r)} = \sum_{i=1}^{q} \EuScript{X}_{t,i} ^{2} \widehat{\tau}_{i}^{\beta,(r-1)}$
         \State \hspace{1cm} $\widehat{s}_{t}^{(r)} = g_{out} \left(\widehat{c}_{t}^{(r)},\widehat{\tau}_{t}^{c,(r)}, y_{t} \right) $
        \State \hspace{1cm} $\widehat{\tau}_{t}^{s,(r)} = - \frac{\partial}{\partial \widehat{c}}g_{out} \left(\widehat{c}_{t}^{(r)},\widehat{\tau}_{t}^{c,(r)}, y_{t} \right)$
        \EndFor
        \State 2) \textsc{\underline{Input Messages Step:}} 
        \For{$i = 1$ \algorithmicto $q$}
        \State  \hspace{1cm} $\widehat{d}_{i}^{(r)} = \widehat{\beta}_{i}^{(r-1)} + \widehat{\tau}_{i}^{d,(r)}\sum_{t=1}^{T} \EuScript{X}_{t,i}\widehat{s}_{t}^{(r)}$
        \State  \hspace{1cm} $\widehat{\tau}_{i}^{d,(r)} = \left( \sum_{t=1}^{T} \EuScript{X}_{t,i}^{2} \widehat{\tau}_{t}^{s,(r)} \right)^{-1}$        
        \State \hspace{1cm} $\widehat{\beta}_{i}^{(r)} = g_{in} \left(\widehat{d}_{i}^{(r)},\widehat{\tau}_{i}^{d,(r)} \right)$
        \State \hspace{1cm} $\widehat{\tau}_{i}^{\beta,(r)}  =\widehat{\tau}_{i}^{d,(r)}  \frac{\partial}{\partial \widehat{d}}g_{in} \left(\widehat{d}_{i}^{(r)},\widehat{\tau}_{i}^{d,(r)} \right)$        
        \EndFor
        \State $r = r + 1$
    \EndWhile
    \State Obtain mean and variance of $\beta$ as $\widehat{\beta} = \left(\widehat{\beta}_1^{(r)},...,\widehat{\beta}_q^{(r)} \right)$ and $\tau^{\beta} = \left(\widehat{\tau}_{1}^{\beta,(r)},...,\widehat{\tau}_{q}^{\beta,(r)} \right)$
\end{algorithmic}
\end{algorithm}

The algorithm above assumes a known regression variance, e.g. normalized to be one. Of empirical interest is the derivation of an update rule for the variance parameter when this is both unknown and time varying. Here I propose a novel, computationally trivial estimator of the variance that builds on approximations used in the Bayesian stochastic volatility estimator of Kim et al. (1998). First, we write the regression model in \eqref{reg1} in the following form
\begin{equation}
y = \EuScript{X} \beta + \Sigma v,
\end{equation}
where $\Sigma$ is a $T \times T$ diagonal matrix with the time-varying standard deviations $\sigma_{t}$ on its main diagonal. Subsequently, conditional on knowing $\beta$ by means of some estimate $\widehat{\beta}$, we can re-write the above model as
\begin{eqnarray}
 \log \left[ \left( y - \EuScript{X} \widehat{\beta}  \right)^{2}\right] & = & \log \left(diag\left(\Sigma\right)^{2} \right) + \log(v^2), \Rightarrow  \\
 \widetilde{y} & = & \widetilde{\sigma}^{2} + \widetilde{v},  \label{vola}
\end{eqnarray}
where $diag\left(\Sigma\right)^{2}$ is a $T \times 1$ vector with elements $\sigma_{t}^{2}$ $\forall t \in [1,T]$, and variables with a $\widetilde{\bullet}$ denote quantities in log-squares. In particular, the distribution of $\widetilde{v}$ is $\log-\chi^{2}$ with one degree of freedom. Following Kim et al. (1998) we can approximate this with a mixture of seven Normal distributions with means $\mu_{i}$, variances $V_{i}$ and component weights $\pi_{i}$, where $i=1,...,7$ and $\sum_{i} \pi_{i}=1$.\footnote{The exact values of $\mu_{i}$, $V_{i}$, $\pi_{i}$ for all seven components is provided in the Online Appendix, Section B.} Then equation \eqref{vola} can be replaced with the following set of seven equations
\begin{equation}
\text{ \ \ \ \ \ \ \ \ \ \ \ } \widetilde{y} =  \widetilde{\sigma}^{2} + u_{i}, \text{ \ \ \ } i=1,...,7,
\end{equation}
where $u_i \sim N\left(\mu_{i}, V_{i} \right)$. An estimator of the $T \times 1$ vector of log-volatilities is of the form $E_{i}\left(\widetilde{\sigma}^{2} \right) = \widetilde{y} - \mu_{i}$, and the final volatility estimate at time $t$ is
\begin{equation}
\widehat{\sigma}_{t}^{2} = \exp \left( \sum_{i=1}^{7} \pi_{i} \left( \widetilde{y}_{t} - \mu_{i} \right) /7 \right). \label{final_vola}
\end{equation}
Similar expressions can also be derived for the posterior variance of $\sigma_{t}^{2}$ if desired, for example, when computing the posterior predictive density via simulation. It turns out that the resulting estimate of volatility is similar to the standard stochastic volatility estimator of Kim et al. (1998), but it is much less persistent due to the lack of dependence of $\sigma_{t}^{2}$ on $\sigma_{t-1}^{2}$. More evidence on the excellent properties of this simple estimator of stochastic volatility is provided in the Online Appendix, Section D.1.

Finally, Online Appendix, Section C, provides detailed Monte Carlo evidence on the usefulness of the proposed econometric specification and algorithm. By simulating artificial data from models with various patterns of time-variation in parameters, it is assessed how good the specification in equation \eqref{reg1}, with the assistance of the sparse Bayesian learning prior, is at recovering the true time-varying parameters. At the same time, a second simulation exercise shows the ability of the GAMP algorithm with shrinkage prior to perform high-dimensional shrinkage even in cases with more predictors than observations. A final simulation exercise discusses the stability of the GAMP algorithm in models with correlated predictors, and assesses numerically the case where the likelihood function is not i.i.d. While the results of the simulated data exercises suggest that the proposed algorithm provides a reasonable balance between computational speed and estimation accuracy, the next section establishes that the proposed algorithm is also very useful in a forecasting application using real macroeconomic data.

\section{Empirical illustration: Forecasting inflation}
This section describes the set-up and results of a comprehensive forecasting exercise that demonstrates the merits of the modeling approach outlined in the previous section. Most applications of time-varying parameter regressions focus in particular on inflation. Of course, this class of models is flexible enough to provide useful forecasts of any other variable of interest; see Bauwens et al. (2015) for assessing structural breaks in several monthly and quarterly macroeconomic time series. Nevertheless, there is ample evidence that structural breaks in inflation are so evident and complex, such that TVP models are particularly useful for forecasting this variable; see Chan et al. (2012), Groen et al. (2013), Koop and Korobilis (2012), Pettenuzzo and Timmermann (2017) and Stock and Watson (2007) among many others. 

The data collected for this exercise are 115 macroeconomic variables from Federal Reserve Economic Data (FRED) of St. Louis Federal Reserve Bank website. The data originally span the period 1959M1 to 2016M6, but the effective sample is smaller after taking stationarity transformations and lags. The stationarity transformations follow standard norms in this literature (see Stock and Watson, 1999) and exact details are provided in the Online Appendix, Section A.

The empirical application builds on the seminal work of Stock and Watson (1999) for forecasting inflation. These authors specify the following benchmark forecasting model
\begin{equation}
\pi_{t+h}^{h} - \pi_{t} = \phi_{0} +  z_{t} \theta (L) +  \Delta \pi_{t} \gamma (L) + e_{t+h}, 
\end{equation}
where $\pi_{t}^{h} = (1200/h)log\left(P_{t}/P_{t-h}\right)$ is the $h$-period inflation in the price level $P_{t}$. As Stock and Watson (1999; Section 2) explain in detail the assumption here is that inflation is $I(1)$ while the exogenous variables in $z_{t}$ are $I(0)$. Two modifications of this basic forecasting model are in order. First, as Stock and Watson (1999, 2002) also suggest, the high-dimensional variables $z_{t}$ are replaced by factors $f_{t}$ estimated using principal components. Second, the forecasting equation is enhanced with time-varying parameters and stochastic volatility. The final forecasting model used in this paper is of the form
\begin{equation}
\pi_{t+h}^{h} - \pi_{t} = \phi_{t,0} +  f_{t} \theta_{t} (L) +  \Delta \pi_{t} \gamma_{t} (L) + e_{t+h}, \label{bencmark_forecasting_eq}
\end{equation}
where $e_{t} \sim N \left(0, \sigma^{2}_{t} \right)$ and $f_{t}$ is a lower-dimensional vector of factors.

The forecasting exercise is run for two measures of inflation, namely the consumer price index for all items (CPIAUCSL) and the personal consumption expenditures price index (PCEPI). The forecast horizons evaluated are $h=1,3,6,12$ which correspond to one-month, one-quarter, one-semester and one-year ahead forecasts, respectively. Following Bauwens et al. (2015) evaluation of forecasts is based on the mean square forecast error (MSFE) for point forecasts, and on the logarithm of the average predictive likelihoods (log APL) for comparing whole forecast densities. Exactly $50\%$ of the sample is used for evaluation of out-of-sample forecasts, leading to a period of $343-h$ months where MSFEs and log APLs are calculated. Note that while estimation entails the spread $\pi_{t+h}^{h} - \pi_{t}$, all forecast evaluations in this Section (see also alternative model in equation \eqref{bencmark_forecasting_eq2}) pertain to $\pi_{t+h}^{h}$.

When applying the proposed GAMP estimation methodology, equation \eqref{bencmark_forecasting_eq} is estimated using two own lags of the dependent variable, the first 20 principal component estimates of the factors $f_{t}$ (updated recursively using only information up to time $t$) and two lags of these factors (that is, their values in periods $t$ and $t-1$). As explained in the main text, this TVP model can be estimated using GAMP by casting it into the form \eqref{tvp_reg1} by setting $y_{t} = \pi_{t+h}^{h} - \pi_{t}$, $x_{t} = \left[1,f_{t},\Delta \pi_{t}\right]$, $\beta_{t} = \left(\phi_{t,0},\theta_{t}(L)^{\prime}, \gamma_{t}(L)^{\prime}\right)^{\prime}$ and $e_{t+h} = \varepsilon_{t}$. Written in this static form and using all available observations, the proposed empirical model has nearly 30000 regression coefficients and another 700 volatility parameters to estimate. The only input that the GAMP algorithm requires is choice of two scalar prior hyperparameters. For the sparse Bayesian learning prior of equations \eqref{SBL1} - \eqref{SBL2} these hyperparameters are set, as explained in Section 3, to the uniform values $\underline{a} = \underline{b} = 1 \times 10^{-10}$. This approach to estimating the TVP regression of \eqref{bencmark_forecasting_eq} using GAMP is abbreviated as \textbf{TVP-GAMP} in the results presented next. 

The benchmark time-varying regression approach estimated with the GAMP algorithm is contrasted  against a range of popular algorithms for inference in models with many predictors and/or stochastic variation in coefficients. The list of competing specifications and estimation algorithms is the following:
\begin{itemize}
\item \textbf{KP-AR}: This is a structural breaks AR(2) model based on Koop and Potter (2007). It only features an intercept and two lags of inflation.
\item \textbf{GK-AR}: This is a structural breaks AR(2) model based on Giordani and Kohn (2008). It only features an intercept and two lags of inflation.
\item \textbf{TVP-AR}: This is a typical TVP-AR(2) model with stochastic volatility, estimated with MCMC methods, similar to Pettenuzzo and Timmerman (2017). It only features an intercept and two lags of inflation.
\item \textbf{UCSV}: The unobserved components stochastic volatility model of Stock and Watson (2007) is a special case of a TVP regression with no predictors - it is a local level state-space model featuring stochastic volatility in the state equation.
\item \textbf{TVD}: The time-varying dimension (TVD) model of Chan et al. (2012) features an intercept, two lags of inflation, and the first three principal components estimates of the factors. The number of factors is restricted to three for computational reasons. Also for computational reasons one cannot do time-varying selection among all possible $2^{p}$ models constructed with $p$ predictors, therefore, I follow Chan et al. (2012) and do dynamic selection of either models with one variable at a time, or the full model with all variables.
\item \textbf{TVS}: The time-varying shrinkage (TVS) algorithm of Kalli and Griffin (2014) features an intercept, two lags and the first three principal components estimates of the factors (also restricted to three factors for computational reasons).
\item \textbf{TVP-BMA}: Introducing a Bayesian model averaging prior in the TVP regression is fairly trivial as Groen et al. (2013) have shown. We can use with this algorithm up to 10 principal component estimates of the factors, an intercept and two lags of inflation.
\item \textbf{BMA}: This is a constant parameter version of the forecasting regression specification that features the stochastic search variable selection (SSVS) of George and McCulloch (1993). Even though this prior can be also used for variable selection, here it is used in a Bayesian model averaging (BMA) setting. For this algorithm we use the same number of predictors as in TVP-GAMP, namely an intercept, two own lags of inflation, and two lags of the first 20 principal components. However, this model is the only one in the comparison that doesn't have time-varying parameters.
\end{itemize}
All these models collapse to being special cases of the benchmark equation \eqref{bencmark_forecasting_eq}, despite the fact that different specifications might imply various additional assumptions about how the coefficients might evolve over time (whereas TVP-GAMP does not rely on such additional assumptions). All models except for the UCSV have in common an intercept and the two own lags of inflation.\footnote{In order to understand better whether forecast gains can be achieved from specifying a model with many predictors, or with flexible time-variation, or both, I only calculate direct multi-step forecasts from all competing models. That way all algorithms are used to estimate different versions of the same regression with $y_{t+h}$ on the left hand side (for each $h$) and information dated $t$ or earlier on the left hand side. However, iterated forecasts can be computed from models with no exogenous predictors (e.g. TVP-AR or UCSV). Direct forecasts are better when the model is misspecified, while iterated forecasting models in general result in more efficient econometric estimates and sharper predictive densities. Examination of $h=12$ month ahead iterated forecasts from the KP-AR, GK-AR, TVP-AR and UCSV models reveals that these are, most times, slightly inferior to respective direct forecasts in terms of MSFE, but they can be in some cases up to $15\%$ better in terms of average log predictive likelihoods. Iterated forecasting results are not presented here, but they are available from the author.} For those algorithms that rely on shrinkage priors (TVP-GAMP, TVD, TVS, TVP-BMA, and BMA) the intercept and the two lags of inflation are never allowed to shrink by using a noninformative prior on them. Therefore, whenever shrinkage (static or dynamic) is implemented this only applies to the exogenous information in the factors. Exact details of the econometric specifications and prior settings associated with the competing models is provided in the Online Appendix, Section E.

A final note is on computation. All of the competing models listed above are based on estimation using MCMC and in particular the Gibbs sampler. Most of these models were originally developed by their respective authors for forecasting inflation. This is due to the fact that time-varying parameter regressions have consistently been found to be superior for this series. However, even though one would normally expect more breaks to be present in higher frequency monthly inflation, all of these papers estimate their models using quarterly data. This is done for computational reasons. Due to the fact that here these models are estimated for monthly data, I follow Bauwens et al. (2015) and base inference only on 5000 samples from the posterior after a burn-in period of 1000 draws, that is, a total of 6000 MCMC iterations. Convergence criteria suggest that such low number of iterations is sufficient for forecasting, even though it might not be satisfactory for other econometric exercises. Despite the low number of MCMC iterations, computation is quite cumbersome taking several hours for some models. In contrast, it takes only minutes to run the full recursive exercise using the TVP-GAMP model that features both time-varying parameters and the full set of available predictors. The GAMP algorithm not only involves simple scalar computations, but also converges fairly quickly after 10 to 100 iterations. Once convergence is achieved, the first two posterior moments are readily available for further inference, rather than having to store thousands of samples from the posterior of a high-dimensional parameter vector.

The results from this forecasting exercise are presented in Tables 1 and 2, and are very encouraging for the proposed TVP-GAMP method. \autoref{table:msfe_results} shows MSFEs relative to an AR(2) benchmark (with an intercept), such that numbers lower than one signify better performance of a competing model relative to that benchmark AR(2) specification. It can be seen that under the specified regression model, point forecasts from TVP-GAMP dominate alternatives by a substantial amount, both for CPI and PCE inflation. The forecast gains are increasing with the horizon. \autoref{table:apl_results} shows the logarithm of the average predictive likelihood (log APL), and this metric is quoted as a spread from the log APL of the simple AR(2) specification. Positive values signify better performance relative to the benchmark AR(2). Using this metric, TVP-GAMP is either the top performing model or among the top, for the four forecast horizons and the two measures of inflation.

\bigskip
\begin{table}[H]
\begin{center}
\caption{Point forecast performance: MSFEs relative to AR(2) benchmark.}  \label{table:msfe_results}
{\footnotesize
\begin{tabular}{llllllllll}\hline
	&			\multicolumn{4}{c}{CPI}					&		&			\multicolumn{4}{c}{PCE deflator}					\\
	&	$h=1$	&	$h=3$	&	$h=6$	&	$h=12$	&		&	$h=1$	&	$h=3$	&	$h=6$	&	$h=12$	\\ \hline
KP-AR	&	0.952	&	0.998	&	0.978	&	0.904	&		&	1.030	&	1.027	&	1.042	&	0.973	\\
GK-AR	&	0.997	&	1.007	&	1.002	&	0.993	&		&	1.005	&	1.006	&	0.999	&	0.997	\\
TVP-AR	&	1.009$^{\ast}$	&	1.047$^{\ast\ast\ast}$	&	1.258$^{\ast\ast\ast}$	&	1.224$^{\ast\ast\ast}$	&		&	1.053$^{\ast\ast\ast}$	&	1.119$^{\ast\ast\ast}$	&	1.141$^{\ast\ast\ast}$	&	1.123$^{\ast\ast\ast}$	\\
UCSV	&	1.032$^{\ast\ast}$	&	1.063$^{\ast\ast\ast}$	&	1.286$^{\ast\ast}$	&	1.312	&		&	1.060$^{\ast\ast}$	&	1.077$^{\ast\ast}$	&	1.234$^{\ast}$	&	1.159$^{\ast\ast\ast}$	\\
TVD	&	1.014	&	0.988	&	0.942	&	0.919	&		&	1.091	&	1.108	&	1.156	&	1.029	\\
TVS	&	1.146$^{\ast\ast}$	&	1.547	&	1.555	&	1.155	&		&	1.084$^{\ast}$	&	1.413	&	2.150	&	1.208$^{\ast}$	\\
BMA	&	0.965$^{\ast\ast\ast}$	&	0.952$^{\ast\ast\ast}$	&	0.883$^{\ast\ast\ast}$	&	0.859$^{\ast\ast\ast}$	&		&	0.993	&	0.971	&	0.94$^{\ast\ast}$	&	0.929$^{\ast\ast\ast}$	\\
TVP-BMA	&	1.089	&	0.981	&	1.099	&	0.825	&		&	1.268	&	1.198	&	1.490	&	1.091	\\
TVP-GAMP	&	0.988$^{\ast}$	&	0.870$^{\ast\ast\ast}$	&	0.749$^{\ast\ast\ast}$	&	0.714$^{\ast\ast\ast}$	&		&	1.020	&	0.965	&	0.915$^{\ast\ast\ast}$	&	0.866$^{\ast\ast\ast}$	\\ \hline
\end{tabular}
}
\end{center}
\par 
{\small \emph{Model acronyms are as follows: 
\textbf{KP-AR:} Koop and Potter (2007) structural breaks AR($p$) model;
\textbf{GK-AR:} Giordani and Kohn (2008) structural breaks AR($p$) model;
\textbf{TVP-AR:} Pettenuzzo and Timmermann (2017) time-varying parameter AR($p$) model;
\textbf{UCSV:} Stock and Watson (2007) unobserved components stochastic volatility;
\textbf{TVD:} Chan et al. (2012) time-varying dimension regression
\textbf{TVS:} Kalli and Griffin (2014) time-varying sparsity regression
\textbf{BMA:} George and McCulloch (1993) stochastic search variable selection regresison
\textbf{TVP-BMA:} Groen et al. (2012) time-varying Bayesian model averaging model
\textbf{TVP-GAMP:} Shrinkage representation of time-varying parameter regression, with generalized approximate message passing estimation}

\bigskip
\emph{Next to MSFE values the results of the Diebold-Mariano statistic are presented, with $^\ast$ significance at the 10\% level; $^{\ast\ast}$ significance at the 5\% level; 
$^{\ast\ast\ast}$ significance at the 1\% level.}
}
\end{table}

\begin{table}[H]
\begin{center}
\caption{Density forecast performance: log APLs relative to AR(2) benchmark.} \label{table:apl_results}
\begin{tabular}{lccccccccc}\hline
	&			\multicolumn{4}{c}{CPI}					&		&			\multicolumn{4}{c}{PCE deflator}					\\
	&	$h=1$	&	$h=3$	&	$h=6$	&	$h=12$	&		&	$h=1$	&	$h=3$	&	$h=6$	&	$h=12$	\\ \hline
KP-AR	&	0.090	&	0.081	&	0.002	&	-0.036	&		&	0.011	&	0.074	&	-0.057	&	-0.035	\\
GK-AR	&	-0.025	&	-0.029	&	0.004	&	0.037	&		&	0.034	&	0.138	&	0.056	&	0.052	\\
TVP-AR	&	0.118	&	0.111	&	0.181	&	0.067	&		&	0.036	&	0.007	&	-0.036	&	-0.029	\\
UCSV	&	0.161	&	0.239	&	0.224	&	0.144	&		&	0.048	&	0.245	&	-0.067	&	0.059	\\
TVD	&	-0.103	&	-0.005	&	-0.339	&	-0.380	&		&	-0.097	&	0.062	&	-0.885	&	-0.262	\\
TVS	&	0.018	&	-0.163	&	-0.660	&	-0.367	&		&	-0.001	&	0.003	&	-0.427	&	-0.246	\\
BMA	&	0.030	&	-0.067	&	0.042	&	0.084	&		&	-0.056	&	-0.002	&	-0.062	&	0.030	\\
TVP-BMA	&	0.121	&	0.313	&	0.413	&	0.399	&		&	-0.026	&	0.227	&	0.205	&	0.219	\\
TVP-GAMP	&	-0.204	&	0.258	&	0.320	&	0.321	&		&	0.061	&	0.260	&	0.045	&	0.191	\\ \hline
\end{tabular}
\end{center}
\par 
{\small \emph{See notes in \autoref{table:msfe_results} for details of model acronyms.}}

\end{table}

It is notable that these results contradict the previous claims that time-variation in parameters is important for inflation. The three models with the largest number of predictors, namely BMA and TVP-GAMP, and to a lesser degree TVP-BMA, seem to be improving a lot over time-varying parameter models with no predictors. The results seem to suggest that information in predictors is more important than the specification of time variation in regression parameters. This observation is not undermined by the fact that point forecasts from TVP-BMA are not significant, and that density forecasts from BMA are quite poor relative to TVP-BMA and TVP-GAMP. First, TVP-BMA is overparametrized\footnote{Shrinkage in TVP-BMA is only across predictors, but this model does not restrict the amount of time-variation in parameters.} its point forecast performance is not as good as the more conservative (in terms of time-variation in parameters, not available number of predictors) BMA and TVP-GAMP specifications. Second, when considering density forecasts, BMA is definitely misspecified since it does not allow for stochastic volatility, and it naturally doesn't perform as well as TVP-BMA and TVP-GAMP that allow for changing variance. Therefore, these findings suggest that TVP-GAMP is overall the best model and that its specification is flexible enough to capture both structural change and utilize information in a large set of predictors at the same time. Most importantly, the SBL prior allows to strike a good balance between these two modeling characteristics by removing irrelevant predictors as well as regularizing time variation.

These results are in stark contrast to existing results for TVP models presented in the papers cited above (see e.g. footnotes in Table 1). The culprit is simply the assumption that inflation is I(1) that Stock and Watson (1999) introduce in their seminal paper, and that it is adopted in equation \eqref{bencmark_forecasting_eq}. Once the random walk dynamics are removed from inflation (i.e. inflation gap becomes the dependent variable), the role of time-varying parameters in forecasting becomes less important and the most significant feature is the information included in exogenous predictors. It would be interesting then, as a robustness check, to specify the forecasting regression for inflation using the following form
\begin{equation}
\pi_{t+h}^{h} = \phi_{t,0} +  f_{t} \theta_{t} (L) +  \pi_{t} \mu_{t} (L) + e_{t+h}. \label{bencmark_forecasting_eq2}
\end{equation}
This equation is more in line with the forecasting model estimated in papers such as Chan et al. (2012), Groen et al. (2013), or Pettenuzzo and Timmermann (2017).

\autoref{table:msfe_results2} shows results based on this alternative specification of equation \eqref{bencmark_forecasting_eq2} for CPI inflation only. The left part of the table presents MSFE results, while the right panel presents log APLs. In this case it is evident that the various variants of TVP models considered improve tremendously over the benchmark. As a matter of fact, models such as the KP-AR, TVP-AR and UCSV also improve a lot relative to the constant parameter BMA. Looking at point forecasts and the associated MSFE results, we can observe many differences among TVP models, especially as the forecast horizon increases. For example, the structural breaks KP-AR specification has the lowest relative MSFE for $h=12$ among all models, but the also structural breaks GK-AR specification is among the worst performing (but still much better than the simple AR model). TVD and TVS estimated with the monthly data are not only cumbersome, but also do not perform as well as TVP models with no predictors. In contrast, the TVP-BMA algorithm is performing quite well, even though it still doesn't beat TVP models with no predictors. In this alternative forecasting regression, TVP-GAMP is not the top forecasting model but its performance is still quite good. If it wasn't for the exceptional performance of the KP-AR model, TVP-GAMP would have been a top model for $h=1,3,6$. 

When looking at density forecast evaluation the results might not comply with the results for the point forecasts. Still good performing models are the KP-AR and the TVP-AR, but now the BMA and TVP-GAMP beat models such as the UCSV. With such diverse set of flexible models it is hard to pin down which exact features help in point and density forecasts. Nevertheless, for the forecasting regression \eqref{bencmark_forecasting_eq2} it seems that the way time variation in parameters is specified is more important than information in exogenous predictors. Further numerical evidence on the relative forecast performance of some of the competing models, is provided in Online Appendix, Section D.2.

\bigskip
\begin{table}[H]
\begin{center}
\caption{Point and density forecast performance using alternative definition of the CPI forecasting regression}  \label{table:msfe_results2}
\begin{tabular}{llllllllll}\hline
	&			\multicolumn{4}{c}{\underline{MSFE}}					&		&			\multicolumn{4}{c}{\underline{log APL}}					\\
	&	$h=1$	&	$h=3$	&	$h=6$	&	$h=12$	&		&	$h=1$	&	$h=3$	&	$h=6$	&	$h=12$	\\ \hline
KP-AR	&	0.901	&	0.706	&	0.756	&	0.544	&		&	0.042	&	0.321	&	0.154	&	0.128	\\
GK-AR	&	0.963	&	0.929	&	0.900	&	0.882	&		&	0.071	&	0.168	&	0.014	&	0.125	\\
TVP-AR	&	0.852	&	0.917	&	0.800	&	0.587	&		&	0.210	&	0.353	&	0.422	&	0.114	\\
UCSV	&	0.911	&	0.898	&	0.817	&	0.638	&		&	0.114	&	0.163	&	0.118	&	0.154	\\
TVD	&	0.902	&	0.851	&	0.863	&	0.873	&		&	-0.041	&	0.150	&	0.022	&	0.021	\\
TVS	&	0.960	&	0.929	&	0.891	&	0.905	&		&	0.033	&	0.134	&	0.033	&	0.037	\\
BMA	&	0.995	&	1.109	&	1.233	&	0.914	&		&	0.118	&	0.273	&	0.151	&	0.187	\\
TVP-BMA	&	0.926	&	0.903	&	0.805	&	0.650	&		&	0.092	&	0.088	&	0.087	&	0.126	\\
TVP-GAMP	&	0.944	&	0.876	&	0.819	&	0.768	&		&	0.190	&	0.276	&	0.264	&	0.136	\\ \hline
\end{tabular}
\end{center}
\par 
{\small \emph{See notes in \autoref{table:msfe_results} for details of model acronyms.}

\bigskip
\emph{Unlike the previous two tables that present results for both CPI and PCE, this table only shows results for CPI, where its left panel focuses on MSFEs and its right panel on log APLs. However, as in the previous two tables, MSFEs and log APLs are relative to an AR(2) benchmark. MSFE entries lower than one mean that the estimation method of the respective row does better than the benchmark. Log APL entries higher than zero mean that the estimation method of the respective row does better than the benchmark.}
}
\end{table}

\section{Conclusions}
This paper evaluates a new methodology for performing Bayesian inference in high-dimensional regression models. The proposed Generalized Approximate Message Passing (GAMP) is a fast algorithm for approximating iteratively the first two moments of the marginal posterior distribution of a high-dimensional vector of coefficients. It is established how effortlessly the GAMP algorithm can be extended with interesting modeling features such as hierarchical shrinkage priors, time-varying coefficients and stochastic volatility, and many predictors. The benefit of the proposed approach is demonstrated using an inflation forecasting exercise that leads to the recursive estimation of regression models with thousands of covariates. Due to the low algorithmic complexity, GAMP could be generalized to much higher dimensions with millions of predictors/covariates, as it is also trivially parallelizable.

The current study opens up new avenues for research. First, the proposed framework for modeling time-vayring parameters using hierarchical shrinkage priors can be extended in interesting ways. For example, shrinkage estimators/priors that apply on group of coefficients (such as the Group Lasso) can be used in this setting so that coefficients are shrunk either in groups of predictors for a given time period or in groups of consecutive time periods for a given predictor. This is because in the TVP setting the vector of regression coefficients $\beta$ has elements that correspond both to predictor $j$, $j=1,...,p$, but also to time period $t$, $t=1,...,T$. One can think of other shrinkage priors in order to perform a more structured approach to uncovering patterns of time-variation in parameters, such as various pooling priors used in the panel data literature. Finally, the paper proposes the framework of factor graphs for designing efficient algorithms. Many macroeconomic problems currently do not typically involve extensive use of Big Data sets, however, they involve multivariate models with possibly thousands of coefficients, such as VAR, factor, and DSGE models. Bayesian estimation of these models is quite cumbersome, many times relying on linear or nonlinear state-space methods. As empirical macroeconomic models become larger and more complex, factor graph inference could help economists come up with novel efficient algorithms and unveil new features in macroeconomic data.

\newpage
\begin{center}
\LARGE{Online Appendix to ``High-dimensional macroeconomic forecasting using message passing algorithms''} \\
\hfill \\
\large{Dimitris Korobilis}
\end{center}

\begin{appendix}
\setcounter{page}{1}
\renewcommand{\theequation}{A.\arabic{equation}} \setcounter{equation}{0} %
\renewcommand{\thetable}{A\arabic{table}} \setcounter{table}{0}
\setcounter{footnote}{0}

\section{Data Appendix}
All series were downloaded from Michael McCracken's FRED-MD database ($https://research.stlouisfed.org/econ/mccracken/fred-databases/$) and cover the period 1959M1 to 2016M6. All series are seasonally adjusted and all variables are transformed to be approximately stationary. In particular, if $w_{i,t}$ is the original un-transformed series in levels,
when the series is used as a predictor the transformation codes (column T of the table) are: 1 - no transformation
(levels), $x_{i,t}=w_{i,t}$; 2 - first difference, $%
x_{i,t}=w_{i,t}-w_{i,t-1} $ ; 3- second difference, $x_{i,t}=\Delta
w_{i,t}-\Delta w_{i,t-1}$ 4 - logarithm, $x_{i,t}=\log w_{i,t}$; 5 - first
difference of logarithm, $x_{i,t}=\log w_{i,t}-\log w_{i,t-1}$; 6 - second
difference of logarithm, $x_{i,t}=\Delta \log w_{i,t}-\Delta \log w_{i,t-1}$.

{\small
\begin{longtable}{llcl}
\caption[Monthly U.S. macro data set (based on FREDMD)]{Monthly U.S. macro data set (based on FREDMD)}\label{datatable} \\
No & Mnemonic	&	T	&	Long description	\\ \hline \hline \endfirsthead
 & & & \\ \hline \endfoot
\multicolumn{4}{l}{Table \ref{datatable} (continued)} \\ \hline \hline \endhead
1	&	RPI	&	5	&	   Real  Personal Income   	\\
2	&	W875RX1	&	5	&	   RPI ex.  Transfers   	\\
3	&	DPCERA3M086SBEA	&	5	&	   Real  PCE   	\\
4	&	CMRMTSPLx	&	5	&	   Real  M\&T  Sales	\\
5	&	RETAILx	&	5	&	   Retail and  Food Services Sales   	\\
6	&	INDPRO	&	5	&	   IP Index   	\\
7	&	IPFPNSS	&	5	&	   IP: Final Products and  Supplies   	\\
8	&	IPFINAL	&	5	&	   IP: Final Products   	\\
9	&	IPCONGD	&	5	&	   IP: Consumer Goods   	\\
10	&	IPDCONGD	&	5	&	   IP: Durable Consumer Goods   	\\
11	&	IPNCONGD	&	5	&	   IP: Nondurable Consumer Goods   	\\
12	&	IPBUSEQ	&	5	&	   IP: Business Equipment   	\\
13	&	IPMAT	&	5	&	   IP: Materials   	\\
14	&	IPDMAT	&	5	&	   IP: Durable Materials   	\\
15	&	IPNMAT	&	5	&	   IP: Nondurable Materials   	\\
16	&	IPMANSICS	&	5	&	   IP: Manufacturing   	\\
17	&	IPB51222S	&	5	&	   IP: Residential Utilities   	\\
18	&	IPFUELS	&	5	&	   IP: Fuels   	\\
19	&	CUMFNS	&	2	&	   Capacity Utilization:  Manufacturing   	\\
20	&	HWI	&	2	&	   Help-Wanted Index for U.S.   	\\
21	&	HWIURATIO	&	2	&	   Help Wanted to Unemployed ratio   	\\
22	&	CLF16OV	&	5	&	   Civilian Labor Force   	\\
23	&	CE16OV	&	5	&	   Civilian Employment   	\\
24	&	UNRATE	&	2	&	   Civilian Unemployment Rate   	\\
25	&	UEMPMEAN	&	2	&	   Average Duration of  Unemployment   	\\
26	&	UEMPLT5	&	5	&	   Civilians Unemployed $\le 5$ Weeks   	\\
27	&	UEMP5TO14	&	5	&	   Civilians Unemployed 5-14  Weeks   	\\
28	&	UEMP15OV	&	5	&	   Civilians  Unemployed $> 15$ Weeks   	\\
29	&	UEMP15T26	&	5	&	   Civilians Unemployed 15-26  Weeks   	\\
30	&	UEMP27OV	&	5	&	   Civilians Unemployed $>27$ Weeks   	\\
31	&	CLAIMSx	&	5	&	   Initial Claims   	\\
32	&	PAYEMS	&	5	&	   All Employees: Total nonfarm   	\\
33	&	USGOOD	&	5	&	   All Employees: Goods-Producing   	\\
34	&	CES1021000001	&	5	&	   All Employees: Mining and  Logging  	\\
35	&	USCONS	&	5	&	   All Employees: Construction   	\\
36	&	MANEMP	&	5	&	   All Employees: Manufacturing   	\\
37	&	DMANEMP	&	5	&	   All  Employees: Durable goods   	\\
38	&	NDMANEMP	&	5	&	   All Employees: Nondurable goods   	\\
39	&	SRVPRD	&	5	&	   All  Employees: Service Industries   	\\
40	&	USTPU	&	5	&	   All Employees: TT\&U	\\
41	&	USWTRADE	&	5	&	   All  Employees: Wholesale Trade   	\\
42	&	USTRADE	&	5	&	   All Employees: Retail Trade   	\\
43	&	USFIRE	&	5	&	   All  Employees: Financial Activities   	\\
44	&	USGOVT	&	5	&	   All Employees: Government   	\\
45	&	CES0600000007	&	1	&	   Hours: Goods-Producing   	\\
46	&	AWOTMAN	&	2	&	   Overtime Hours: Manufacturing   	\\
47	&	AWHMAN	&	1	&	   Hours: Manufacturing   	\\
48	&	HOUST	&	4	&	   Starts: Total   	\\
49	&	HOUSTNE	&	4	&	   Starts: Northeast   	\\
50	&	HOUSTMW	&	4	&	   Starts: Midwest   	\\
51	&	HOUSTS	&	4	&	   Starts: South   	\\
52	&	HOUSTW	&	4	&	   Starts: West   	\\
53	&	AMDMNOx	&	5	&	   Orders: Durable Goods   	\\
54	&	AMDMUOx	&	5	&	   Unfilled Orders: Durable Goods   	\\
55	&	BUSINVx	&	5	&	   Total Business Inventories   	\\
56	&	ISRATIOx	&	2	&	   Inventories to Sales  Ratio   	\\
57	&	M1SL	&	6	&	   M1  Money Stock   	\\
58	&	M2SL	&	6	&	   M2  Money Stock   	\\
59	&	M2REAL	&	5	&	   Real  M2  Money Stock   	\\
60	&	BUSLOANS	&	6	&	   Commercial and  Industrial Loans   	\\
61	&	REALLN	&	6	&	   Real  Estate Loans   	\\
62	&	NONREVSL	&	6	&	   Total Nonrevolving Credit   	\\
63	&	CONSPI	&	2	&	   Credit to PI ratio   	\\
64	&	S\&P 500	&	5	&	   S\&P 500   	\\
65	&	S\&P: indust	&	5	&	   S\&P Industrial 	\\
66	&	S\&P div yield	&	2	&	   S\&P Divident yield   	\\
67	&	S\&P PE ratio	&	5	&	   S\&P Price/Earnings  ratio   	\\
68	&	FEDFUNDS	&	2	&	   Effective Federal Funds Rate   	\\
69	&	CP3Mx	&	2	&	   3-Month AA Comm. Paper Rate   	\\
70	&	TB3MS	&	2	&	   3-Month T-bill   	\\
71	&	TB6MS	&	2	&	   6-Month T-bill   	\\
72	&	GS1	&	2	&	   1-Year T-bond   	\\
73	&	GS5	&	2	&	   5-Year T-bond   	\\
74	&	GS10	&	2	&	   10-Year T-bond   	\\
75	&	AAA	&	2	&	   Aaa  Corporate Bond Yield   	\\
76	&	BAA	&	2	&	   Baa  Corporate Bond Yield   	\\
77	&	COMPAPFFx	&	1	&	   CP  - FFR spread   	\\
78	&	TB3SMFFM	&	1	&	   3 Mo. - FFR spread   	\\
79	&	TB6SMFFM	&	1	&	   6 Mo. - FFR spread   	\\
80	&	T1YFFM	&	1	&	   1 yr. - FFR spread   	\\
81	&	T5YFFM	&	1	&	   5 yr. - FFR spread   	\\
82	&	T10YFFM	&	1	&	   10 yr. - FFR spread   	\\
83	&	AAAFFM	&	1	&	   Aaa  - FFR spread   	\\
84	&	BAAFFM	&	1	&	   Baa  - FFR spread   	\\
85	&	EXSZUSx	&	5	&	   Switzerland / U.S.  FX Rate   	\\
86	&	EXJPUSx	&	5	&	   Japan / U.S.  FX Rate   	\\
87	&	EXUSUKx	&	5	&	   U.S.  / U.K. FX Rate   	\\
88	&	EXCAUSx	&	5	&	   Canada / U.S.  FX Rate   	\\
89	&	WPSFD49207	&	6	&	   PPI: Final demand less energy	\\
90	&	WPSFD49502	&	6	&	   PPI: Personal cons	\\
91	&	WPSID61	&	6	&	   PPI: Processed goods	\\
92	&	WPSID62	&	6	&	   PPI: Unprocessed goods	\\
93	&	OILPRICEx	&	6	&	   Crude Oil Prices: WTI   	\\
94	&	PPICMM	&	6	&	   PPI: Commodities   	\\
95	&	CPIAUCSL	&	6	&	   CPI: All Items   	\\
96	&	CPIAPPSL	&	6	&	   CPI: Apparel   	\\
97	&	CPITRNSL	&	6	&	   CPI: Transportation   	\\
98	&	CPIMEDSL	&	6	&	   CPI: Medical Care   	\\
99	&	CUSR0000SAC	&	6	&	   CPI: Commodities   	\\
100	&	CUUR0000SAD	&	6	&	   CPI: Durables   	\\
101	&	CUSR0000SAS	&	6	&	   CPI: Services   	\\
102	&	CPIULFSL	&	6	&	   CPI: All Items Less  Food   	\\
103	&	CUUR0000SA0L2	&	6	&	   CPI: All items less shelter   	\\
104	&	CUSR0000SA0L5	&	6	&	   CPI: All items less medical care   	\\
105	&	PCEPI	&	6	&	   PCE: Chain-type Price Index   	\\
106	&	DDURRG3M086SBEA	&	6	&	   PCE: Durable goods   	\\
107	&	DNDGRG3M086SBEA	&	6	&	   PCE: Nondurable goods   	\\
108	&	DSERRG3M086SBEA	&	6	&	   PCE: Services   	\\
109	&	CES0600000008	&	6	&	   Ave. Hourly Earnings: Goods   	\\
110	&	CES2000000008	&	6	&	   Ave. Hourly Earnings: Construction   \\
111	&	CES3000000008	&	6	&	   Ave. Hourly Earnings: Manufacturing   \\
112	&	MZMSL	&	6	&	   MZM Money Stock   	\\
113	&	DTCOLNVHFNM	&	6	&	   Consumer Motor Vehicle Loans   	\\
114	&	DTCTHFNM	&	6	&	   Total Consumer Loans and  Leases   	\\
115	&	INVEST	&	6	&	   Securities in Bank Credit   	\\
\end{longtable}
}

\bigskip

\newpage
\renewcommand{\theequation}{B.\arabic{equation}} \setcounter{equation}{0} %
\renewcommand{\thetable}{B\arabic{table}} \setcounter{table}{0}
\renewcommand{\thealgorithm}{B\arabic{algorithm}} \setcounter{algorithm}{0}

\section{Technical Appendix}
\subsection{Generic Derivation of Generalized Approximate Message Passing Algorithm}
The basic signal extraction problem in engineering (but using traditional regression notation) involves observations $y$, a known ``transform matrix'' $x$ and the ``signal'' $\beta$. For the purpose of making the algorithm usable in traditional regression problems, we also assume an ``additive white Gaussian noise'' (AWGN), as the disturbance term is called in signal processing. That is, we assume the following model
\begin{equation}
y = x \beta + \varepsilon,
\end{equation}
where $y$ is $T \times 1$, $x$ is $T \times q$, $\beta$ is $q \times 1$ and $\varepsilon \sim N(0,\sigma \times I_{T})$ is also $T \times 1$. The TVP regression problem falls into this general form, but with $x$ replaced with $\EuScript{X}$, which is defined in the main text. In order to be consistent with Rangan (2011) and I denote the likelihood function as $p\left( y | z \right)$ where $z=x \beta$. Rangan (2011) has proposed two variants of GAMP, one for maximum a-posteriori (MAP) estimation of the signal $\beta$ and one for minimum mean square error (MMSE) estimation. Both have different properties, but the focus of this paper is on MMSE estimation due to its modularity and usability in regression problems familiar to macroeconomists. The derivation of the algorithm is based on the factor graph in the main text. In the following the function $\mu_{a \rightarrow b}$ will denote a message from $a$ to $b$, and $m_{a \rightarrow b}$ its logarithm. 

The starting point are the iterations of sum-product / loopy Belief Propagation derived in the main text, which can be written in logarithmic form\footnote{For notational simplicity and clarity I am ignoring any normalizing constants that enter the logarithmic expressions additively. Note that for the same reason I am ignoring any hyperparameters that $p \left( \beta \right)$ might rely on, as is the case with the hierarchical shrinkage prior used in this paper.} as
\begin{eqnarray} \label{eqn:BP1_App1}
m_{ p(y_{t} | z_{t}) \rightarrow \beta_i }^{(r+1)} & = & \log \left( \int p \left(y_t | z_{t} \right) \prod_{j=1,j \neq i} ^{q} \mu_{ \beta_j \rightarrow p(y_{t} | z_{t}) }^{(r)} d\beta_{j \neq i} \right) = \log E_{z_{t}} \left[ p \left( y_{t} \vert z_{t} \right) \right],  \\ 
\label{eqn:BP2_App1}
m_{ \beta_i \rightarrow p(y_{t} | z_{t}) }^{(r+1)} & = & \log \left( p \left( \beta_i \right) \right) + \sum_{\mathcal{s}=1, \mathcal{s} \neq t}^{T}  m_{ p(y_{\mathcal{s}} | \beta) \rightarrow \beta_i }^{(r)},
\end{eqnarray}
where $E_{z_{t}} \left[ p \left( y_{t} \vert z_{t} \right) \right]$ is the expectation of $p \left( y_{t} \vert z_{t} \right)$ over $z_{t}$ (assuming the $\beta_j$ are distributed independently according to $\mu_{ \beta_j \rightarrow p(y_{t} | z_{t}) }$). Under this scheme, the marginal posterior of $\beta_i$ can be approximated with
\begin{equation}
p \left( \beta_{i} \vert y \right) \propto exp \left[ m_{i} \right],
\end{equation}
where $m_{i}$ is defined as
\begin{equation}
m_{i} \equiv \log \left( p \left( \beta_{i} \right) \right) + \sum_{t=1}^{T} m_{ p(y_{t} | z_{t}) \rightarrow \beta_i }.  \label{eqn:m_i}
\end{equation}

There are several algorithms for approximating the iterations of the sum-product loopy Belief Propagation, for instance, Wand (2017) shows the benefits of Mean Field Variational Bayes (MFVB) methods in semiparametric regression that can apply to a wide class of hierarchical regression models. All GAMP does at this point is to introduce Gaussian approximations to the BP iterations. In order to pin down all the necessary proofs we need the following result replicated from Rangan (2011)
\begin{lemma}
Consider a random variable $U$ with a conditional probability density function of the form
\begin{equation}
p \left( u \vert v \right) \equiv \frac{1}{Z\left( v \right)} exp \left( \phi(0) + uv \right),
\end{equation}
where $Z\left( v \right)$ is a normalization constant (called the partition function). Then
\begin{eqnarray}
\frac{\partial}{\partial v} \ln Z\left( v \right) & = & E \left( U \vert V=v \right), \\
\frac{\partial^{2}}{\partial^{2} v} \ln Z\left( v \right) & = & \frac{\partial}{\partial v} E \left( U \vert V=v \right) \\
 & = & var \left( U \vert V=v \right).
\end{eqnarray}
Proof: The relations are standard properties of exponential families. \QEDA
\end{lemma}
\bigskip
Additionally, let
\begin{eqnarray}
    \widehat{\beta}_{i} & \equiv & E \left( \beta_{i}  \vert m_{i} \right), \label{def1} \\
    \widehat{\tau}^{\beta}_{i} & \equiv & var \left(  \beta_{i}  \vert m_{i} \right), \label{def2} \\
    \widehat{\beta}_{i \rightarrow t} & \equiv & E \left(  \beta_{i}  \vert m_{ \beta_j \rightarrow p(y_{t} | z_{t}) } \right), \label{def3} \\    
    \widehat{\tau}^{\beta}_{i \rightarrow t} & \equiv & var \left(  \beta_{i}  \vert m_{ \beta_j \rightarrow p(y_{t} | z_{t}) } \right), \label{def4} 
\end{eqnarray}
where $E \left( x \vert f \right)$ denotes the expectation of random variable $x$ conditional on the function (density) $f$, and $m_{i}$ and $m_{ \beta_j \rightarrow p(y_{t} | z_{t}) }$ are defined in eqs. \eqref{eqn:m_i} and \eqref{eqn:BP2_App1}, respectively. 

\subsubsection*{Part 1: Messages from factor nodes to variable nodes}
Fist we approximate the messages emitted by the function (``output'') nodes, found in equation \eqref{eqn:BP1_App1}. Notice that this equation is equivalently interpreted as the (logarithm of the) expectation of $p\left( y_{t} \vert z_{t}\right)$ with respect to $z_{t}$ being distributed as $\prod_{j=1,j \neq i} ^{q} \mu_{ \beta_j \rightarrow p(y_{t} | z_{t}) }$. Given that $p\left( y_{t} \vert z_{t}\right)$ is the regression likelihood, we can write
\begin{eqnarray}
m_{ p(y_{t} | z_{t}) \rightarrow \beta_i } & \propto & \log E_{z_{t}} \left[ p \left( y_{t} \vert z_{t} \right) \right] \\
& = & \log E_{z_{t}} \left( -\frac{1}{2} \frac{\left(y_{t} - z_{t}\right)^{2}}{\sigma^{2}}\right) \\
& = &  \log E_{z_{t}} \left( -\frac{1}{2} \frac{\left(y_{t} - x_{t,i}\beta_{i} - \sum_{j \neq i} x_{t,j}\beta_{j}\right)^{2}}{\sigma^{2}}\right),
\end{eqnarray}
where $z_{t} = x_{t}\beta = \sum_{j=1}^{q} x_{t,j}\beta_{j} = x_{t,i}\beta_{i} + \sum_{j \neq i} x_{t,j}\beta_{j}$. Based on the Barry-Esseen central limit theorem, $z_{t}$ conditional on $\beta_{i}$ can be approximated by a Normal distribution with mean and variance
\begin{eqnarray}
E \left( z_{t} | \beta_{i} \right) = x_{t,i} \beta_{i} + \sum_{j\neq i} x_{t,j} E \left(  \beta_{i}  \vert m_{ \beta_j \rightarrow p(y_{t} | z_{t}) } \right), \label{z_message} \\ 
var \left( z_{t} | \beta_{i} \right) = x_{t,i}^{2} var(\beta_{i}) +  \sum_{j\neq i} x_{t,j}^{2} var \left(  \beta_{i}  \vert m_{ \beta_j \rightarrow p(y_{t} | z_{t}) } \right), \label{var_z_message}
\end{eqnarray}
where the variance of $\beta_{i}$ in equation \eqref{var_z_message} above is zero simply because we condition on $\beta_{i}$. Given the definitions in \eqref{def3} and \eqref{def4} we can then write
\begin{equation}
z_{t} \vert \beta_{i} \sim N \left( x_{t,i} \beta_{i} + \sum_{j\neq i} x_{t,j}\widehat{\beta}_{i \rightarrow t},   \sum_{j\neq i} x_{t,j}^{2}\widehat{\tau}^{\beta}_{i \rightarrow t}    \right),
\end{equation}
and the output message of the BP iteration is of the form
\begin{eqnarray}
m_{ p(y_{t} | z_{t}) \rightarrow \beta_i } & \propto & \log E_{z_{t}} \left[ p \left( y_{t} \vert z_{t} \right) \right] \\
& \approx & \log \int p \left(y_t | z_{t} \right)  N \left( x_{t,i} \beta_{i} + \sum_{j\neq i} x_{t,j}\widehat{\beta}_{j \rightarrow t},   \sum_{j\neq i} x_{t,j}^{2}\widehat{\tau}^{\beta}_{j \rightarrow t}    \right) dz_{t}.
\end{eqnarray}
In order to proceed further we will use the following definitions:
\begin{eqnarray}
\widehat{c}_{t \rightarrow i}  &=   \sum_{j\neq i} x_{t,j}\widehat{\beta}_{j \rightarrow t},  \text{ \ \ \ \ \ \ \ \ \ \ \ \ }    \widehat{\tau}_{t \rightarrow i}^{c} &=  \sum_{j\neq i} x_{t,j}^{2}\widehat{\tau}^{\beta}_{j \rightarrow t},
\\
\widehat{c}_{t} & = \sum_{i=1}^{q} x_{t,i} \widehat{\beta}_{i \rightarrow t},   \text{ \ \ \ \ \ \ \ \ \ \ \ \ \ \ }   \widehat{\tau}_{t}^{c}  &=  \sum_{i=1}^{q} x_{t,i}^{2}\widehat{\tau}^{\beta}_{i \rightarrow t}, \\
 H \left( \widehat{c},\widehat{\tau}^{c},y \right) & \equiv  \log E \left( p(y \vert z)\right), \text{ \ \ } g_{out}\left( \widehat{c},\widehat{\tau}^{c},y \right) & = \frac{\partial}{\partial \widehat{c}} H\left( \widehat{c},\widehat{\tau}^{c},y \right), \\
\widehat{s}_{t} & = g_{out}\left( \widehat{c}_{t},\widehat{\tau}_{t}^{c},y \right), \text{ \ \ \ \ \ \ \ \ \ \ \ \ \ }  \tau_{t}^{s} & = \frac{\partial}{\partial \widehat{c}} g_{out}\left( \widehat{c}_{t},\widehat{\tau}_{t}^{c},y \right).
\end{eqnarray}
Using these definitions we can now rewrite the message $m_{ p(y_{t} | z_{t}) \rightarrow \beta_i }$ as
\begin{eqnarray}
m_{ p(y_{t} | z_{t}) \rightarrow \beta_i } & = & H\left( \widehat{c}_{t \rightarrow i} + x_{i,t}\beta_{i}, \widehat{\tau}, y  \right) \\
& = & H\left( \widehat{c}_{t} + x_{i,t} \left( \beta_i - \widehat{\beta}_i \right), \widehat{\tau}, y \right). \label{H_fun_app}
\end{eqnarray} 
A second order approximation of equation \eqref{H_fun_app} gives
\begin{equation}
m_{ p(y_{t} | z_{t}) \rightarrow \beta_i }  \approx \widehat{s}_{t}x_{i,t} \left( \beta_i - \widehat{\beta}_i \right) - \frac{1}{2} \widehat{\tau}^{s}\left( x_{i,t} \left( \beta_i - \widehat{\beta}_i \right) \right)^{2}. \label{Taylor}
\end{equation}
Therefore this new approximation implies that calculation of $m_{ p(y_{t} | z_{t}) \rightarrow \beta_i }$ relies on knowledge of $\widehat{s}$ and $\widehat{\tau}^{s}$, that is, knowledge of the function $g_{out}$ and its derivative. After some algebra it can be shown (Rangan, 2011) that
\begin{eqnarray}
g_{out}\left( \widehat{c}, \widehat{\tau}^{c},y \right) & = & \frac{1}{\tau^{c}} \left( \widehat{z} - \widehat{c} \right), \\
\frac{\partial}{\partial \widehat{c}} g_{out}\left( \widehat{c}, \widehat{\tau}^{c},y \right) & = & \frac{1}{\widehat{\tau}^{c}} \left( \frac{1}{\widehat{\tau}^{c}} \widehat{\tau}^{z} - 1 \right),
\end{eqnarray}
where $\widehat{z} = E \left(z \vert \widehat{c}, \widehat{\tau}^{c} \right) \equiv \frac{\int z p \left( y\vert z \right) N\left(z \vert \widehat{c},\widehat{\tau}^{c} \right) dz}{\int p \left( y\vert z \right) N\left(z \vert \widehat{c},\widehat{\tau}^{c} \right) dz}$ and $\widehat{\tau}^{z} = var \left(z \vert \widehat{c} , \widehat{\tau}^{c} \right) \equiv \frac{\partial}{\partial \widehat{c} } \widehat{z}$.

\subsubsection*{Part 2: Messages from variable nodes to factor nodes}
We now need to consider the messages $m_{ \beta_j \rightarrow p(y_{t} | z_{t}) }^{(r+1)}$ and expand equation \eqref{eqn:BP2_App1}. Based on the second order Taylor expansion in equation \eqref{Taylor}, we can now write
{\footnotesize
\begin{eqnarray}
m_{ \beta_i \rightarrow p(y_{t} | z_{t}) } & = & \log \left( p \left( \beta_i \right) \right) + \sum_{\mathcal{s}=1, \mathcal{s} \neq t}^{T}  m_{ p(y_{\mathcal{s}} | \beta) \rightarrow \beta_i } \\
& \approx & \log \left( p \left( \beta_i \right) \right) + \sum_{\mathcal{s}=1, \mathcal{s} \neq t}^{T} \left[ \widehat{s}_{\mathcal{s}}x_{i,\mathcal{s}} \left( \beta_i - \widehat{\beta}_i \right) - \frac{1}{2} \widehat{\tau}_{\mathcal{s}}^{s}\left( x_{i,\mathcal{s}} \left( \beta_{i} - \widehat{\beta}_{i} \right) \right)^{2} \right]. \label{BP2_new}
\end{eqnarray}    }
Define next the quantities
\begin{eqnarray}
\widehat{d}_{i \rightarrow t} &= \widehat{\tau}_{i \rightarrow t}^{d} \sum_{\mathcal{s}=1, \mathcal{s} \neq t}^{T}  \left( \widehat{s}_{\mathcal{s}}x_{i,\mathcal{s}} + \widehat{\tau}_{\mathcal{s}}^{s}x_{i,\mathcal{s}}^{2} \widehat{\beta}_{i} \right),  \text{ \ \ \ \ } \widehat{\tau}_{i \rightarrow t}^{d} &= \left[ \sum_{\mathcal{s}=1, \mathcal{s} \neq t}^{T}  \widehat{\tau}_{\mathcal{s}}^{s}x_{i,\mathcal{s}}^{2} \right]^{-1}, \\
\widehat{d}_{i}  &= \widehat{\beta}_{i} + \widehat{\tau}_{i}^{d} \sum_{t=1}^{T} x_{i,t}\widehat{s}_{t} ,  \text{ \ \ \ \ \ \ \ \ \ \ \ \ \ \ \ \ \ \ \ \ \ \ \ \ } \widehat{\tau}_{i}^{d} &= \left[ \sum_{t=1}^{T}  \widehat{\tau}_{t}^{s}x_{i,t}^{2} \right]^{-1}, \\
\widehat{d}_{i \rightarrow t} &=\widehat{d}_{i} - \widehat{\tau}_{i}^{d}x_{i,t} \widehat{s}_{t},  \text{ \ \ \ \ \ \ \ \ \ \ \ \ \ \ \ \ \ \ \ \ \ \ \ \ \ \ \ \ \ \ } \widehat{\tau}_{i \rightarrow t}^{d} &\approx \widehat{\tau}_{i}^{d}. \\
\end{eqnarray}
We can write equation \eqref{BP2_new} as
\begin{eqnarray}
m_{ \beta_i \rightarrow p(y_{t} | z_{t}) }  &  = & \log \left( p \left( \beta_i \right) \right) - \frac{1}{2 \widehat{\tau}^{d}_{i \rightarrow t}} \left( \widehat{d}_{i \rightarrow t} - \beta_{i}\right)^{2}\\
&  = & \log \left( p \left( \beta_i \right) \right) - \frac{1}{2 \widehat{\tau}^{d}_{i}} \left( \widehat{d}_{i} - \widehat{\tau}_{i}^{d}x_{i,t} \widehat{s}_{t} - \beta_{i}\right)^{2}.
\end{eqnarray}
Therefore the input scalar function $g_{in}$ and its derivative are of the form
\begin{eqnarray}
g_{in} \left( \widehat{d}_{i}, \widehat{\tau}_{i}^{d} \right) &=& E \left(\beta_{i} \vert \widehat{d}_{i}, \widehat{\tau}_{i}^{d}\right), \\
\frac{\partial}{\partial \widehat{d}_{i}} g_{in} \left( \widehat{d}_{i}, \widehat{\tau}_{i}^{d} \right) &=& \frac{1}{\widehat{\tau}_{i}^{d}} var \left(\beta_{i} \vert \widehat{d}_{i}, \widehat{\tau}_{i}^{d}\right).
\end{eqnarray}
Exact functional forms obviously depend on the form of the prior. After outlining this generic, and rather tedious proof of GAMP, in the next Section I provide exact numerical details for the case of the model and prior presented in the main text.

\subsection{GAMP algorithm for time-varying parameter regression with sparse Bayesian learning prior}

Consider the regression model
\begin{equation*}
y_{t}=x_{t}\beta +\varepsilon _{t},
\end{equation*}%
where $\varepsilon _{t}\sim N\left( 0,\sigma_{t}^{2}\right) ,y_{t}$ is scalar, $%
x_{t}$ is $1\times q$ vector, and consider the prior distribution $\beta
\sim N_{p}\left( 0,\underline{V} \right) $. In the main paper the matrix of predictors was $\EuScript{X}_{t}$ and it had a certain block-diagonal structure, but the algorithm below holds for any non-sparse or non-block-diagonal matrix $x_{t}$. The prior is of the form
\begin{eqnarray}
p \left( \beta_i \vert \alpha_i \right) & \sim & N \left( 0,\alpha_{i} \right), \\
p \left( \alpha_{i}^{-1} \right) & \sim & Gamma \left( \underline{a},\underline{b} \right).
\end{eqnarray}
As also shown in Zou et al. (2016), the core GAMP algorithm presented in the main text can now be augmented to accommodate EM-like updates for the hyperparameters $\alpha_{i}$. Optimizing with respect to $\alpha$ in iteration $(r+1)$ means finding the maximum of the following Q-function
\begin{equation}
\alpha^{\left( r+1 \right)} = \arg \max_{\alpha} Q \left(\alpha \vert \alpha^{\left( r \right)} \right) \equiv E_{\alpha^{(r)}} \left[ \log p \left(\alpha \vert y, \beta^{(r)} \right) \right].
\end{equation}
Taking the derivative of the Q-function w.r.t. $\alpha_i$ and setting it to zero gives the usual formula found also in variational Bayes and Gibbs-sampler updates of $\alpha$
\begin{equation}
\alpha_{i}^{\left( r+1 \right)} = \frac{2\underline{a}-1}{2\underline{b} + \left( \widehat{\beta}_{i}^{(r)} \right)^{2}}, \label{Malp1}
\end{equation}
where $\widehat{\beta}_{i}^{(r)}$ is some estimate of $\beta_{i}$ in the previous iteration $(r)$.

The algorithm below is an extension of the original Generalized Approximate Message Passing (GAMP) algorithm of Rangan et al. (2016) that incorporates a step for updating stochastic volatility and an EM-like step for updating the prior hyperparameters of the sparse Bayesian learning prior. Inside this algorithm I denote with $\widehat{x}^{(r)}$ and $\widehat{\tau}^{x,(r)}$ the estimates of the mean and variance of quantity $x$, respectively, at the $r$-th iteration of the algorithm. For presentational simplicity in the algorithm below the expressions for the means of various parameters are presented before the expressions for their variances, even though in practice variances have to be calculated first in order for means to be subsequently calculated. Finally note that, as explained in the main text, the values of  $\mu_{i}$, $V_{i}$, $\pi_{i}$ for each of the seven components of the mixture approximation are those given in \autoref{table:mix_comp}.

\begin{table}[H]
\centering
\caption{Seven-component mixture approximation to a $\log \chi^2$ distribution with one degree of freedom} \label{table:mix_comp}
\begin{tabular}{cccc}\hline
component & $\pi_{i}$ &  $\mu_{i}$ & $V_{i}$ \\ \hline
1	&	0.00730	&	-10.12999	&	5.79596	\\
2	&	0.10556	&	-3.97281	&	2.61369	\\
3	&	0.00002	&	-8.56686	&	5.17950	\\
4	&	0.04395	&	2.77786	&	0.16735	\\
5	&	0.34001	&	0.61942	&	0.64009	\\
6	&	0.24566	&	1.79518	&	0.34023	\\
7	&	0.25750	&	-1.08819	&	1.26261	\\ \hline
\end{tabular}
\end{table}

\begin{algorithm}[H]
\caption{\textit{Full Generalized Approximate Message Passsing (GAMP) with stochastic volatility and sparse Bayesian learning prior}}\label{algorithm:GAMP_full}
\begin{algorithmic}[1]
\State Initialize $\widehat{\beta}_{i}^{(0)} = 0$ and $\widehat{\alpha}_{i}^{(0)}= (100)^{-1}$ $\forall i=1,...,q$, and set $\left(\sigma_{t}^{2}\right)^{(r)}=1$ and $\widehat{s}_{t}^{(0)}=0$ $\forall
t=1,...,T$.
    \State $r=1$
    \While{$ \Vert \widehat{\beta}^{\left(r\right)} - \widehat{\beta}^{\left(r-1\right)} \Vert \rightarrow 0$}
    \State 1) \textsc{\underline{GAMP for estimation of $\beta$}} 
    \For{$t = 1$ \algorithmicto $T$}
        \State  \hspace{1cm} $\widehat{c}_{t}^{(r)}  = \sum_{i=1}^{q} x_{t,i} \widehat{\beta}_{i}^{(r-1)} - \widehat{s}_{t}^{(r-1)}\widehat{\tau}_{t}^{c,(r)}$
        \State  \hspace{1cm} $\widehat{\tau}_{t}^{c,(r)} = \sum_{i=1}^{q} x_{t,i} ^{2} \widehat{\tau}_{i}^{\beta,(r-1)}$
        \State \hspace{1cm} $\widehat{z}_{t}^{(r)} = \widehat{\tau}_{t}^{z,(r)}\left( \frac{y_{t}}{\left(\sigma_{t}^{2}\right)^{(r-1)}} + \frac{\widehat{c}_{t}^{(r)}}{\widehat{\tau}_{t}^{c,(r)}}\right) $
        \State \hspace{1cm} $\widehat{\tau}_{t}^{z,(r)} = \frac{\widehat{\tau}_{t}^{c,(r)} \left(\sigma_{t}^{2}\right)^{(r-1)}}{\widehat{\tau}_{t}^{c,(r)} + \left(\sigma_{t}^{2}\right)^{(r-1)}}$
        \State \hspace{1cm} $\widehat{s}_{t}^{(r)} = \frac{\left(\widehat{z}_{t}^{(r)} -  \widehat{c}_{t}^{(r)}  \right)}{\widehat{\tau}_{t}^{c,(r)}} $
        \State \hspace{1cm}  $\widehat{\tau}_{t}^{s,(r)} = \frac{\left(1 - \frac{\widehat{\tau}_{t}^{z,(r)}}{\widehat{\tau}_{t}^{c,(r)}} \right)}{\widehat{\tau}_{t}^{c,(r)}}$
        \EndFor
        \For{$i = 1$ \algorithmicto $q$}
        \State  \hspace{1cm} $\widehat{d}_{i}^{(r)} = \widehat{\beta}_{i}^{(r-1)} +  \frac{\sum_{t=1}^{T} x_{t,i}\widehat{s}_{t}^{(r)}}{\widehat{\tau}_{i}^{d,(r)}}$
        \State  \hspace{1cm} $\widehat{\tau}_{i}^{d,(r)} = \sum_{t=1}^{T} x_{t,i}^{2} \widehat{\tau}_{t}^{s,(r)} $        
        \State \hspace{1cm} $\widehat{\beta}_{i}^{(r)} = \frac{\widehat{d}_{i}^{(r)}  \widehat{\tau}_{i}^{d,(r)}}{\widehat{\tau}_{i}^{\beta,(r)} }$ \hfill This is $E(\beta_i \vert y)$
        \State \hspace{1cm} $\widehat{\tau}_{i}^{\beta,(r)}  = \left(\alpha_i^{(r-1)} +  \widehat{\tau}_{i}^{d,(r)}\right) $        \hfill This is $var(\beta_i \vert y)$
        \EndFor
        \State
        \State 2) \textsc{\underline{Update prior hyperameter $\alpha$}}
        \For{$i = 1$ \algorithmicto $q$}
        \State $\alpha_{i}^{\left( r \right)} = \frac{2\underline{a}-1}{2\underline{b} + \left(\widehat{\beta}_{i}^{(r)}\right)^{2}}$  \hfill This is $E(\alpha_i \vert y)$
        \EndFor
        \State
        \State 3) \textsc{\underline{Update volatility $\sigma_{t}^{2}$}}
        \State $\widetilde{y}_{t} = \log\left( \left( y_{t} - x_{t}\widehat{\beta}^{(r)} \right)^{2} + 1\times 10^{-10} \right)$
        \State $\left(\sigma_{t}^{2}\right)^{(r)} = \exp \left( \sum_{i=1}^{7} \pi_{i} \left( \widetilde{y}_{t} - \mu_{i} \right) /7 \right)$ \hfill This is $E(\sigma_{t}^{2} \vert y)$
        \State
        \State $r = r + 1$
    \EndWhile
\end{algorithmic}
\end{algorithm}

\newpage
\renewcommand{\theequation}{C.\arabic{equation}} \setcounter{equation}{0} 
\renewcommand{\thefigure}{C.\arabic{figure}} \setcounter{figure}{0} 
\renewcommand{\thetable}{C\arabic{table}} \setcounter{table}{0}
\section{Simulation study}
This Section presents the results of three simulation exercises using artificial data. The first exercise focuses on the ability of the proposed econometric specification in \eqref{reg1}, in combination with the SBL prior in \eqref{SBL1}-\eqref{SBL2}, to recover the dynamics of stochastic regression parameters under various scenarios about the true nature of the underlying time-variation. For that reason, this exercise focuses on a regression with a single time-varying intercept (i.e. time-varying trend or local-level model). By doing so, we can focus on the aspect of time-variation by switching off the additional estimation challenges implied by the presence of many predictors, and at the same time use established benchmarks for comparison (MCMC methods for TVP models).

The second simulation exercise focuses on the numerical precision of the GAMP algorithm, and the ability of the SBL prior to shrink a high-dimensional vector of regression parameters. For that reason, in this second simulation exercise data are generated from a regression model with many predictors. In this exercise the true and estimated regression parameters are all constant, in order to control for the large effect that time-variation has on estimation accuracy and focus only on the aspect of modeling many covariates.

Finally, this Section concludes by assessing whether the approximations introduced by GAMP are detrimental for time-series inference. In particular, numerical stability of GAMP in the presence of persistent and highly correlated variables is discussed. A third Monte Carlo exercise also shows that GAMP is able to estimate accurately coefficients from an AR(4) model, despite the fact that the GAMP assumptions do not explicitly account for correlation between $y_{t}$ and its lags $y_{t-1}$, $y_{t-2}$ etc (i.e. GAMP is derived under the assumption that the likelihood function is i.i.d.).

\subsection{Monte Carlo Exercise 1: Estimation of time-variation}
In the first simulation exercise artificial data are generated from a simple regression with only a time-varying intercept (local-level model) and variance fixed to one, which is of the form
\begin{eqnarray}
y_{t} & = & c_{t} + \varepsilon_{t}, \label{ss_eq1} \\
c_{t} & = & F(z_{t}) + u_{t}, \label{ss_eq2}
\end{eqnarray}
where $\varepsilon_{t} \sim N(0,1)$ and both $y_{t}$ and $c_{t}$ are scalar. The first equation above specifies the regression of $y_{t}$ on the time-varying intercept $c_{t}$, and the second equation specifies the time series dynamics of $c_{t}$ which depend on some function $F(\bullet)$ of variables $z_{t}$. Due to the fact that the TVP-GAMP algorithm estimates the time-varying intercept $c_{t}$ in a generic way (i.e. relying only on shrinkage principles) this simulation study considers three different functions $F(z_{t})$ that may drive the time evolution of $c_{t}$. That is, given that with real macroeconomic and financial data the econometrician never knows the true function $F(z_{t})$ that generated the time-varying intercept, this exercise helps find out whether estimating equation \eqref{ss_eq1} combined with a shrinkage prior on $c_{t}$ is a broad enough parametric approach that can always capture fairly well the true underlying process of $c_{t}$. The three cases of data generating processes (DGPs) are:
\begin{itemize}
\item[DGP 1:] \textbf{Random Poisson jumps}. The evolution of $c_{t}$ is of the form
\begin{equation}
c_{t} = \mu + sign(\delta_{t})\times \mu \times k_{t} + u_{t},  u_{t} \sim N(0,T^{-\frac{3}{4}}),
\end{equation}
where $\mu$ is the unconditional mean of the process, $k_{t} \sim Poisson(\lambda)$ is a Poisson distributed variable, $sign(\delta_{t})$ is the sign operator giving values $1$ if $\delta_{t}>0$ and $-1$ if $\delta_{t}<0$, with approximately $50\%$ probability for each event, something that is achieved by randomly sampling $\delta_{t}$ from $U(-1,1)$. Therefore, $c_{t}$ is equal to $\mu$ when $k_{t}=0$, while it is equal to $2\mu$ if $k_{t}=1$ and $\delta_t>0$ or equal to $0$ if $k_{t}=1$ and $\delta_t<0$. The idea here is to allow $c_{t}$ to be constant and equal to $\mu$ for most time periods, but then have randomly  expressed negative or positive jumps of varying intensity. In order to achieve this, I set $\lambda=0.1$ and $\mu \sim U(0,4)$.
\item[DGP 2:] \textbf{Regression effects}. The evolution of $c_{t}$ is that of a regression-like model with exogenous predictors $z_{t}$ and is of the form
\begin{equation}
c_{t} = \beta_{0} + \beta_{1}z_{1t}+... + \beta_{10}z_{10t} + u_{t},  u_{t} \sim N(0,T^{-\frac{3}{4}}),
\end{equation}
where $z_{i,t} \sim N(0,1)$ for $i=1,...,10$ are the exogenous regressors, and $\beta_{j} \sim U(-1,1)$ for $j=0,1,...,10$ are randomly generated regression coefficients.
\item[DGP 3:] \textbf{Random walk evolution}. The evolution of $c_{t}$ has the typical form
\begin{equation}
c_{t} = c_{t-1} + u_{t},  u_{t} \sim N(0,T^{-\frac{2}{4}}),
\end{equation}
given initial condition $c_{0} \sim U(-1,1)$.
\end{itemize}

In this simulation study artificial datasets are generated from each of the three specifications above, comprising equation \eqref{ss_eq1} and each of the three parametrizations above that should replace the generic representation of equation \eqref{ss_eq2}. The purpose of using a simple specification with just a time-varying intercept is to allow numerically stable comparisons between TVP-GAMP and the standard, off-the-shelf MCMC algorithm for time-varying parameter models used in papers such as Primiceri (2005) and Pettenuzzo and Timmermann (2017).\footnote{The empirical section does more in-depth comparisons with recent algorithms for TVP models that also allow some form of shrinkage. Because these algorithms rely on several subjectively chosen tuning parameters, they cannot be considered as appropriate ``default'' benchmarks for a simulation study.} This algorithm relies on the third specification above, that is, it is an estimator of the following model
\begin{eqnarray}
y_{t} & = & c_{t} + \varepsilon_{t}, \varepsilon \sim N(0,\sigma_{t}^{2}), \label{MCMC1} \\
c_{t} & = & c_{t-1} + u_{t}, u_{t} \sim N(0,\alpha^{-1}), \label{MCMC2} \\
\log \sigma_{t}^{2} & = & \log \sigma_{t-1}^{2} + v_{t}, v_{t} \sim N(0,\omega) \label{MCMC3},
\end{eqnarray}
where -- even though in the data-generating process (DGP) it holds that $\sigma_{t}^{2}=1$ for all $t=1,...,T$ and $q=T^{-\frac{2}{4}}$ -- when estimating the model above, volatility is an unknown parameter that varies with time (as is the case with the TVP-GAMP algorithm). Therefore, due to the particular parametric form assumed by the MCMC algorithm the expectation is that this algorithm will fit well data generated from the third DGP which is also a random walk on $c_{t}$. However, the specification is flexible enough to approximate well the other two DGPs.

In particular, as Primiceri (2005, Section 4) explains in detail, and is a well-established fact in the literature on time-varying parameter models (see the discussion in Amir-Ahmadi et al., forthcoming), the prior on $\alpha^{-1}$ in equation \eqref{MCMC2} determines in a direct way the amount of time-variation in $c_{t}$: Looser (diffuse) priors on $\alpha^{-1}$ allow for the possibility of more time-variation, while tighter priors allow for a slow-moving and more persistent process.\footnote{Given our interpretation in this paper of $c_{t}  =  c_{t-1} + u_{t}$ as the conditional hierarchical prior $p(c_{t} \vert c_{t-1},\alpha^{-1}) \sim N(c_{t-1}, \alpha^{-1})$, it turns out that a very diffuse hyperprior on $q$ would essentially ``kill-off'' the random walk evolution, or put differently the $u_{t}$ component would dominate over $c_{t-1}$ on the RHS of equation \eqref{MCMC2}. In this case, the MCMC algorithm would become numerically identical to TVP-GAMP algorithm since their assumed conditionally Normal priors, $p(c_{t} \vert c_{t-1},\alpha) \sim N(c_{t-1}, \alpha^{-1})$ and $p (c_{t} \vert \alpha) \sim N(0,\alpha^{-1})$ respectively, would tend to become Uniform. Of course in practical situations that involve high-dimensional parameters and $\alpha^{-1}$ becomes a large matrix, we cannot allow a diffuse prior on it for numerical reasons (see again discussion in Primiceri, 2005, Section 4).  Nevertheless, in this simple example with scalar $\alpha^{-1}$ and fixed measurement variance it is possible numerically to use a loose enough prior on this parameter in order to build better intuition.} For that reason, the prior on $\alpha^{-1}$, in the MCMC algorithm with random walk evolution, is inverse Gamma of the form $\alpha^{-1} \sim iGamma(\underline{a},\underline{b})$, and the TVP model is estimated using two alternative choices of prior hyperparameters: a loose choice which corresponds to $\underline{a},\underline{b}=0.4$ and a tight choice which corresponds to $\underline{a},\underline{b}=1$.\footnote{In comparison to the choices of Primiceri (2005), who discusses in detail selection of the scale parameter $k_{q}$ in an inverse Wishart prior (his $\alpha^{-1}$ parameter was a matrix, not scalar as it is here), the tight choice roughly corresponds to $k_{q} = 0.1$ and the loose choice roughly corresponds to $k_Q=1000$ (note that MATLAB allows to generate draws for a scalar $\alpha^{-1}$ from both an inverse Wishart and an inverse Gamma generator, even though the former is not defined theoretically for scalars).} Therefore, in the following I present results for three estimators: the approximate GAMP estimator of the TVP model (abbreviated \textbf{TVP-GAMP}), the MCMC estimator with tight prior on $\alpha^{-1}$ (called \textbf{MCMC, tight prior}), and the MCMC estimator with loose prior on $\alpha^{-1}$ (called \textbf{MCMC, loose prior}). Notice at this point that even though we could also try loose and tight version of the Sparse Bayesian Learning prior of TVP-GAMP, for simplicity the default hyperparameter choices described in Section 3.1 are used.

\begin{figure}[H]
\centering
\includegraphics[scale=0.7]{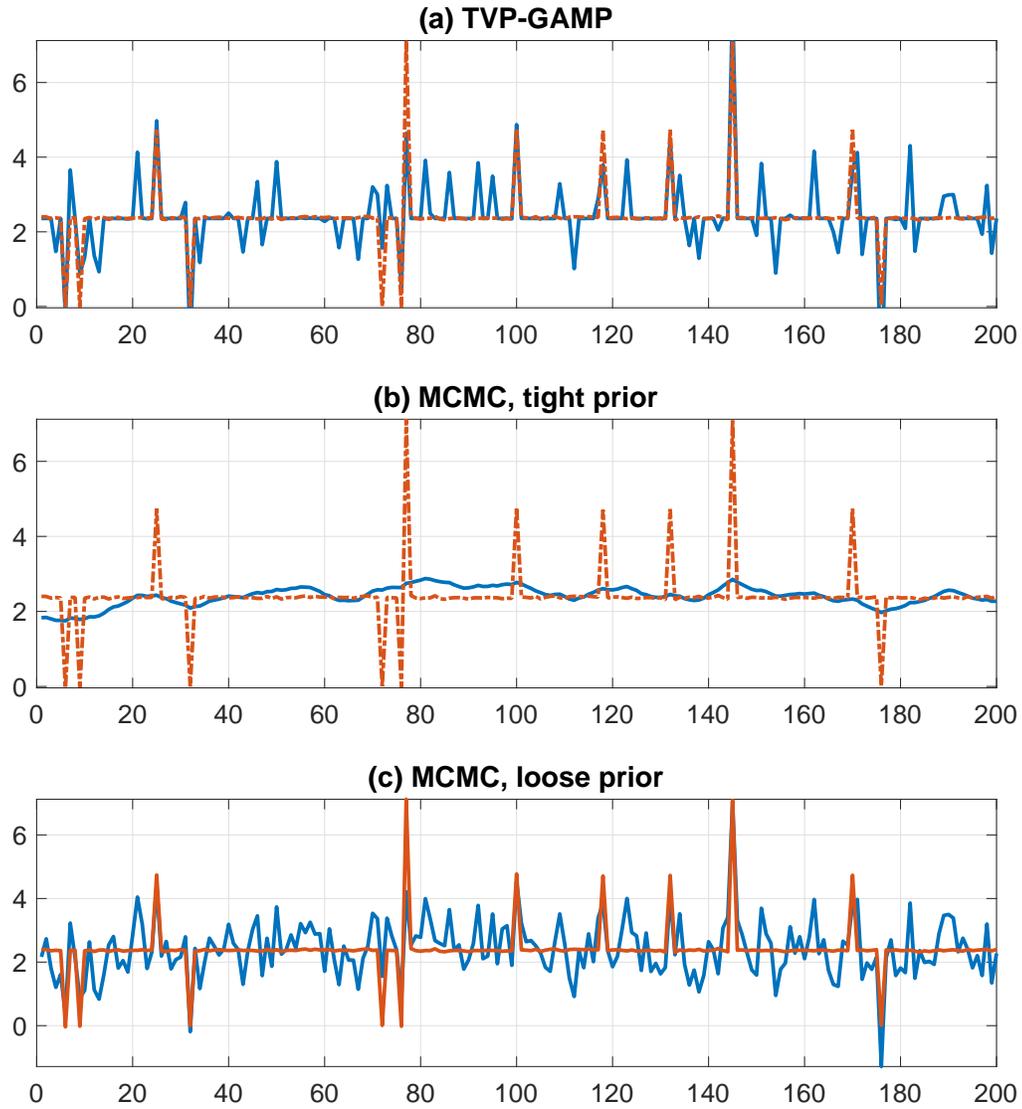}
\caption{\emph{Time-varying intercept generated by the first DGP (random Poisson jumps) for $T=200$ observations. The dashed line in each of the three panels shows the true time path of the generated time-varying intercept parameter, $c_{t}$, and the solid lines the posterior mean estimates of this intercept using TVP-GAMP, MCMC with a tight prior and MCMC with a loose prior.}}
\label{fig:MC1}
\end{figure}

\begin{figure}[H]
\centering
\includegraphics[scale=0.7]{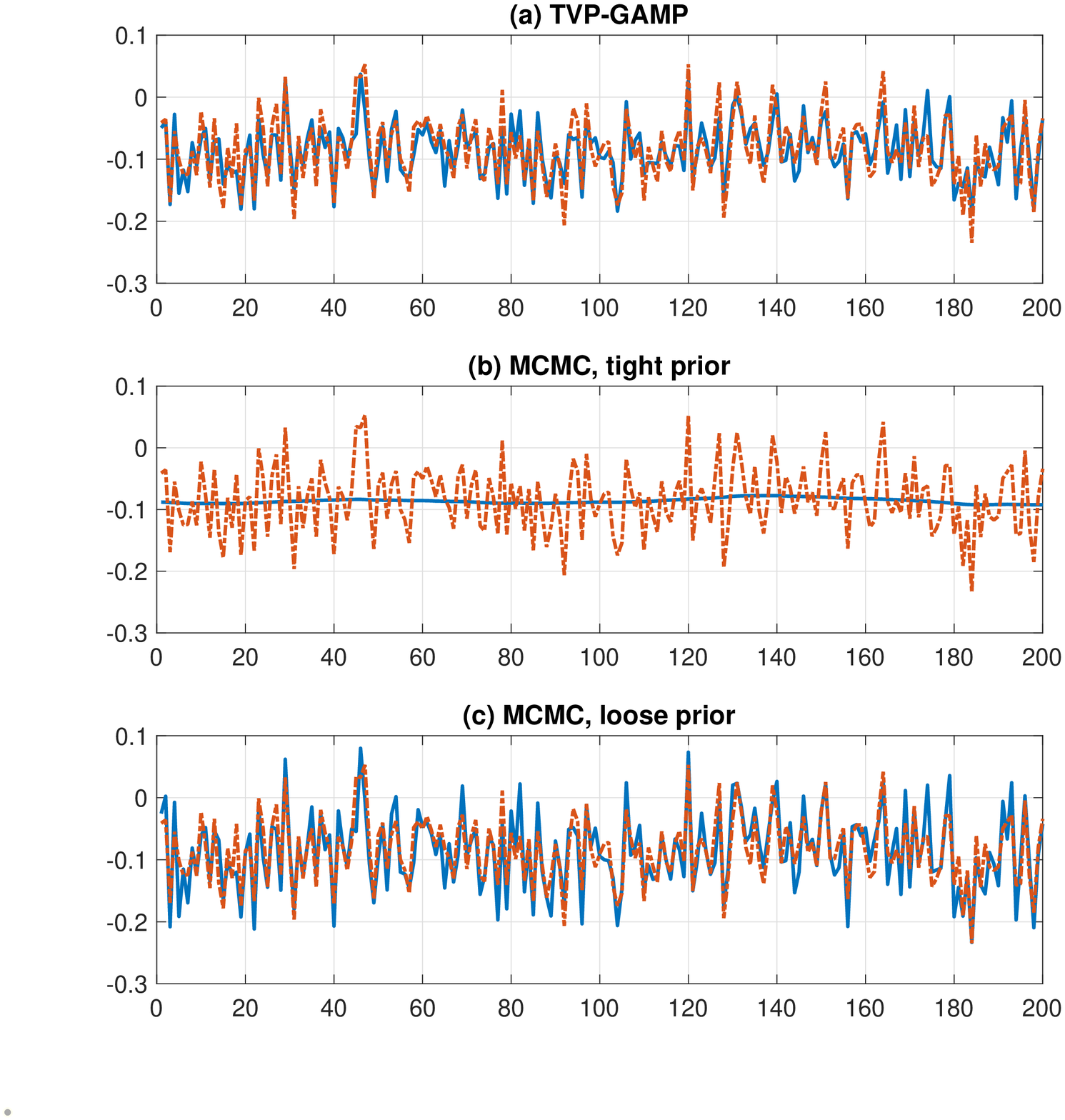}
\caption{\emph{Time-varying intercept generated by the second DGP (regression effects) for $T=200$ observations. The dashed line in each of the three panels shows the true time path of the generated time-varying intercept parameter, $c_{t}$, and the solid lines the posterior mean estimates of this intercept using TVP-GAMP, MCMC with a tight prior and MCMC with a loose prior.}}
\label{fig:MC2}
\end{figure}

\begin{figure}[H]
\centering
\includegraphics[scale=0.7]{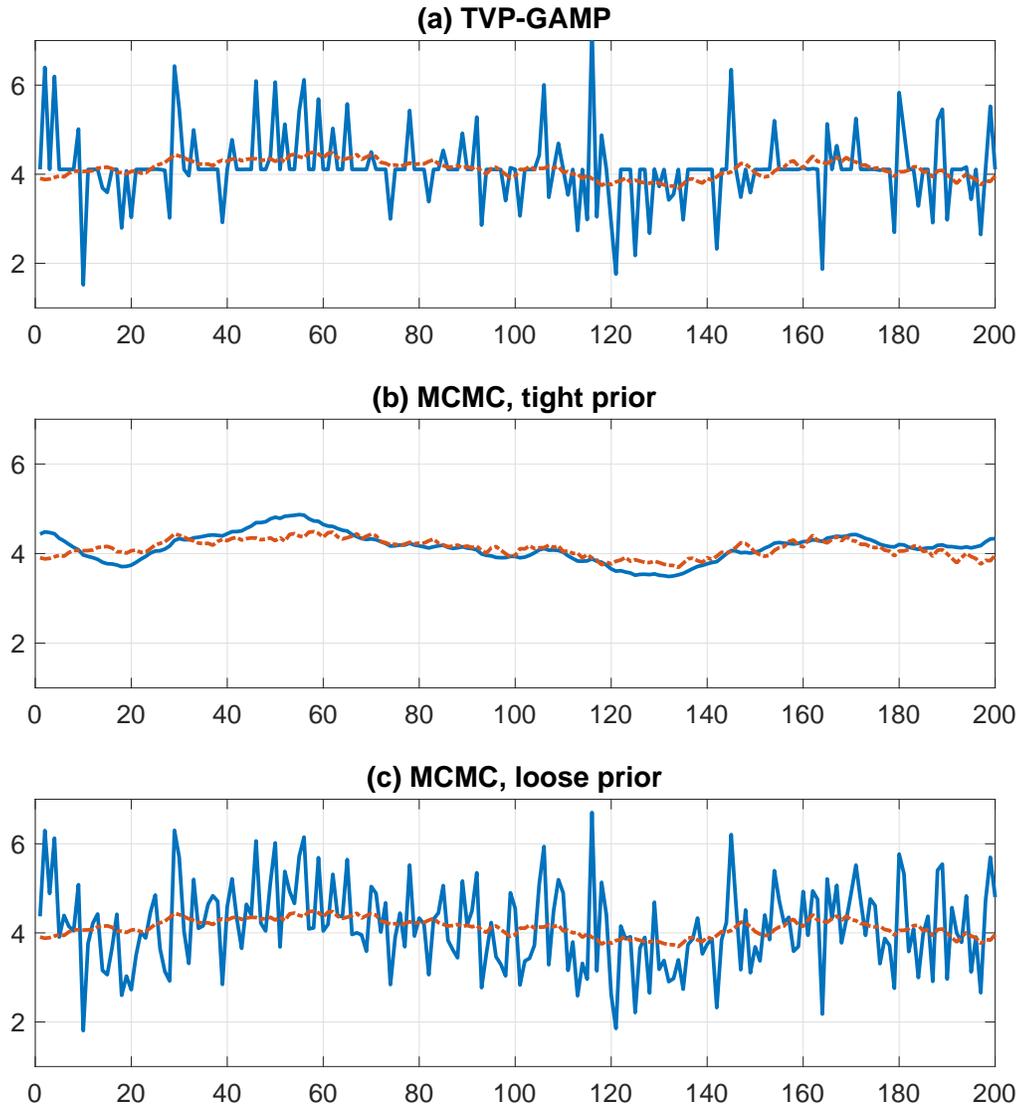}
\caption{\emph{Time-varying intercept generated by the first DGP (random walk evolution) for $T=200$ observations. The dashed line in each of the three panels shows the true time path of the generated time-varying intercept parameter, $c_{t}$, and the solid lines the posterior mean estimates of this intercept using TVP-GAMP, MCMC with a tight prior and MCMC with a loose prior.}}
\label{fig:MC3}
\end{figure}

Before proceeding to the numerical results of this Monte Carlo exercise, \Cref{fig:MC1,fig:MC2,fig:MC3} help build intuition about how well the different estimation algorithms approximate the three data generating processes. The red dashed line in all three panels of \autoref{fig:MC1} is the generated coefficient $c_{t}$ coming from a single simulation using the setting of DGP 1 (random Poisson jumps) with $T=200$. Subsequently, in each of the three panels of this figure the solid blue lines show the estimates from the respective three algorithms (TVP-GAMP, and MCMC with tight and loose prior). The TVP-GAMP having a parametric structure with minimal assumptions on the way $c_{t}$ evolves over time, is able to successfully shrink this coefficient to its constant value that dominates for most of the $200$ observations. At the same time, the algorithm successfully captures with high accuracy all 12 abrupt jumps in $c_{t}$. However, there a few more cases where such jumps are estimated when they shouldn't, that is, when the true parameter is constant. Nevertheless, the performance of the estimated $c_{t}$ from TVP-GAMP is quite reasonable, since only one observation is available to estimate these jumps, so the estimation problem is quite noisy. Moving to panel (b) of \autoref{fig:MC1}, it can be seen that the MCMC algorithm that estimates $c_{t}$ using a random walk evolution using a tight prior (similar to what Primiceri, 2005, used in his seminal paper) is not a good choice. It allows for a very smooth evolution that completely misses the abrupt jumps in the coefficients. In contrast, the exact same algorithm with a loose prior (panel (c)) does a much better job at capturing the abrupt jumps. However, this improvement comes at a cost of estimating time variation in $c_{t}$ even in those periods where the true $c_{t}$ is constant.

Similarly, the dashed red lines in \autoref{fig:MC2} show the estimated $c_{t}$ from a single, random run of DGP 2 (regression effects) using $T=200$. In this case, both TVP-GAMP and MCMC with a loose prior (panels (a) and (c) of the Figure) capture quite well the true $c_{t}$, although the MCMC algorithm produces some more pronounced peaks (see for example how the blue line ``overshoots'' relative to the red line around observation 80). The MCMC with a tight prior simply underestimates the amount of time variation in $c_{t}$. 

Lastly, \autoref{fig:MC3} repeats exactly the same exercise using one random run from DGP 3 (random walk) with $T=200$. The prior expectation here is that MCMC will do better in general than TVP-GAMP, since the MCMC algorithm explicitly assumes that $c_{t}$ evolves as a random walk (which is the exact assumption also in the DGP). However, this statement is far from true, as the performance of the MCMC algorithm depends greatly on the prior used. We see in panel (b) that the MCMC with tight prior gives indeed the best estimate of $c_{t}$ (solid line), since the true $c_{t}$ (dashed line) was generated assuming a modest amount of time-variation. However, the MCMC with loose prior fails completely to fit the true $c_{t}$. In this case, even if the TVP-GAMP doesn't assume random walk evolution when estimating $c_{t}$ it is able to produce less estimation error than MCMC with loose prior.

\newpage

Therefore, there are three major observations to take away from this graphical analysis:
\begin{enumerate}
\item TVP-GAMP does a good job at approximating time-varying parameters in a wide range of scenarios regarding the exact nature of time-variation.
\item The traditional MCMC algorithm for time-varying parameter models is extremely flexible as well, but its performance heavily relies on the choice of prior.
\item The TVP-GAMP using the sparse Bayesian learning prior is fully automatic and needs no tuning. Instead, the MCMC algorithm needs a subjective input of prior. While papers such as Amir-Ahmadi et al (forthcoming) have worked to towards solving this issue, in practice numerical instability restrictions do not allow researchers to specify looser or diffuse priors. As a consequence, most applications involving TVP models use tight priors that assume that variation in parameters over time is mildly stronger than what one can obtain by applying OLS recursively.\footnote{This is what Cogley and Sargent (2005) call a ``business as usual'' prior.} This exercise shows that a TVP model with mild/restricted random walk evolution of parameters, might not be always the ideal solution for capturing nonstandard patterns of time-variation.
\end{enumerate}

\begin{figure}[H]
\centering
\includegraphics[scale=0.48,trim={2cm 1cm 2cm 1cm}]{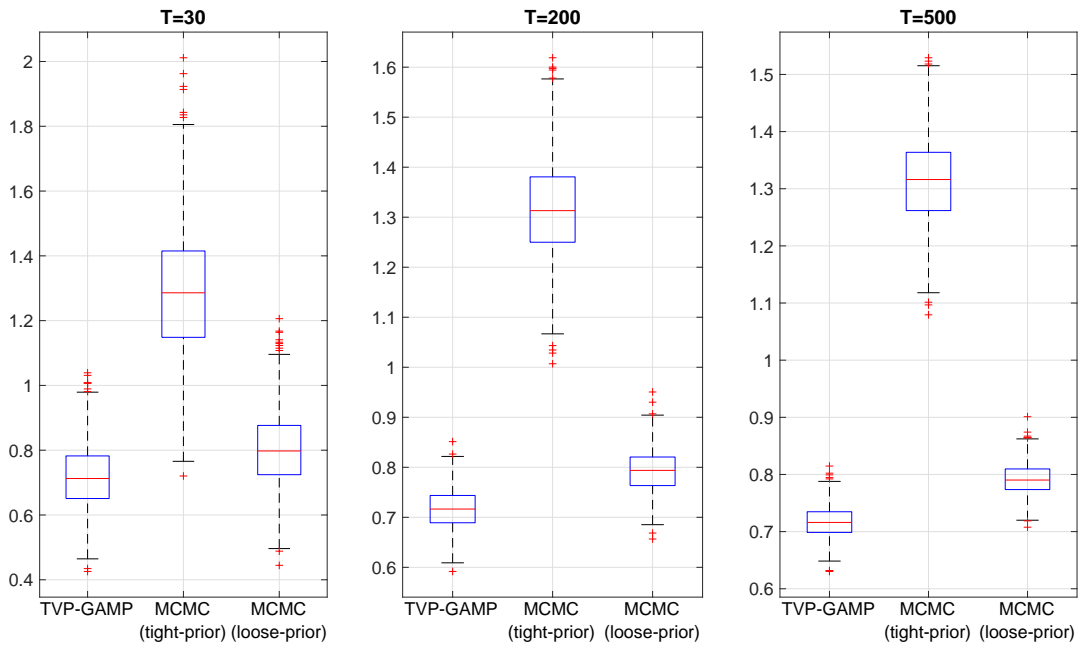}
\caption{\emph{Boxplots of absolute deviation (AD) statistics over the 1000 Monte Carlo samples from DGP1.}}
\label{fig:MCbox1}
\end{figure}

The observations made from this initial graphical screening can be confirmed numerically by calculating the distribution of the absolute deviation (AD) statistics over a sample of 1000 datasets generated from the three DGPs. These are defined as the absolute deviations of the estimated $c_{t}$ from each of the three algorithms relative to the true value of $c_{t}$ generated from each of the three DGPs. These absolute deviations refer to $c_{t}$ $\forall t$, so a scalar value is obtained by averaging over all time periods $t=1,...,T$. Lower values of the AD statistic signify that the respective estimator is much closer to the true value of the parameter $c_{t}$. \autoref{fig:MCbox1} shows  boxplots of the AD statistic over the 1000 iterations using DGP 1 for three different sample sizes of the generated data, namely $T=30,200,500$, that allow us to assess both the small and large sample performance of the estimators. In each of the panels it is obvious that for the case of the random Poisson jumps, TVP-GAMP and MCMC with loose prior generate lower error relative to MCMC with the tight prior. It is also the case that TVP-GAMP is the undisputed best performer relative to MCMC with loose prior. 

\begin{figure}[H]
\centering
\includegraphics[scale=0.48,trim={2cm 1cm 2cm 1cm}]{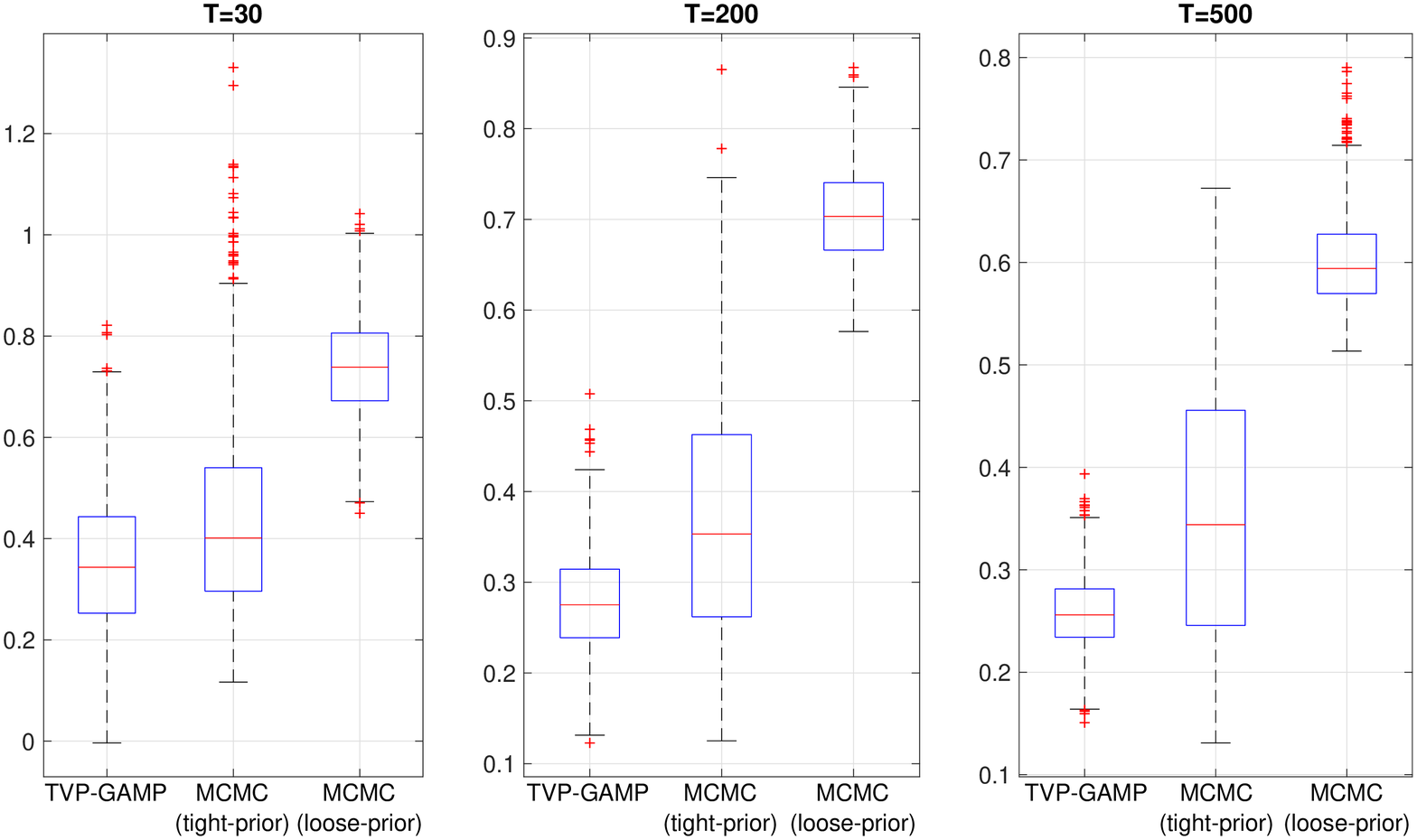}
\caption{\emph{Boxplots of absolute deviation (AD) statistics over the 1000 Monte Carlo samples from DGP2.}}
\label{fig:MCbox2}
\end{figure}

\begin{figure}[H]
\centering
\includegraphics[scale=0.48,trim={2cm 1cm 2cm 1cm}]{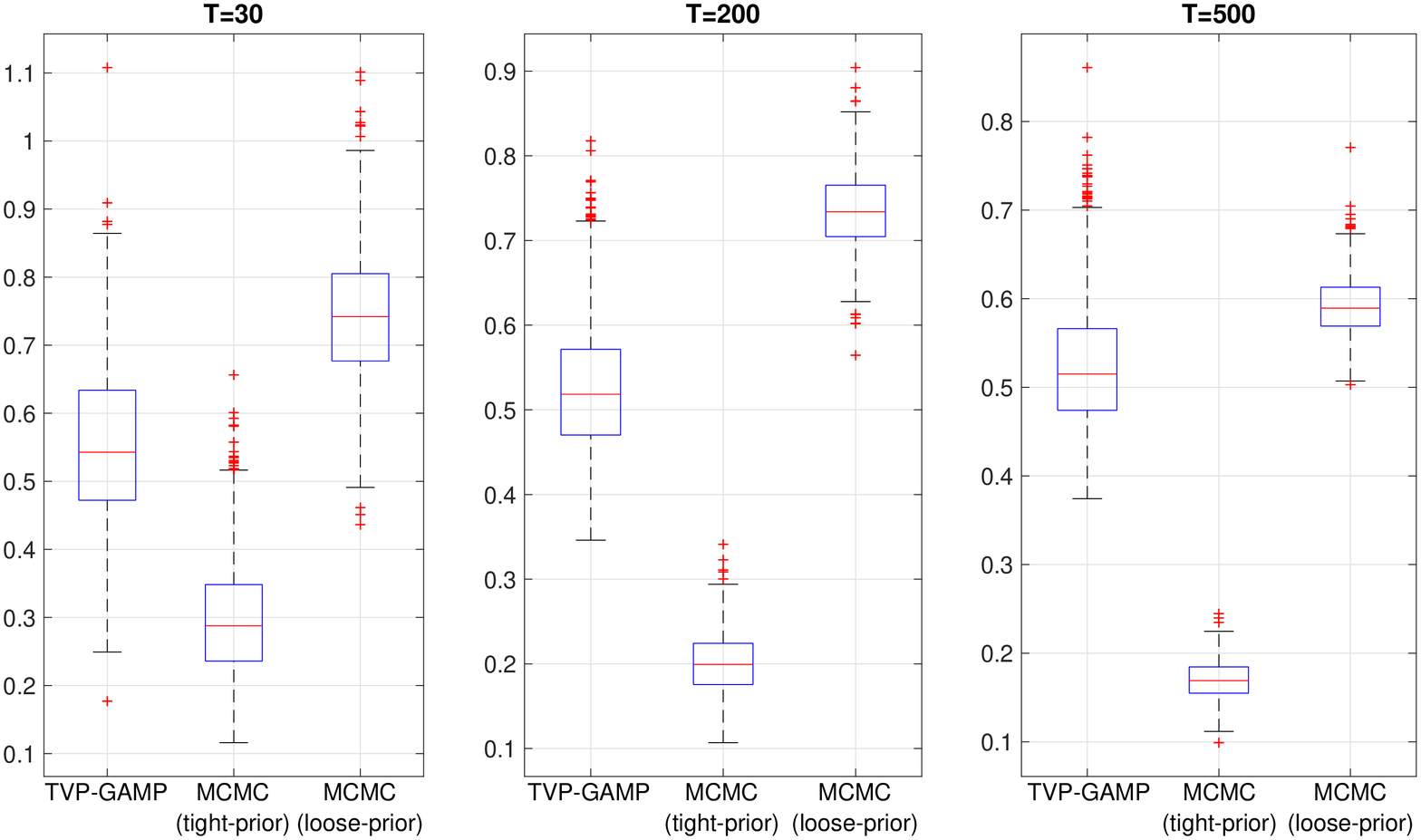}
\caption{\emph{Boxplots of absolute deviation (AD) statistics over the 1000 Monte Carlo samples from DGP3.}}
\label{fig:MCbox3}
\end{figure}

The results are qualitatively similar when using the second DGP, as depicted in \autoref{fig:MCbox2}. The only difference now is that the MCMC with a tight prior does better than the MCMC with loose prior, although the boxplots show that the former case results in much more dispersed distribution of absolute deviation statistics relative to the latter.\footnote{Notice that using the single random run plotted in \autoref{fig:MC2}, it looks like the MCMC with loose prior is far more accurate that the MCMC with tight prior. When replicating this exercise 1000 times and calculate the AD statistics over all $T$ values of the parameter $c_{t}$, it is obvious that this result is reversed.} Finally, \autoref{fig:MCbox3} shows that MCMC with tight prior is by far the most accurate estimation algorithm, since its assumptions about the way $c_{t}$ evolves and the amount of its time-variation (i.e. prior) perfectly match the assumptions in DGP 3. However, the main point of this third case is that TVP-GAMP does significantly better than MCMC with loose prior. This is particularly important because with real data we never know the exact DGP and our choice of prior may not be optimal given the observed data. Assuming the collected sample of any time series (GDP, inflation etc) is a realization from an unknown, ``true'' DGP, it is important that TVP-GAMP approximates fairly well all three diverse cases of DGPs examined in this simulation exercise. In contrast, there is a high amount of estimation risk associated with the traditional MCMC algorithm for TVP models due to the effect of prior selection. Such issues are not relevant for TVP-GAMP the way specified in this paper, since the algorithm works well with default noninformative values on the second layer of its hierarchical sparse Bayesian learning prior.

\vskip 1cm

\subsection{Monte Carlo Exercise 2: High-dimensional regression and shrinkage}
In this subsection I compare the performance of GAMP using data generated from high-dimensional, sparse regression models with constant parameters\footnote{The constant parameter regression is a special case of the TVP-GAMP algorithm presented in the main text, since the proposed algorithm relies on writing the TVP regression in static form. However, because in this Monte Carlo exercise variances are also constant, step 3 in Algorithm \autoref{algorithm:GAMP_full} (i.e. the step that updates stochastic volatilities) is replaced with the approximate posterior mode of the Bayesian estimator of the constant variance $\sigma^{2}$
\begin{eqnarray}
\left( \sigma^{2} \right)^{\left(r+1\right)} & = &  \arg \max_{\sigma^{2}} E_{\beta^{(r)}} \left[ \log  p\left(y \vert x, \beta^{(r)}, \left( \sigma^{2} \right)^{\left(r\right)} \right) p \left(\sigma^{2} \right) \right] \\
& = & \frac{2\underline{c}_2 + \sum_{t=1}^{T} \left(y_t - x_t \widehat{\beta}^{(r)} \right)^{2}} {T + 2\underline{c}_1 - 2}, \label{Msig1}
\end{eqnarray}
using an inverse Gamma prior with scale and rate parameters $\underline{c}_1=0.01$ and $\underline{c}_2=0.01$, respectively.}, and contrast them to MCMC-based algorithms that can handle such large dimensions. I generate $p$ predictors with $T$ observations each, from a Normal distribution with correlation $corr(x_i,x_j) = \rho ^{\vert i-j \vert}$ for $\rho \in \left[0,1\right]$. I assume that only $q$ columns of the predictors $x$ are important for $y$, and the remaining $p-q$ columns are excluded from the regression. That is, the coefficient vector is of the form $\beta = \left[ \beta_1,...,\beta_q, 0, 0 ,...,0,0 \right]$, where $q = \lfloor c \times p \rceil$, $0<c<1$ and $\lfloor \bullet \rceil$ denotes round-off to the nearest integer. I generate $\beta_i$ for $i=1,...,q$ from a continuous $U \left(-4, 4 \right)$ distribution.

Various combinations of $p$, $T$ and $\rho$ cases are considered, in order to assess how sparsity, correlation, and changing number of samples impact performance:
\begin{enumerate}
\item[DGP 1:] $T=50$, $p=100,200,500$ and $\rho=0.3$. I assume moderate correlation, and the case of a small sample $T$ that allows us to fully understand how the algorithm works in the large $p$ - small $T$ limit. I set $c=0.01$ meaning only one, two and five predictors out of $p=100,200,500$, respectively, are responsible for having generated $y$. 
\item[DGP 2:] $T=200$, $p=100,200,500$ and $\rho=0.3$. This is like Model 1, but with larger number of observations, to reflect approximately the $T$ found in quarterly macroeconomic data. I set $c=0.05$ meaning only five, 10 and 25 predictors out of $p=100,200,500$, respectively, are responsible for having generated $y$. 
\item[DGP 3:] $T=200$, $p=100$ and $\rho = 0.9$. In this case I only evaluate the case where $T > p$ so that orthogonalization of predictors is possible. For such high correlation, the GAMP algorithm would collapse and the MCMC-based estimation algorithms would also have a hard time converging. Orthogonalization is implemented by simply taking the sample covariance of the predictors $x$, $\widehat{\Omega}$ and defining $\widetilde{x} = x \widehat{W}^{-1}$ where $W$ is the Cholesky factor of $\widehat{\Omega}$. Note that if $p>T$ then $W$ is rank deficient and orthogonalization of the original space spanned by the columns of $x$ is not possible. I also set $c=0.05$ in this case implying that only five predictors have generated $y$ and the remaining 95 predictors have zero coefficients.
\end{enumerate}
Each of the three Monte Carlo simulations is repeated 500 times. Precision of each algorithm is measured by means of absolute deviations of estimated coefficients from the true, generated ones, in all 500 cases. I evaluate the following three algorithms: 1) the constant parameter version of the TVP-GAMP algorithm with Normal Gamma prior, which I denote as sparse Bayesian learning (abbreviated \textbf{GAMP (SBL)}); 2) the Bayesian least absolute shrinkage and selection operator (abbreviated  \textbf{MCMC (LASSO)}) estimated using the Gibbs sampler as in Park and Casella (2008); and 3) the stochastic search variable selection (abbreviated  \textbf{MCMC (SSVS)}) algorithm of George and McCullogh (1993), also based on the Gibbs sampler.

\begin{figure}[H]
\centering
\includegraphics[scale=0.5,trim={1cm 1cm 1cm 0cm}]{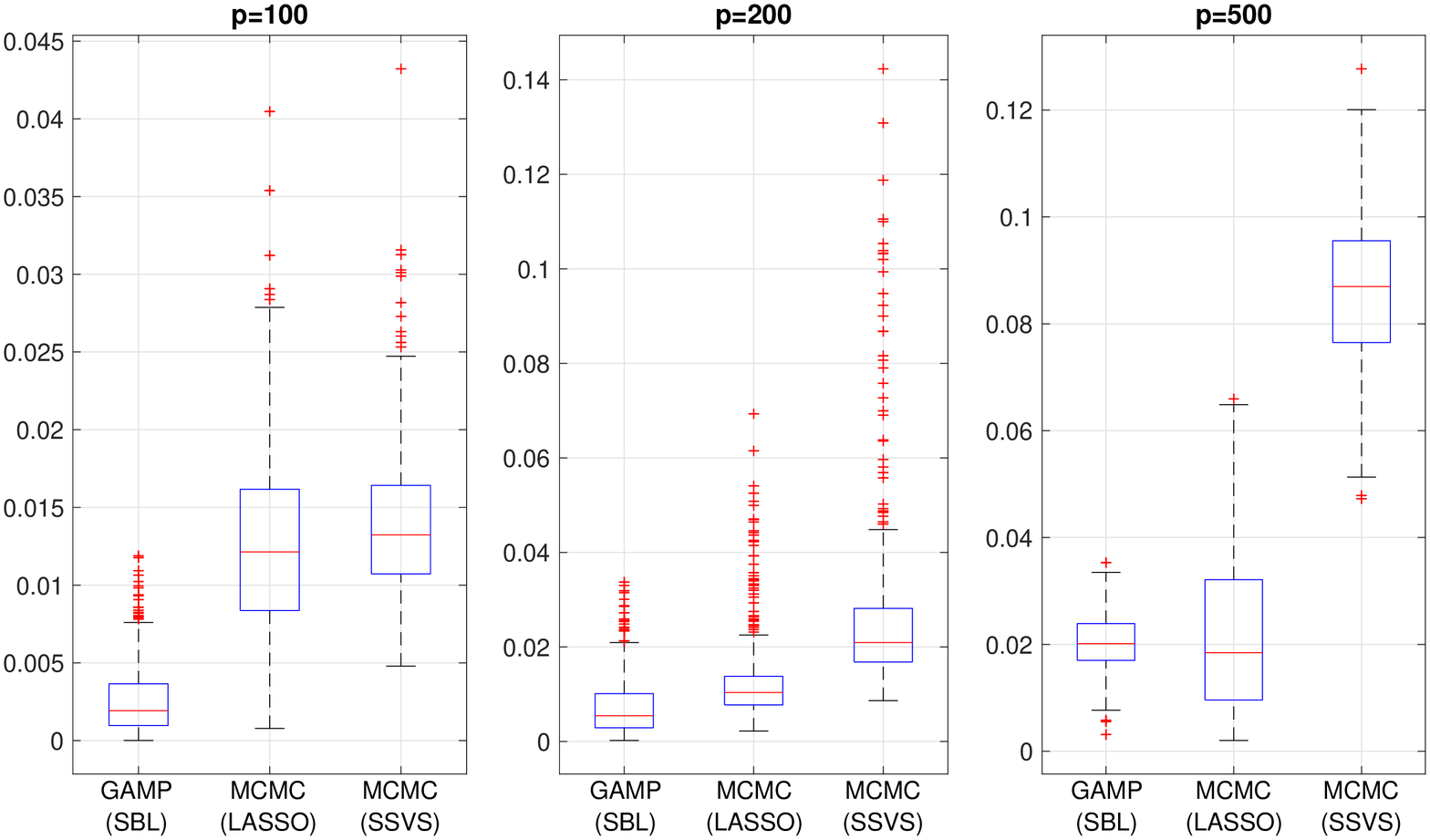}
\caption{\emph{Boxplots of absolute deviation (AD) statistics over the 500 Monte Carlo samples from DGP 1 ($T=50$, $p=100,200,500$, $\rho=0.3$).}} \label{fig:MCFig1}
\end{figure}

\begin{figure}[H]
\centering
\includegraphics[scale=0.5,trim={1cm 1cm 1cm 1cm}]{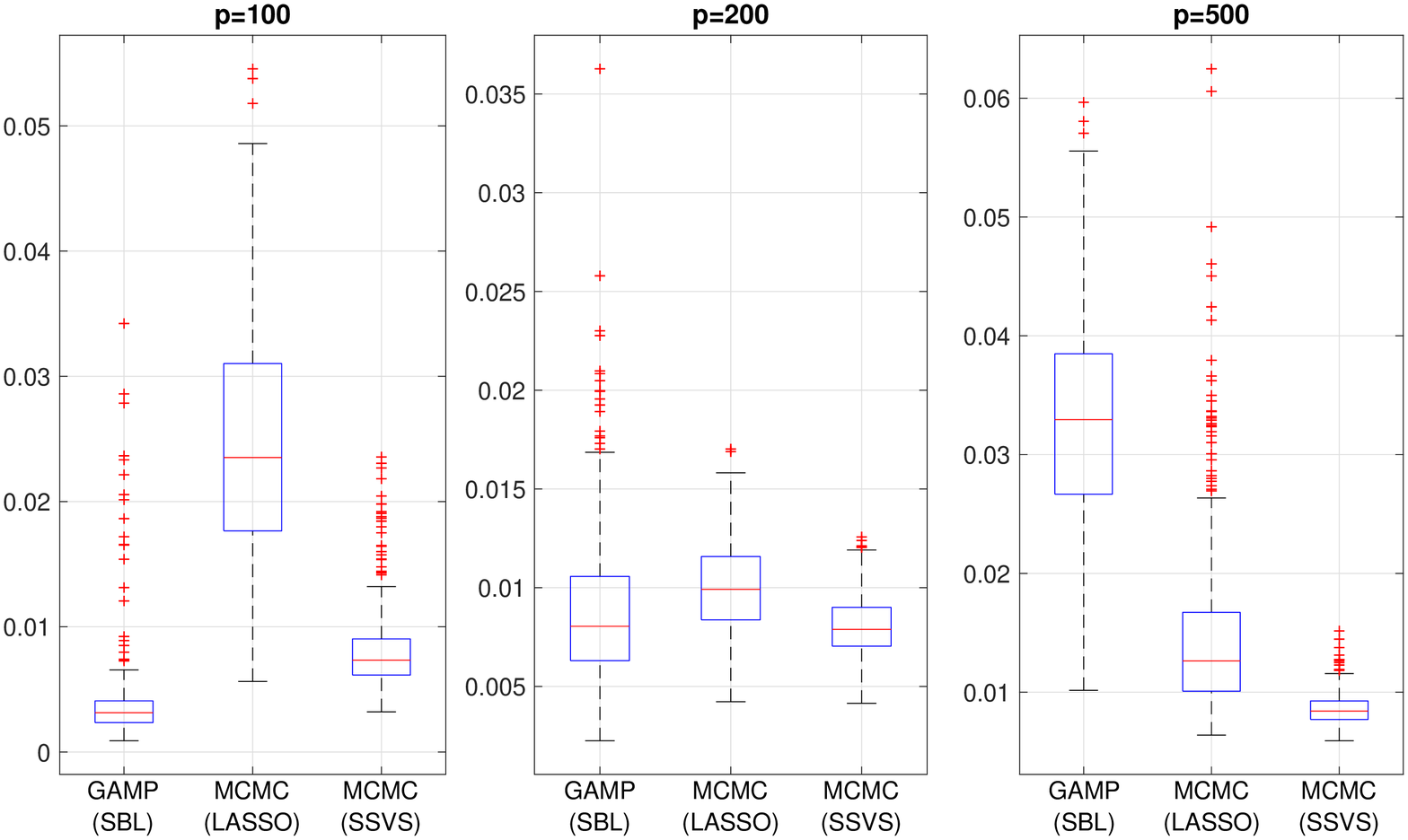}
\caption{\emph{Boxplots of absolute deviation (AD) statistics over the 500 Monte Carlo samples from DGP 2 ($T=200$, $p=100,200,500$, $\rho=0.3$).}} \label{fig:MCFig2}
\end{figure}

\begin{figure}[H]
\centering
\includegraphics[scale=0.5,trim={1cm 1cm 1cm 1cm}]{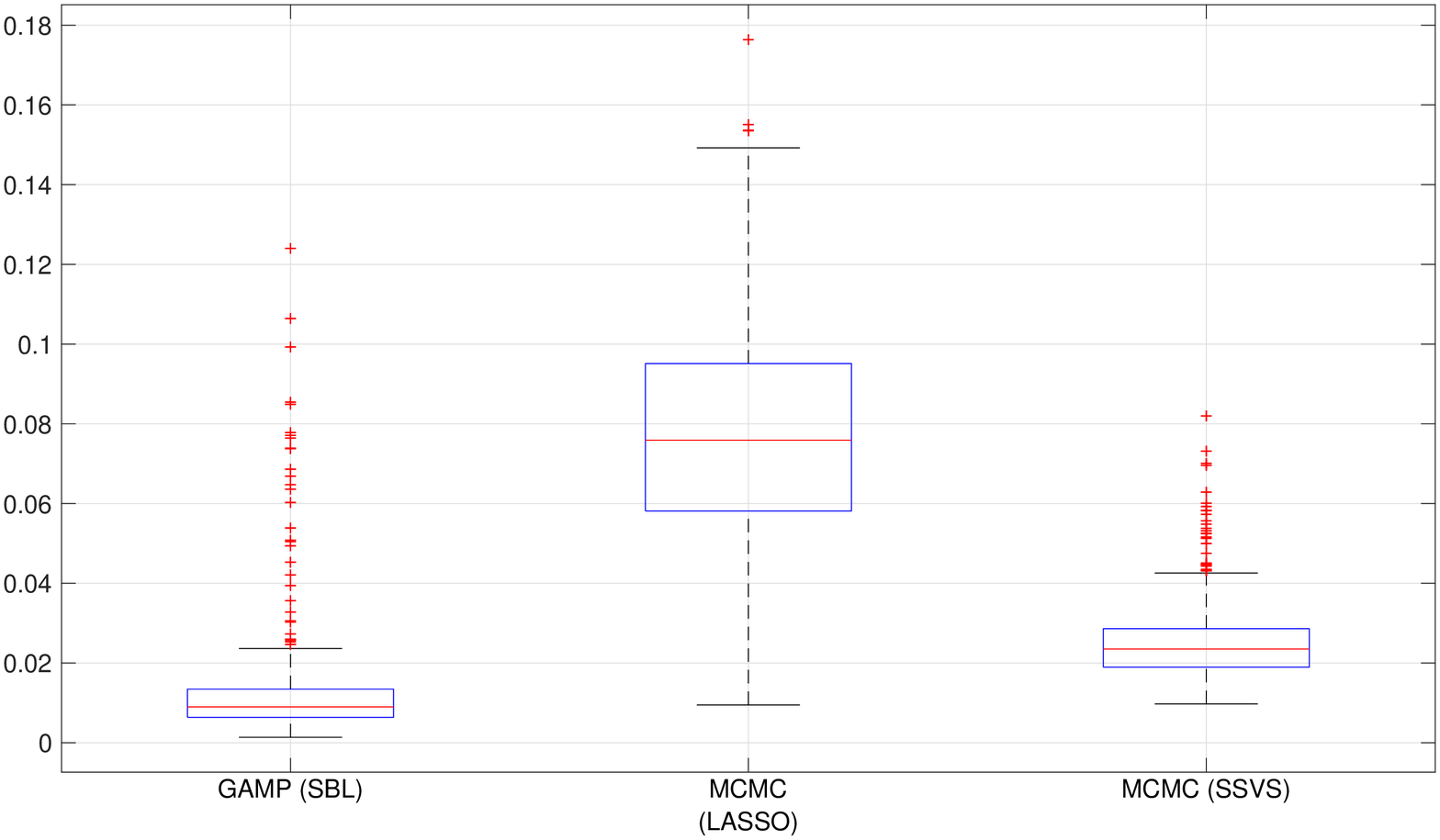}
\caption{\emph{Boxplots of absolute deviation (AD) statistics over the 500 Monte Carlo samples from DGP 3 ($T=200$, $p=100$, $\rho = 0.9$).}} \label{fig:MCFig3}
\end{figure}

The following \Cref{fig:MCFig1,fig:MCFig2,fig:MCFig3} show boxplots of the absolute deviation statistics for the three algorithms run over the 500 Monte Carlo iterations for the three DGPs, respectively. Note first that boxplots of the AD statistics of a naive, unrestricted estimator (OLS applied using one predictor at a time) is omitted from the three graphs, because these are typically five to 20 times larger in magnitude (depending on the values of $T,p$) than the AD values of the three shrinkage algorithms\footnote{In particular, for the mildly correlated predictors that are generated, the OLS applied predictor-by-predictor performs comparably well (in terms of ADs) only for the $q$ coefficients that are non-zero. It is for the case of the zero coefficients where the shrinkage estimators generate substantially lower error compared to OLS.}. The first general observation is that, even if there are visual differences among the distribution of AD statistics for the three estimators, these are typically small if the AD of OLS applied predictor-by-predictor is used as a reference point. When $T$ is small relative to $p$, as depicted in \autoref{fig:MCFig1} for the case $T=50$, GAMP (SBL) performs very well. On average it gives sharper results compared to MCMC (LASSO) and MCMC (SSVS), with lower averages and more concentrated distributions of AD statistics. When $T$ is larger as in \autoref{fig:MCFig2} then performance of the three algorithms is comparable when $p \leq T$, i.e. for $p=100,200$. For the case $p=500$ the GAMP-based algorithm performs slightly worse than the two MCMC-based algorithms, but differences can be considered small (the average AD of OLS in this case is 0.67, while for the three shrinkage algorithms the medians of their ADs are less than 0.05). In general, under mild correlation among predictors, the GAMP-based algorithm performs very well in variable selection and shrinkage relative to established MCMC-based algorithms.

An interesting case is the one where one wants to do inference with correlated predictors. \autoref{fig:MCFig3} shows what happens in the case $T=200, p=100$. While the assumption $p<T$ seems quite restrictive for general Big Data applications, the example in Model 3 is a quite realistic representation of many modern macroeconomic applications. In this instance as well, the performance of GAMP is excellent and considerably better than that of the LASSO.

Having established that GAMP-based shrinkage performs comparably well when the data generating process is that of a sparse regression model, I proceed with establishing the vast computational gains from the GAMP approximation. \autoref{table:computing_times} demonstrates a striking feature of GAMP, that is, the fact that runs much faster than both MCMC-based algorithms used in the simulations above. Entries in this Table are computing times measured in seconds (using the \texttt{tic} and \texttt{toc} commands in MATLAB). The differences are vast, since computation using GAMP (SBL) remains below 0.01 seconds, for all choices of $T$ and $p$. Therefore, this result should make clear to the reader the potential of GAMP-based algorithms to scale to thousands or even millions of predictors. In contrast, MCMC based algorithms have to run for a fixed number of iterations\footnote{Note that for the purposes of the Monte Carlo exercise in particular, I run both MCMC-based estimators using only 2000 iterations after discarding an initial 1000 burn-in iterations. This low number ensures satisfactory numerical precision, but in practical situations one would want to run thousand times more MCMC iterations. Thus, the high computing times of LASSO and SSVS are actually favorable due to the limited number of iterations.} and also to run sequentially. While the GAMP algorithm can be further enhanced by trivially modifying it to run in multiple CPU cores (the times in \autoref{table:computing_times} are NOT based on parallelizing GAMP), MCMC-based algorithms are not easily parallelizable unless further approximations are introduced.

\begin{table}[H]
\centering
\caption{Computing time (seconds) per Monte Carlo iteration for each of the four algorithms.} \label{table:computing_times}
\begin{tabular}{llccc}
	&		&	SBL	&	LASSO	&	SSVS	\\ \hline\hline
	&	$p=100$	&	$<0.01$	&	$4.03$	&	$1.81$	\\
$T=50$	&	$p=200$	&	$<0.01$	&	$12.88$	&	$5.29$	\\
	&	$p=500$	&	$<0.01$	&	$92.99$	&	$38.59$	\\
	&           &         &           &           \\
	&	$p=100$	&	$<0.01$	&	$6.28$	&	$1.98$	\\
$T=200$	&	$p=200$	&	$<0.01$	&	$12.54$	&	$5.11$	\\
	&	$p=500$	&	$<0.01$	&	$54.89$	&	$18.29$	\\ \hline
\multicolumn{5}{l}{ \scriptsize{\emph{Notes: The reference machine is a 64 bit Windows 7 PC with Intel}} } \\
\multicolumn{5}{l}{ \scriptsize{\emph{Core 7 4770K CPU, 32GB DDR3 RAM, MATLAB 2016a.}} }
\end{tabular}
\end{table}

\subsection{Monte Carlo Exercise 3: (Auto) correlation, and numerical stability of GAMP}

Up to this point, the assumption was that the GAMP algorithm will always converge. However, the nature of the GAMP approximations have two implications: 1) models with likelihood of the form $p(y_{t}| y_{t-1},x, \theta)$, i.e. regression models with both exogenous predictors and own lags, are not accounted for in the GAMP approximation; and 2) when the correlation of exogenous predictors is very high, the approximation to the marginal posterior becomes poor. Both cases would result in numerical instability of the original GAMP algorithm, and in this case convergence is not guaranteed. 

The second case is somewhat addressed in the previous Monte Carlo exercise, although not fully. That is, for mild correlation in a large number of predictors -- even in the case of more predictors than observations -- the algorithm has very high numerical accuracy and converges quickly. When correlation was quite high between predictors (0.9) the third case of the previous exercise addressed this issue by orthogonalizing first the predictors. Orthogonalization of predictors guarantees that the GAMP algorithm will converge. However, if this is not possible (e.g. when the number of predictors is larger than the number of observations), there are still ways to guarantee that the GAMP algorithm will converge. A large literature exists that extends GAMP in order to guarantee convergence and the reader is referred to Al-Shoukairi et al. (2018) and references therein. In a nutshell, an \emph{adaptive dumping} can be introduced that guarantees that the algorithm will converge, even though convergence might not be to the global optimum of the parameter values. In the specific case of the GAMP algorithm with sparse Bayesian learning prior, Al-Shoukairi et al. (2018) show that their proposed algorithm (which shares lots of common components with the algorithm used in this paper), will converge both in terms of parameter estimation as well as optimal shrinkage fixed points; the reader is referred to this paper for detailed discussion.

\begin{figure}[H]
\centering
\includegraphics[scale=0.6,trim={1cm 1cm 1cm 1cm}]{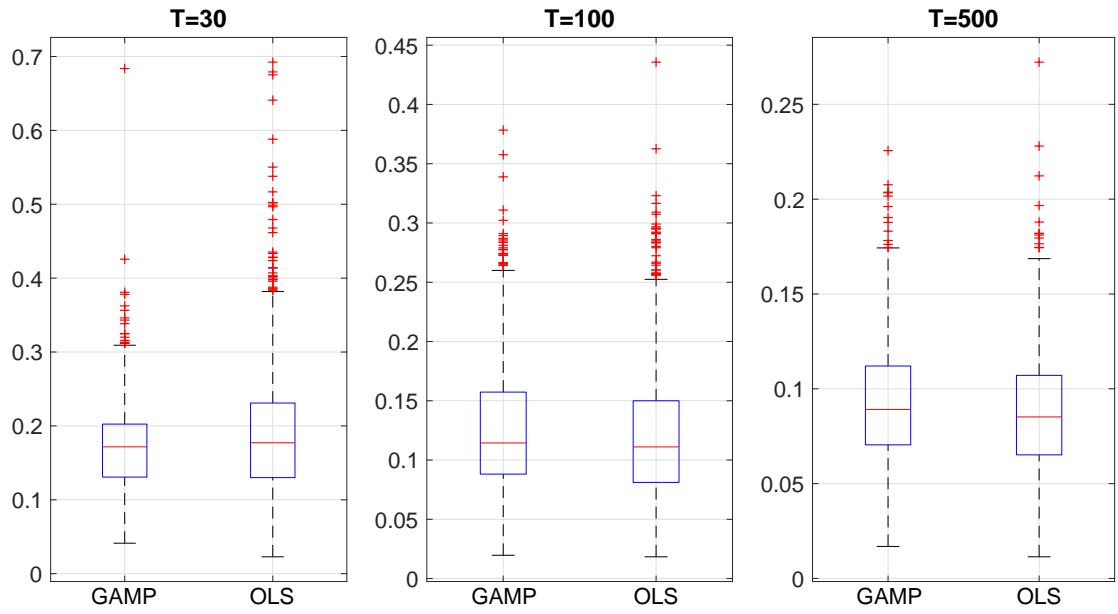}
\caption{\emph{Boxplots of absolute deviation (AD) statistics over the 1000 Monte Carlo samples from and AR(4) data generating process.}} \label{fig:MCFig4}
\end{figure}

In contrast, the first implication discussed above -- that is the fact that likelihoods with autoregressive components can only be approximated by the GAMP algorithm -- needs further examination since the vast majority of economic and financial time series have a strong autoregressive structure. For that reason a third Monte Carlo exercise examines what is the implication of having a likelihood of the form $p(y_{t}| y_{t-1},x, \theta)$, and estimating the parameters $\theta$ using GAMP under the approximation that $y_{t}$ and $y_{t-1}$ are independent. The best way to understand this issue, is to focus on the case of a simple autoregressive model. The setup of this Monte Carlo exercise is trivial: generate 1000 samples of length $T$ from the following AR(4) model without intercept
\begin{equation}
y_{t} = 0.40y_{t-1}  + 0.22y_{t-2} + 0.05y_{t-3} + 0.14y_{t-4} + \varepsilon_{t}
\end{equation}
where $\varepsilon_{t} \sim N(0,1)$ and the values of the autoregressive coefficients in the DGP are obtained by estimating an AR(4) for US GDP growth over the sample 1960Q1-2016Q4. Data are generated for $T=30,100,500$ and then an AR(4) model is fitted using two methods: GAMP (with constant, unknown variance as in the second Monte Carlo exercise of the previous subsection) and OLS. The results of this simple exercise are summarized in \autoref{fig:MCFig4} by means of boxplots of AD statistics as in the previous two exercises. GAMP not only converges in all 3000 cases (1000 iterations for each of $T=30,100,500$), but its numerical performance is identical to OLS. In addition, for the small-sample case $T=30$ GAMP performs a little bit better only because it relies on a shrinkage prior relative to OLS that is an unrestricted estimator. Therefore, this simple exercise provides sufficient support for the choice in the empirical exercise to include two lags of inflation when forecasting with the TVP-GAMP algorithm.
 
 \vskip 1cm

\newpage
\renewcommand{\theequation}{D.\arabic{equation}} \setcounter{equation}{0} 
\renewcommand{\thefigure}{D.\arabic{figure}} \setcounter{figure}{0} 
\renewcommand{\thetable}{D\arabic{table}} \setcounter{table}{0}
\section{Additional results using real data}
\subsection{Empirical estimates from TVP regression using GAMP}
We can use the GAMP algorithm derived in Appendix B in order to find estimates of $\beta$, $\sigma_{t}^{2}$ and $\alpha$. In this paper interest lies in the particular TVP model of equation \eqref{tvp_reg1}. Once we write this model in static form, then the Algorithm \autoref{algorithm:GAMP_full} can be readily applied. Note that while in the main text, for simplicity in notation, the reference equation is \eqref{tvp_reg1}, in the empirical exercise the non-centered parametrization (Fr\"{u}hwirth-Schnatter and Wagner, 2010) of the TVP regression is used. This parametrization has the form
\begin{equation}
y_{t} = x_{t} \widetilde{\beta} + x_{t} \widetilde{\beta}_{t} + \varepsilon_{t}, \label{tvp_nc}
\end{equation}
where using this notation the $\beta_t$ of \eqref{tvp_reg1} is now equivalent to $\beta_{t} = \widetilde{\beta} + \widetilde{\beta}_t$. That is, the equivalent representation in equation \eqref{tvp_nc} splits the coefficient into a constant part, and an add-on time-varying part (i.e. on top of the level of the constant coefficient). In practice, the estimated constant part $\widetilde{\beta}$ is equivalent to the initial condition $\beta_0$ of the coefficient $\beta_{t}$ in equation \eqref{tvp_reg1}. The non-centered representation is numerically more stable. Additionally, as shown in Belmonte et al. (2014) -- albeit, in a state-space model setting -- this parametrization allows to shrink flexibly a coefficient either towards zero or towards a constant value (by shrinking only $\widetilde{\beta}_t$ but not $\widetilde{\beta}$). The static form of this non-centered parametrization is
\begin{equation}
y = \EuScript{X} \beta + \varepsilon, \label{tvp_expanded}
\end{equation}
where $\beta = \left[\widetilde{\beta}^{\prime},\widetilde{\beta}_{1}^{\prime},...,\widetilde{\beta}_{T}^{\prime} \right]^{\prime}$. It is the case then that estimates of $\beta_{t}$ can be recovered from the elements of $\beta$ once this parameter vector is estimated using GAMP.

\begin{figure}[H]
\centering
\includegraphics[scale=0.7]{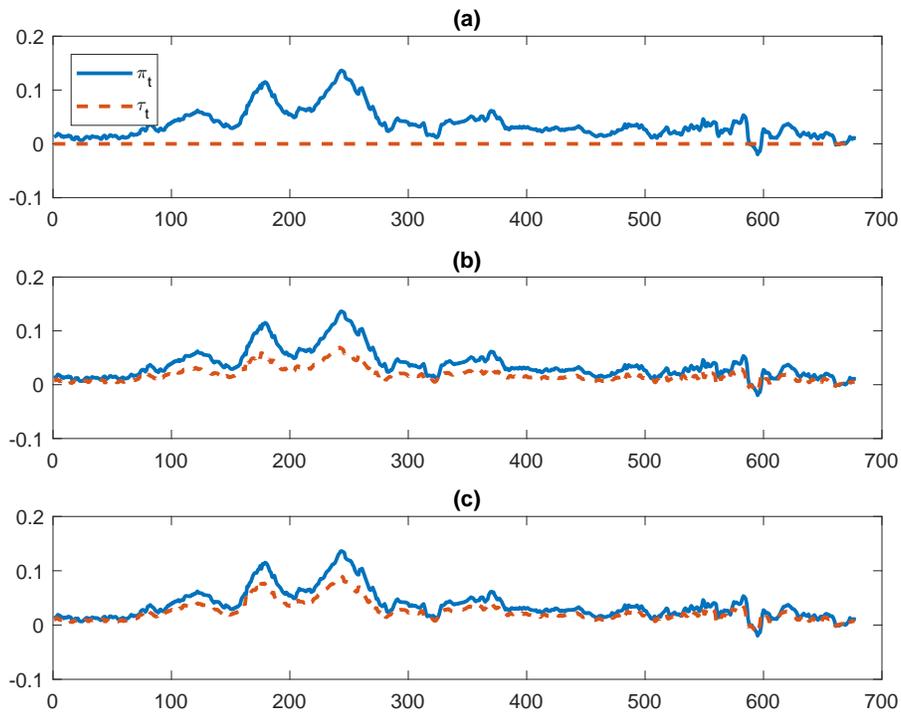}
\caption{\emph{Time-varying estimates of the trend, $\tau_{t}$, implied by various choices of the shrinkage hyperparameters $\underline{a},\underline{b}$. The values used in each of the three panels of this figure are (a) $\underline{a}=1$, $\underline{b}=1\times 10^{-10}$ , (b) $\underline{a}=1$, $\underline{b}=0.1$ , and (c) $\underline{a}=1$, $\underline{b}=10$. In all three panels the solid line is the data for annual inflation rates, $\pi_{t}$, and the dashed line is the estimate of the trend. The horizontal axis denotes the monthly observations of inflation, corresponding to the sample period 1959M1 to 2016M6, minus observations lost from transforming CPI to inflation, and taking lags.}}
\label{fig:trend_estimates}
\end{figure}

An interesting empirical question is how do estimates of time-varying parameters look like when estimating the formulation above that does not allow for persistence in these parameters (in contrast to the majority of macroeconomic TVP models, that specify persistent AR/RW processes for stochastic parameters). This question holds both for the time-varying coefficients $\beta_{t}$ and the time-varying volatility $\sigma_{t}^{2}$. The specified forecasting models in the main paper rely on many predictors that result in thousands of parameters that would be hard to summarize using plots. To simplify things, in this Appendix I estimate a time-varying intercept (trend) model for inflation, that is, what would have been a local-level model in the traditional state-space form. Therefore, the following specification is assumed:
\begin{equation}
\pi_{t} = \tau_{t} + \phi \pi_{t-1} + \varepsilon_{t},
\end{equation}
where $\pi_{t}$ is the annual (12-month) difference in the logarithm of CPIAUSL (CPI, All items). This specification is similar to that of Stock and Watson (2007) with a few key differences, e.g. $\tau_{t}$ in their paper follows a random walk process with stochastic volatility, while persistence here is captured using the AR component $\phi \pi_{t-1}$ and the prior variance for $\tau_{t}$ is not time-varying. For estimation purposes the equation above is written in the form \eqref{tvp_expanded}:
\begin{equation}
\pi = \mathcal{i}\tau^{c} + \tau^{tvp} + \pi_{(-1)}\phi^{c} +  \varepsilon_{t},
\end{equation}
where $\mathcal{i}$ is a $T \times 1$ vector of $1$'s, $\tau^{c}$ is the constant part of $\tau_{t}$ and $\tau^{tvp}$ is its time-varying part (also a $T \times 1$ vector), such that $\tau_{t}$ can be recovered as $\tau_{t} = \tau^{c} + \tau_{t}^{tvp}$. Additionally, $\pi = \left[\pi_{2},...,\pi_{T}\right]^{\prime}$ and $\pi_{(-1)} = \left[\pi_{1},...,\pi_{T-1}\right]^{\prime}$. Using this specification, we can apply to each element of $\tau^{c}$ and $\tau_{t}^{tvp}$ the same independent Normal-inverse Gamma prior used for the coefficients $\beta$ in the TVP regression, that is, the prior in equations \eqref{SBL1} and \eqref{SBL2}. In the main document, forecasting results have been produced using the default, noninformative choice $\underline{a}=\underline{b}=1\times 10^{-10}$. However, it is important to demonstrate how the amount of time-variation in the model above depends on $\alpha_{i}$ which is regulated from the choices of $\underline{a},\underline{b}$. This is illustrated in \autoref{fig:trend_estimates} for three pairs of values of these hyperparameters, namely $\underline{a}=1$, $\underline{b}=1\times 10^{-10}$ (panel (a)), $\underline{a}=1$, $\underline{b}=0.1$ (panel (b)), $\underline{a}=1$, $\underline{b}=10$ (panel (c)). The first choice overshrinks the trend not only to time-invariance but also to a value of zero, while the other two priors gradually allow for more time-variation. Due to the lack of a random walk evolution for the trend in this specification, in the case of zero shrinkage the trend will become identical to the original inflation series.

\begin{landscape}
\begin{figure}
\centering
\includegraphics[scale=0.5,trim={4cm, 0, 0, 0,}]{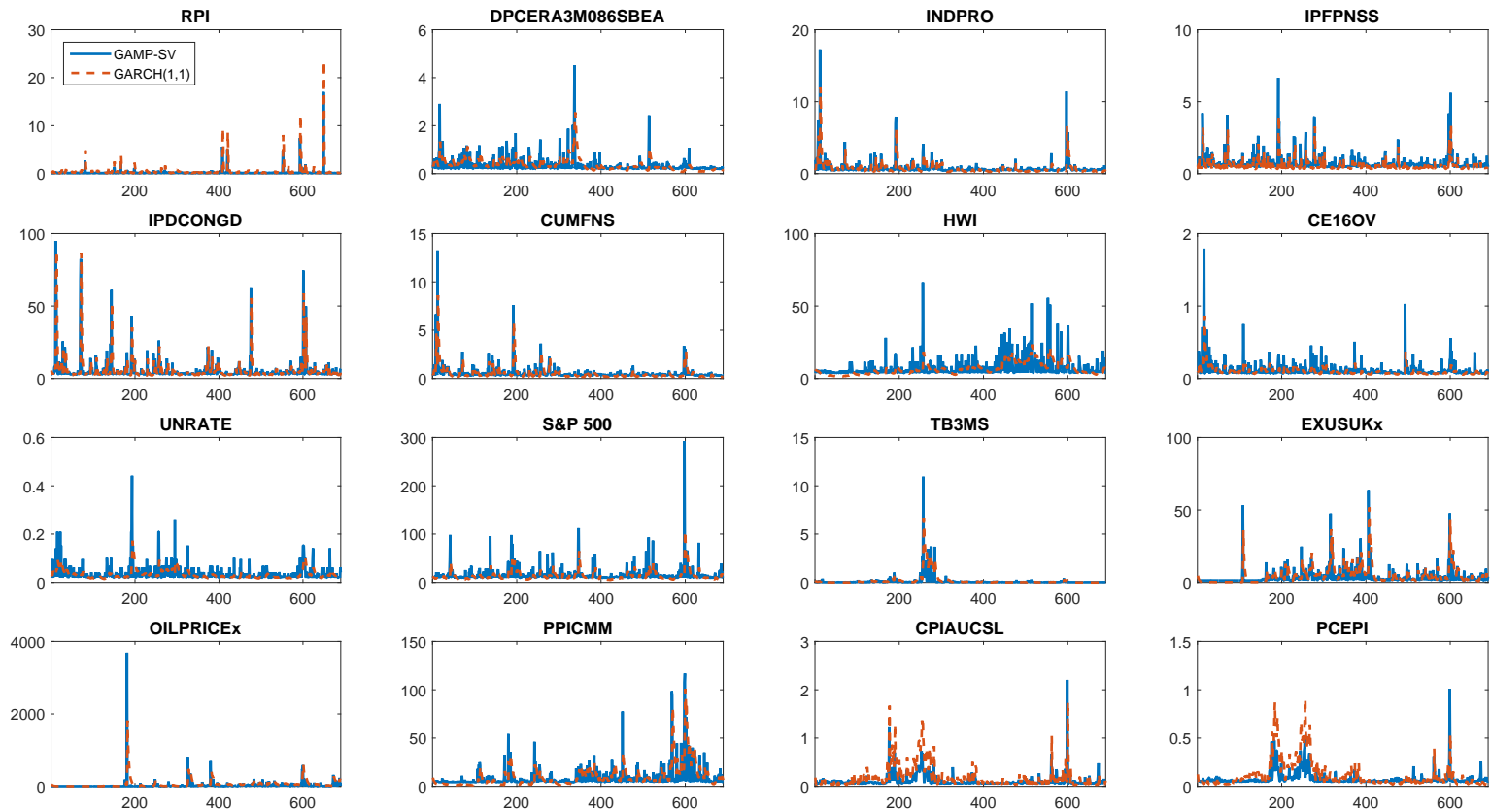}
\caption{\emph{Estimates of volatility implied by the GAMP updates of stochastic volatility (blue solid line), and a traditional GARCH(1,1) model estimated using maximum likelihood (red dashed line), for 16 selected variables from the monthly U.S. data set. All models feature only an intercept (no own lags, or other regression effects) as well Gaussian errors, in order to focus purely on the volatility estimates from the two methods.}}
\label{fig:volatility_estimates}
\end{figure}
\end{landscape}

A similarly important question is how good the volatility estimates are based on the proposed GAMP algorithm; see equation \eqref{final_vola}. These are expected to be less persistent than a GARCH(1,1) or SV model, and closely related to the ``squared returns'' estimator of volatility typically used in financial econometrics. \autoref{fig:volatility_estimates} compares the GAMP volatility estimates with a simple GARCH(1,1) model estimated with maximum likelihood for the case of 16 important series from the U.S. data set described in Appendix A. For simplicity, and given the focus on estimation of volatility, the modeling assumption for the mean of $y_{t}$ in both the GAMP and GARCH(1,1) cases is that of a constant intercept and Gaussian errors. For all 16 variables the two estimates are quite close, especially in those cases where volatility doesn't seem persistent (such as price series; see RPI, OILPRICE and CPIAUCSL). In other cases where a variable has very persistent volatility (notably the HWI variable), the GAMP volatility estimator provides a very noisy approximation to GARCH(1,1). Note that current results using the GARCH(1,1) as a benchmark, are qualitatively identical when comparing GAMP with a Bayesian SV specification similar to Kim, Shephard and Chib (1998). 

\newpage

\subsection{Identifying inflation forecast gains}
\begin{figure}[H]
\centering
\includegraphics[scale=0.55,trim={2cm, 0cm, 1cm, 1cm}]{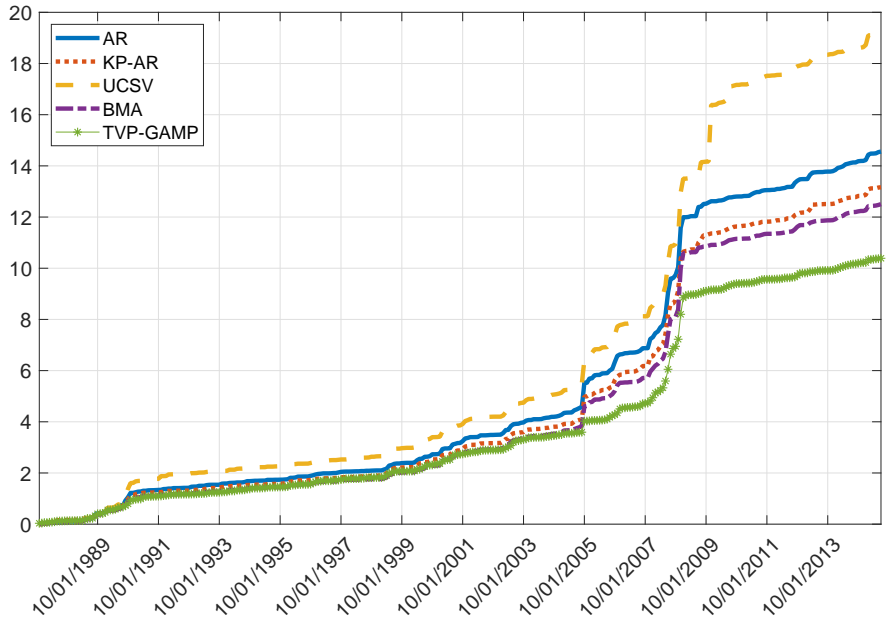}
\caption{\emph{Cumulative sum of squared forecast errors over the forecast evaluation period, CPI inflation for $h=12$ using the specification in equation \eqref{bencmark_forecasting_eq}.}}
\label{fig:cumsumMSFE_CPIUR_h12}
\end{figure}

\begin{figure}[H]
\centering
\includegraphics[scale=0.55,trim={2cm, 0cm, 1cm, 1cm}]{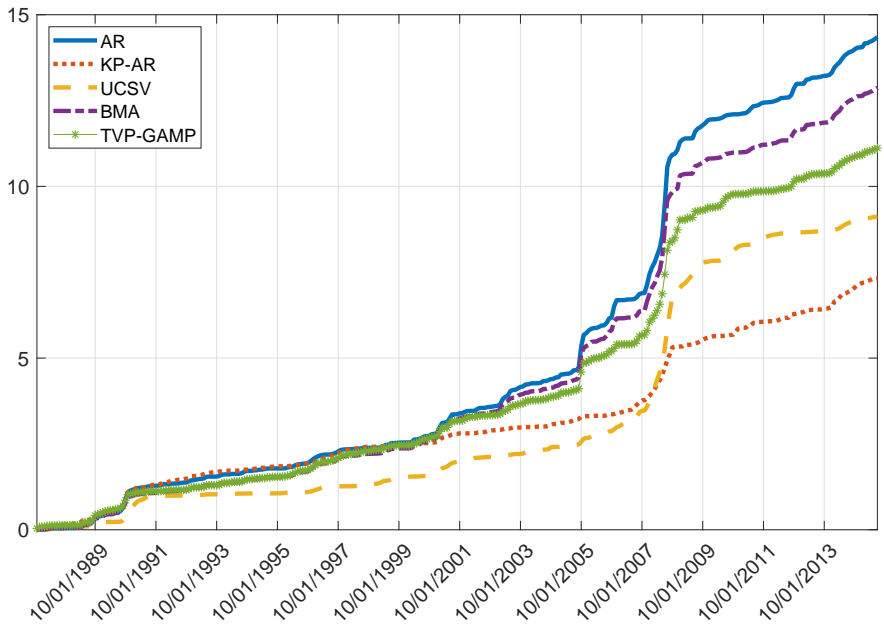}
\caption{\emph{Cumulative sum of squared forecast errors over the forecast evaluation period, CPI inflation for $h=12$ using the specification in equation \eqref{bencmark_forecasting_eq2}.}}
\label{fig:cumsumMSFE_CPI_h12}
\end{figure}

While Tables 1-3 in the main text show the ranking of different models based on their average forecast performance during the evaluation period, it is of interest to identify those episodes that contribute the most in their relative performance. For that reason, the following four figures plot cumulative squared forecast errors, rather than just presenting the mean squared forecast error over the sample. \autoref{fig:cumsumMSFE_CPIUR_h12} presents results for forecast horizon $h=12$ using the specification in equation \eqref{bencmark_forecasting_eq}, while \autoref{fig:cumsumMSFE_CPI_h12} presents results for forecast horizon $h=12$ using the specification in equation \eqref{bencmark_forecasting_eq2}. The dates in the x-axis of the graphs represent the time period where the forecast was made (hence, for $h=12$ the date 01/01/2000 on the x-axis means that the squared forecast error is based on the observation of 01/01/2001).

These Figures show that there were a few critical dates/episodes in the evaluation sample that determine differences in forecast performance. In \autoref{fig:cumsumMSFE_CPIUR_h12} the TPV-GAMP dominates in all periods and the differences become more obvious after 1999, 2005 and 2008. In \autoref{fig:cumsumMSFE_CPI_h12} the UCSV dominates until early 2008 (which refers to inflation forecasts for early 2009), and after this date the KP-AR is the best model. TVP-GAMP's performance over time looks closer to that of the BMA, even though TVP-GAMP improves a lot over BMA. The general message is that, while the global financial crisis has lead to severe deterioration in performance of all forecasting models, it is not the sole most important episode of \emph{relative} performance. It turns out that ``inflection points'' also exist in the early 1990s, late 1990s, and 2005. Nevertheless, the largest gaps showing divergence in forecast performance are observed during the last 10 years in the sample.

\newpage
\renewcommand{\theequation}{E.\arabic{equation}} \setcounter{equation}{0} %
\renewcommand{\thetable}{E\arabic{table}} \setcounter{table}{0}
\section{Competing specifications}
This Section sheds light on the specification of benchmark models used in the forecasting illustration. These are mainly heavily parametrized TVP regression models that involve selection of a wide array of prior hyperparameters and initial conditions. In order to make the comparison fair I try only to use what can be thought of as ``default'' values for all priors and initial conditions. These are values that have minimal impact, and are typically the default values used in several other papers in this literature. 

In particular, many of the models presented in this Section (KP-AR, GK-AR, UC-SV, TVP-AR) are the benchmark models used in a large-scale forecasting exercise for many monthly and quarterly U.S. macroeconomic time series published in Bauwens et al. (2015). Therefore, instead of replicating all tedious details regarding posterior computation as well as selection of priors, the reader is mostly referred to this paper and its long online Appendix for these specific benchmark models. For other recently proposed TVP models (TVD and TVS) that are not explicitly included in the forecasting exercise of Bauwens et al. (2015), I try to adopt default values suggested by the authors who created them (and provide relevant citations to their papers). If this is not feasible, as is the case with BMA and TVP-BMA, only then I provide exact details about computation and prior selection.

\subsection{KP-AR, Koop and Potter (2007)}
The Koop and Potter (2007) specification is a structural breaks model that builds on the more general time-varying parameter (TVP) specification but doesn't allow breaks to occur necessarily in each time period. The KP-AR model is of the form
\begin{eqnarray}
y_{t+h} & = & x_{t} \beta_{s_{t}} + \sigma_{s_{t}}\varepsilon_{t+h}, \\
\beta_{s_{t}} & = & \beta_{s_{t-1}} + \eta_{s_{t}}, \\
\log \sigma_{s_{t}} & = & \log \sigma_{s_{t-1}} + \zeta_{s_{t}},
\end{eqnarray}
where $x_{t}$ includes only an intercept and lags of $y_{t}$, $\varepsilon_{t+h}$ is an error following the standard Normal distribution and $s_{t} \in \left\lbrace 1,2,...,K \right\rbrace$ is a Markov switching process with $K$ states. This specification follows much of the Bayesian structural breaks literature and assume that the transition probabilities matrix is block diagonal, such that we can move from one regime to the next and never come back (which is the distinguishing feature of structural breaks compared to standard regime-switching specifications). I follow Bauwens et al (2015) and specify a maximum number of $K_{max} = 10$ and allow the Gibbs sampler to determine how many structural breaks are relevant (up to the maximum of $K_{max}$). I also use fairy reasonable priors and initial conditions as in Bauwens et al. (2015), and the reader is referred to this paper and its Appendix for all the tedious computational details.

\subsection{GK-AR, Giordani and Kohn (2008)}
The Giordani and Kohn (2008) model is also a structural breaks model based on state-space form, however, the way breaks occur are very different from the KP specification outlined above. The GK-AR model is a dynamic mixture model of the form
\begin{eqnarray}
y_{t+h} & = & x_{t} \beta_{t} + \varepsilon_{t+h}, \\
\beta_{t} & = & \beta_{t-1} + K_{t} \eta_{t},
\end{eqnarray}
where $x_{t}$ includes only an intercept and lags, $K_{t} \in \left\lbrace 0,1 \right\rbrace$. Gerlach, Carter and Kohn (2000) outline an algorithm for sampling efficiently $K_{t}$. Conditional on knowing $K_{t}$ (e.g. conditional on having a sample from its posterior), the remaining model parameters can be updated using standard expressions for their conditional posteriors, leading in a straightforward Gibbs sampling scheme. Estimation and prior settings are again those used in Bauwens et al. (2015) and the reader is referred to their detailed Appendix.

\subsection{UC-SV, Stock and Watson (2007)}
The Stock and Watson (2007) unobserved components stochastic volatility (UC-SV) model only allows for a time-varying intercept, that is, it is a local level specification of the form
\begin{eqnarray}
y_{t+h} & = & \tau_{t} +  \sigma^{\varepsilon}_{t}\varepsilon_{t+h}, \\
\tau_{t} & = & \tau_{t-1} + \sigma^{\eta}_{t}\eta_{t}, \\
\log \sigma^{\varepsilon}_{t} & = & \log \sigma^{\varepsilon}_{t-1} + \zeta_{t}, \\
\log \sigma^{\eta}_{t} & = & \log \sigma^{\eta}_{t-1} + \xi_{t},
\end{eqnarray}
where we observe that not only the measurement error $\varepsilon_{t+h}$ features stochastic volatility, but also the variance of state error $\eta_{t}$. This model has been specifically proposed for forecasting inflation, but it is a parsimonious and flexible time-varying parameter specification that can fit other series as well. This model is the most parsimonious among all other time-varying parameter specifications presented in this Section, as it only requires specification of initial conditions and priors for the scalar variances of the volatility parameters. In any case, selection of these hyperparameters needed for estimation follows again the implementation of Bauwens et al. (2015).

\subsection{TVP-AR, Pettenuzzo and Timmermann (2017)}
The time-varying parameter autoregression has not been obviously first proposed by Pettenuzzo and Timmermann (2017), however, this is a recent of many representative studies that find that the TVP-AR model beats a large number of alternative models when forecasting inflation in particular. It is a regression with an intercept and autoregressive lags where regression coefficients and variances are all time-varying. By allowing for autoregressive terms the TVP-AR generalizes the UC-SV model, however, it typically only features stochastic volatility in the measurement equation and not in any of the state equations. All priors used for estimation of this model also follow the ``default'' values specified in Bauwens et al. (2015), and the reader is referred to this paper for more details.

\subsection{TVP-BMA, Groen, Paap and Ravazzolo (2013)}
Even though this specification builds on Groen, Paap and Ravazzolo (2013), it is actually only a simplified version of their proposed model.\footnote{In particular, their model also features a dynamic mixture specification as in Giordani and Kohn (2008), but since the GK specification is estimated separately (see GK-AR model above), I do not add the dynamic mixture part in the Groen, Paap and Ravazzolo (2013) specification for computational reasons.} In practice, the TVP-BMA specification takes the following form
\begin{eqnarray}
y_{t+h} & = & \sum_{j=1}^{p} x_{jt} s_{j} \beta_{jt} + \varepsilon_{t+h}, \\
\beta_{t} & = & \beta_{t-1} + \eta_{t},
\end{eqnarray}
where $s_{j}$ is an indicator variable such that when $s_{j}=0$ then the $j^{th}$ predictor is removed from the regression in all periods, while when $s_{j}=1$ the predictor is included. The full Bernoulli posterior of each parameter $s_{j}$ is a sequence of zero and one values, such that the posterior mean can be interpreted as a well-defined probability of inclusion in the regression model of each predictor $j$. This probability can be used for model selection or Bayesian model averaging, hence the acronym TVP-BMA for this specification. We assume that the probabilities $s_j$ have a Bernoulli prior with prior inclusion probability of each predictor equal to $0.5$. Given this parameter, all other initial conditions and priors are identical to the TVP-AR case.

\subsection{TVD, Chan et al. (2012)}
The time-varying dimension (TVD) model of Chan et al. (2012) takes the following form
\begin{eqnarray}
y_{t+h} & = & \sum_{j=1}^{p} x_{j,t} s_{j,t} \beta_{j,t} + \varepsilon_{t+h}, \\
\beta_{t} & = & \beta_{t-1} + \eta_{t},
\end{eqnarray}
where $s_{j,t}$ is an indicator variable such that when $s_{j,t}=0$ the $j^{th}$ predictor is removed from the regression model in period $t$ only, and when $s_{j,t}=1$ it is included in the regression. This is a very flexible specification that generalizes the Groen, Paap and Ravazzollo (2013) model to allow a predictor to exit the regression only for certain periods. I have used the default settings and priors suggested by Chan et al. (2012)\footnote{Joshua Chan kindly provided code for their model, which I gratefully acknowledge.}. Following the exact implementations by the authors, also means that for computational reasons, the $s_{j,t}$ are not allowed to index all possible $2^{p}$ models available in each time period. Instead only models with one variable at a time, the full model, and the model with no predictors are estimated and then the optimal model is chosen among this reduced number of specifications. Finally, the authors specify three different ways of specifying a TVD model. The specification used here is the first model presented in that paper. Priors, initial conditions and posterior computation for this first model can be found in Section 1.1 of the online Appendix of Chan et al. (2012).

\subsection{TVS, Kalli and Griffin (2014)}
The time varying sparsity (TVS) model of Kalli and Griffin is of the form
\begin{eqnarray}
y_{t+h} & = & \sum_{j=1}^{p} x_{j,t} \beta_{j,t} + \varepsilon_{t+h}, \\
\beta_{j,t} & = & (1-\alpha_{j})\rho_{j,t} \beta_{j,t-1} + \alpha_j\eta_{j,t},
\end{eqnarray}
where $\rho_{j,t} = \sqrt{\frac{\psi_{j,t}}{\psi_{j,t-1}}}$ and $var\left(\eta_{j,t} \right) = \psi_{j,t}$. In this specification, $\alpha_{j} \in [0,1]$ is a parameter controlling the temporal correlation, and $\psi_{j,t}$ is an autoregressive gamma process. That way, the implied prior for $\beta_{j,t}$ is of normal-gamma autoregressive process form, which generalizes the traditional normal-gamma priors in linear regression. Such priors have very good shrinkage properties, and in the model above the coefficient of each predictor can be shrunk flexibly only in some periods, while be unrestricted in others. Note that these authors specify a Gamma autoregressive process for the variance, instead of the typical stochastic volatility process that all previous methods build upon. All initial conditions are identical to those suggested by the authors. Also following the suggestion of Kalli and Griffin when forecasting inflation (see their Section 5) I set $s^{\ast}=b^{\ast}=0.1$, although it should be acknowledged at this point that their application involves quarterly data instead of the monthly data used in this paper.

\subsection{Bayesian Model Averaging (BMA)}
The BMA approach is the only one in the list of competing specifications that is based on a constant parameter regression with many predictors of the form
\begin{equation}
y_{t+h}  = x_{t} \beta + \varepsilon_{t+h} \label{cons_reg_App}
\end{equation}
where $\beta$ is a $p \times 1$ vector of time-varying parameters and $\varepsilon_{t+h} \sim N \left( 0,\sigma^2 \right)$ with $\sigma^2$ the error variance term. This approach is implemented via the stochastic search variable selection (SSVS) prior of George and McCulloch (1993), with the following hiearchical representation
\begin{subequations} \label{SSVS_prior}
\begin{eqnarray}
p \left( \beta_i |\gamma_i \right)  & \sim & (1-\gamma_i)N\left( 0,\tau_0^2\right) + \gamma_i N\left( 0,\tau_1^2\right), \\
p \left( \gamma_i | \pi \right) & \sim & Bernoulli(\pi),
\end{eqnarray}%
\end{subequations}
where I set $\pi=0.5$, $\tau_0=0.001$ and $\tau_1=4$, and the regression variance parameter $\sigma^2$ has a diffuse prior.

Given these prior settings, the posterior can be obtained by sampling recursively
from the conditional posteriors
\begin{subequations}
\begin{eqnarray}
\beta | \sigma ^{2},\gamma, y & \sim & N\left( \left( x^{\prime }x/\sigma ^{2} + V^{-1}\right) ^{-1}x^{\prime }y/\sigma ^{2},\left( x^{\prime }x/\sigma ^{2} + V^{-1}\right) ^{-1}\right), \\
\gamma_i | \beta_i, \pi, y & \sim & Bernoulli \left( \frac{N(0|\beta_{i},\tau_1^2)\pi}{N(0|\beta_{i},\tau_0^2)(1-\pi) + N(0|\beta_{i},\tau_1^2)\pi} \right), \\
\sigma ^{2}|\beta, y &\sim & iGamma\left( \frac{T}{2},\frac{1}{2} \sum_{t=1}^{T}(y_t-x_t\beta)^2 \right),
\end{eqnarray}
\end{subequations}
where $V$ is a diagonal matrix with $i$-th element $\tau_0^2$ if $\gamma_i=0$ or $\tau_1^2$ if $\gamma_i=1$.

\subsection{The Bayesian lasso}
The full hierarchical representation of the lasso prior\footnote{Note that the Bayesian lasso is only used as a benchmark in the second Monte Carlo exercise, and not in the empirical section (BMA is only used there as a representative method for a constant parameter regression with many predictors).} is 
\begin{subequations}
\label{Lasso_prior}
\begin{eqnarray}
p \left( \beta |\sigma ^{2},\tau _{1}^{2},...,\tau _{p}^{2}\right) &\sim
&N\left( 0,\sigma ^{2}V\right), \\
p \left( \tau _{j}^{2}\right) &\sim & Exponential\left( \frac{\lambda ^{2}}{%
2}\right) \text{, for }j=1,...,p, \\
p \left( \lambda ^{2}\right) &\sim &Gamma\left( r,\delta \right), \\
p \left( \sigma^2 \right) & \propto &  1/\sigma^2,
\end{eqnarray}%
\end{subequations}
where $V=diag\left\{ \tau _{1}^{2},...,\tau _{p}^{2}\right\} $. I follow the recommendations in Park and Casella (2008) and I choose $r=1$ and $\delta=3$.

Given these priors, the posterior can be obtained by sampling recursively
from the conditional posteriors 
\begin{subequations}
\begin{eqnarray}
\beta |\sigma ^{2},\left\{ \tau _{j}^{2}\right\} _{j=1}^{p}, y &\sim
&N\left( \left( x^{\prime }x+V^{-1}\right) ^{-1}x^{\prime }y,\sigma ^{2}\left( x^{\prime }x+V^{-1}\right) ^{-1}\right), \\
\frac{1}{\tau _{j}^{2}}|\beta ,\sigma ^{2},y &\sim &IG\left( \sqrt{\frac{%
\lambda ^{2}\sigma ^{2}}{\beta _{j}^{2}}},\lambda ^{2}\right) \text{, for }%
j=1,...,p, \\
\lambda ^{2}|\beta ,\sigma ^{2},\left\{ \tau _{j}^{2}\right\} _{j=1}^{p},y
&\sim &Gamma\left( p+r,\frac{1}{2}\sum\nolimits_{j=1}^{p}\tau
_{j}^{2}+\delta \right), \\
\sigma ^{2}|\beta ,\left\{ \tau _{j}^{2}\right\} _{j=1}^{p},y &\sim
&iGamma\left( \frac{T-1}{2}+\frac{p}{2},\frac{1}{2}\Psi +\frac{1}{2}\beta
^{\prime }V^{-1}\beta \right),
\end{eqnarray}
\end{subequations}
where $IG$ denotes the Inverse-Gaussian distribution.

\end{appendix}

\end{document}